\documentclass[a4paper,11pt]{article}
\usepackage{jheppub} 
\usepackage{lineno}
\usepackage{tikz}
\usepackage{mathrsfs}
\usepackage{extarrows}

\newcommand{\Sky}{\textbf{Sky}}
\newcommand{\Rep}{\textbf{Rep}}
\newcommand{\QC}{\textbf{QC}}

\newcommand{\tr}{\textrm{tr}}

\newcommand{\CC}{\mathscr{C}}
\newcommand{\Hilb}{\textbf{Hilb}}
\newcommand{\Vect}{\textbf{Vect}}

\usetikzlibrary{decorations.pathmorphing}
\usetikzlibrary{decorations.markings}
\usetikzlibrary{shapes}
\tikzset{line/.style={line width=0.25mm},
curve/.style={line,smooth,tension=1},
->-/.style={decoration={
  markings,
  mark=at position #1 with {\arrow[>=stealth]{>}}},postaction={decorate}},
-<-/.style={decoration={
  markings,
  mark=at position #1 with {\arrow[>=stealth]{<}}},postaction={decorate}},
}

\title{\boldmath Candidate Gaugings of Categorical Continuous Symmetry}

\author[b]{Qiang Jia}
\author[a]{Cheng Ma}
\author[a]{Jiahua Tian}
\affiliation[a]{School of Physics, East China Normal University,\\
500 Dongchuan Road, Shanghai, 200241, China}
\affiliation[b]{Department of Physics, Korea Advanced Institute of Science and Technology,\\
Daejeon, 34141, Korea}

\emailAdd{qjia1993@kaist.ac.kr, jtian1905@gmail.com}

\abstract{Different gaugings of the global symmetry of a quantum field theory are closely related to its various phases. In this work, we study candidate gaugeable symmetries by analyzing candidate Lagrangian algebra data in the Drinfeld center of a symmetry category $\CC^k(G)$ associated to a QFT with continuous global $G$-symmetry and possible 't Hooft anomaly labeled by an integer $k$. We use the combination of the $BF$ theory and the level-$k$ Chern-Simons theory with gauge group $G$ as a semiclassical kernel-theoretic model for the corresponding SymTFT. Under two explicit assumptions, namely that this $BF{+}k$CS theory provides the relevant SymTFT model and that the common $+1$ eigenspaces of the resulting modular kernels detect candidate Lagrangian algebra data in the continuous setting, we derive candidate modular $S$- and $T$-kernels from Hopf-link and framing correlators in $S^3$ semi-classically. We then use these kernels to obtain candidate modular invariants and candidate gaugings. The resulting formulas recover the established cases and suggest a possible extension of this kernel-theoretic picture to compact Lie groups.}

\begin{document}
\maketitle
\flushbottom

\section{Introduction}

The study of the phases of matter has been pursued for a long time. It has long been believed that the phenomenon of phase transition is governed by the Landau paradigm~\cite{landau1937theory, doi:10.1142/0215, HOHENBERG20151}. However, it was noticed that certain aspects of the so-called non-abelian fractional quantum Hall (FQH) liquids cannot be described by the Ginzburg-Landau theory~\cite{Moore:1991ks, Wen:1995qn}. The FQH liquids are instead described by \emph{topological orders}~\cite{Wen:1989iv, Wen:1992vi, Wen:1995qn}, whose key physical data, e.g., the fusions and braidings between the topological lines, are encoded algebraically in the corresponding \emph{modular tensor category} (MTC)~\cite{Moore:1988qv, Kitaev:2005hzj}. The Reshetikhin-Turaev (RT) theory further establishes a bridge between MTC and topological quantum field theory (TQFT)~\cite{Witten:1988hf, RT1991}. Therefore, one is able to study the somewhat abstract algebraic data of MTC through the lens of TQFT, and it would be desirable to understand the classification of the phases of matter via concrete field-theoretical constructions.

It has proven fruitful to view a quantum system $\mathcal{T}$ with (generalized) global $G$-symmetry and 't Hooft anomaly $k$~\cite{Gaiotto:2014kfa} as living on the boundary of a bulk in one-higher dimension~\cite{Kong:2014qka, Kong:2015flk, Kong:2017hcw, Apruzzi:2021nmk, Freed:2022qnc}. The symmetry of $\mathcal{T}$, its anomaly and various gaugings and phases of the symmetry are then encoded in the \emph{symmetry topological field theory} (SymTFT) living in the bulk~\cite{Apruzzi:2021nmk, Schafer-Nameki:2023jdn, Bhardwaj:2023kri}. Furthermore, it has been demonstrated that for finite group $G$ the SymTFT, as an MTC, is equivalent to the \emph{Drinfeld center} $Z(\CC^k(G))$ of the \emph{symmetry category} $\CC^k(G)$ associated to $\mathcal{T}$~\cite{Kong:2015flk, Freed:2022qnc}. The anyon lines are then expected to correspond to various objects of $Z(\CC^k(G))$, and their fusions and braidings correspond to algebraic operations between such objects. Recently, some progress has been made in~\cite{Jia:2025vrj, Stockall:2025ngz} to understand analogous relationships for continuous symmetries based on the seminal work~\cite{Freed:2009qp}. In this work, we adopt the corresponding relation between the SymTFT and $Z(\CC^k(G))$ for continuous symmetries only as a working assumption, in particular when $G$ is an algebraic group or a compact Lie group. Within that framework, candidate anomalies and candidate gaugings --- hence equivalently candidate Lagrangian algebra data --- of $Z(\CC^k(G))$ are expected to be encoded in its algebraic data. Since a SymTFT is, after all, a TQFT, the problem then becomes one of identifying a suitable working TQFT model and extracting from it the physical data that should correspond to algebraic data in the would-be center.

It has been proposed in~\cite{Bonetti:2024cjk, Jia:2025jmn} that the 3d TQFT associated to a 2d QFT $\mathcal{T}$ with anomaly-free global $G$-symmetry for simple and simply-connected compact Lie group $G$ is the $BF$ theory with gauge group $G$~\footnote{In this work we will focus on either $G = U(1)$ or simple and simply-connected compact Lie groups and their complexifications unless explicitly stated otherwise.}. Moreover, when there exists a 't Hooft anomaly labeled by an integer $k\in H^4(BG, \mathbb{Z})$, the TQFT becomes the combination of the $BF$ theory and the level-$k$ Chern-Simons theory~\cite{Jia:2025uun}. In this paper, we use this TQFT as a working model for the SymTFT associated to $Z(\CC^k(G))$. Since this TQFT admits a Lagrangian description, one can attempt a semiclassical evaluation of the relevant physical data, for example, topological correlators of links in $S^3$, and use the resulting kernels to extract candidate information about candidate Lagrangian algebra data in $Z(\CC^k(G))$.

However, it is a nontrivial task to rigorously identify the category associated to the $BF + k\text{CS}$ TQFT. According to the Cobordism Hypothesis~\cite{Baez:1995xq, Lurie:2009keu, Freed:2012hx}, the fundamental datum associated to a TQFT is a fully-dualizable object in a higher category, which is itself a category from which the higher structure can in principle be reconstructed. If the relevant SymTFT is indeed the $BF + k\text{CS}$ theory, then one expects this point-assigned category to realize the symmetry category $\CC^k(G)$. It was proposed in~\cite{Freed:2009qp} that $\CC^k(G)$ is the category of skyscraper sheaves of finite-dimensional $G$-modules with finite support over $G$ for $G = U(1)$, which is denoted by $\Sky(U(1))$. With anomaly $k\in H^4(BU(1), \mathbb{Z})\cong \mathbb{Z}$, the category becomes $\Sky^k(U(1))$ with a twist which geometrically corresponds to a multiplicative bundle gerbe on $U(1)\times U(1)$~\cite{Jia:2025vrj}. One can further compute $Z(\Sky^k(U(1)))$, whose algebraic data matches nicely with the physical data derived from $BF + k\text{CS}$ TQFT for $G = U(1)$~\cite{Freed:2009qp, Jia:2025vrj}. However,  since the finiteness of the support makes $Z(\Sky^k(G))$ extremely small to accommodate all the physically desired anyon lines as its objects~\cite{Henriques:2015xxa, Jia:2025vrj}, one is forced to look for a generalization of $\Sky$ for non-abelian $G$. For algebraic group $G$ it was proposed in~\cite{Jia:2025vrj} and further explored in~\cite{Stockall:2025ngz} that the corresponding category is $\QC^k(G)$, the category of quasi-coherent sheaves of $G$-modules over $G$ twisted by $k\in H^4(BG, \mathbb{Z})$. Therefore, it is natural to expect that the gaugings of the $G$-symmetry of $\mathcal{T}$ should be related to the algebraic data of $Z(\QC^k(G))$. However, we should emphasize that $\QC$ is strictly speaking only well-defined for algebraic $G$, whereas the physical global symmetry group is usually a \emph{real} Lie group. Therefore, the problem of what $\CC$ really is remains unsettled.

Though in this work we will briefly discuss and provide several pieces of evidence on what $\CC$ should be for general $G$, we argue that the kernel-theoretic analysis of candidate Lagrangian algebra data in $Z(\CC^k(G))$ does not really depend on fixing $\CC$ uniquely. Rather, it is the convolution monoidal structure of $\CC$ and the induced monoidal structure on the Drinfeld center that are crucial for our purpose. However, even the analysis of this monoidal structure is a bit too abstract, and we take a more physical approach based on two key assumptions. First, we assume that the SymTFT as a TQFT $\mathcal{Z}$ that assigns to a point the category $\CC^k(G)$ (namely $\mathcal{Z}(\text{pt}) = \CC^k(G)$) is the $BF+k$CS TQFT as we have mentioned earlier. Second we assume that candidate Lagrangian algebra data of $Z(\CC^k(G))$ are detected by modular invariants of the candidate $S$- and $T$-kernels of $Z(\CC^k(G))$, generalizing a known theorem for MTCs~\cite{Ji:2019eqo, Kaidi:2021gbs, You:2023gay}. Since the candidate $S$- and $T$-kernels of $Z(\CC^k(G))$ can be obtained by evaluating the topological correlator between loop operators of $\mathcal{Z}$ on a Hopf link in $S^3$ when $\mathcal{Z}(\text{pt}) = \CC^k(G)$, the above two assumptions allow us to extract candidate invariant data in $Z(\CC^k(G))$ without delving into the categorical details of $\CC$ and its monoidal structure.

In the present paper, we use $\CC^k(G)$ as a \emph{working shorthand} for the category that is actually probed by the kernels and algebra objects studied below, namely, objects are labeled by the conjugacy-class data and centralizer-representation data that enter the $S$- and $T$-matrices (which become integral kernels for continuous $G$), and the only categorical structure used explicitly is the corresponding convolution/fusion behavior encoded by those kernels. We do not attempt here to identify $\CC^k(G)$ with a unique operator-algebraic realization, nor to resolve in this paper the comparison between (Borel or Haar) measurable fields of Hilbert spaces or related analytic models of the same symmetry data. Those model-dependent questions are important, but they are logically separate from the kernel computations and modular-invariant analysis carried out in this work, and will be addressed elsewhere. Accordingly, all statements in the present work should be read as statements about this kernel-theoretic working model and the structures it determines, rather than as depending on a completed categorical equivalence. We will discuss more on this issue in Section~\ref{sec:Symmetry_Category}.

In view of the above two key assumptions, the logical status of the present work is the following. We do not claim here to have established from first principles either that the relevant SymTFT is $BF+k\,CS$ with point assignment $\CC^k(G)$, or that the criterion $S\cdot V=V$, $T\cdot V=V$ continues to characterize candidate Lagrangian algebra data in the non-semisimple continuous setting. Rather, under these assumptions, we derive explicit formulas for the modular kernels and use them to obtain candidate gaugings and modular invariants. The main contribution of this paper is therefore conditional but concrete: we show that this framework is internally consistent, that it reproduces the known examples, and that it yields nontrivial new predictions for general compact Lie groups. Moreover, we adopt the following convention throughout the paper. Whenever we refer to the $S$- and $T$-matrices (or viewing them as integral kernels), the Drinfeld center $Z(\CC^k(G))$, its simple objects, modular invariants, or Lagrangian algebra objects, these terms are to be understood within the working model fixed above unless explicitly stated otherwise. In particular, the genuinely derived statements in this paper are the explicit kernel formulas and the resulting analysis of the modular invariants under the two assumptions stated above. Any interpretation of these data as determining the true symmetry category of an arbitrary QFT, or as establishing a full non-semisimple analogue of the familiar MTC theorems, should be read only as evidence for a broader conjectural picture. Throughout the paper, we will accordingly distinguish between statements that are proved within the assumed framework and statements that should be understood as evidence for the underlying conjectural picture. 

To clarify the scope of novelty relative to recent work, we emphasize that the aim of this paper is not to formulate the general theory of continuous SymTFTs or to prove a full categorical characterization of continuous symmetry categories from first principles; important structural progress in those directions has already appeared in~\cite{Bonetti:2024cjk, Jia:2025jmn, Jia:2025vrj, Stockall:2025ngz}. The concrete advance here is narrower and more explicit. Once the $BF{+}k$CS bulk theory and the modular-invariant criterion are adopted as working assumptions, we derive candidate modular kernels by semiclassical link computations and use them to analyze candidate gaugings in known examples, especially for $U(1)$ and $SU(2)$. In this sense, the contribution of the present paper is the explicit kernel extraction and analysis within the working model, rather than a complete categorical classification.

The work is organized as follows. In section~\ref{sec:Symmetry_Category} we review previous results on continuous symmetry categories and explain the limited kernel-theoretic role that $\CC^k(G)$ will play in this paper. We point out the important subtlety in defining categories such as $\QC^k(G)$ for non-algebraic smooth Lie groups, but we do not attempt to settle that problem here. We then discuss the relation between the SymTFT as a \emph{fully extended} TQFT and the symmetry category, and between loop operators of the SymTFT and the working labels for simple objects in the Drinfeld center. In section~\ref{sec:U(1)_k} we test the framework for $G = U(1)$ with anomaly $k\in H^4(BU(1),\mathbb{Z})$. In section~\ref{sec:Modular_Trans} we derive semiclassical candidate modular $S$- and $T$-kernels by evaluating Hopf-link correlators in the working $BF{+}kCS$ model and we further use these kernels to study candidate invariant-vector data in non-abelian examples, with particular attention to $SU(2)$. We conclude in section~\ref{sec:Conclusion} and point out several directions for future study. Some details of the calculations are presented in the Appendices.

\section{Symmetry Category of Continuous Symmetry and Its Center}\label{sec:Symmetry_Category}

It is known that the symmetry category of continuous $U(1)$ symmetry with anomaly labeled by $k \in H^4(BU(1),\mathbb{Z}) \cong \mathbb{Z}$ is $\Sky^k(U(1))$, the category of skyscraper sheaves of finite-rank modules with finite support on $U(1)$ with convolution tensor product twisted by $k$~\cite{Freed:2009qp, Jia:2025vrj}. The global $U(1)$-symmetry is anomaly-free when $k = 0$. The Drinfeld center of $\Sky^k(U(1))$ is given by:
\begin{equation}
    Z(\Sky^k(U(1))) = \Sky^k(C)\,,
\end{equation}
where the simple objects of $C$ are labeled by $(x,z)\in U(1)\times \mathbb{Z}$. Gauging a sub-symmetry of $U(1)$ is equivalent to identifying a Lagrangian algebra object of $\Sky^k(C)$.

It has been pointed out in~\cite{Henriques:2015xxa} that the category $\Sky$ is not suitable for non-abelian continuous global symmetry due to the finiteness of the support. A natural algebraic replacement is $\QC$, the category of quasi-coherent sheaves of finite-rank modules with convolution tensor product twisted by $k \in H^4(BG,\mathbb{Z})$~\cite{Jia:2025vrj, Stockall:2025ngz}. For compact Lie groups, however, the precise analytic realization of the symmetry category is not yet settled. In the present paper, we therefore use $\CC^k(G)$ only as a working model with the following features: its center has simple labels $([g],\rho)$, its monoidal structure is the convolution fusion structure twisted by $k$, and the modular kernels studied below act on this label set. We will not attempt here to choose uniquely between quasi-coherent, measurable, Hilbert-space-based, or other analytic realizations. Whenever we use notation such as $Z(\CC^k(G))$, simple objects, or Lagrangian algebra objects, it is understood in this restricted sense.

A useful continuous analogue of the finite-group category $\Vect(G)$ is obtained by replacing the direct sum $\bigoplus_{g\in G}V_g$ by a direct integral over a locally compact Hausdorff group $G$ with Haar measure $\mu$:
\begin{equation}\label{eq:Direct_Integral}
    \int_G^\oplus d\mu\ V_g\,.
\end{equation}
These objects form the category of measurable fields of Hilbert spaces over $G$ where each $V_g$ is a Hilbert space~\cite{yetter2005measurable}. The convolution monoidal structure is then
\begin{equation}\label{eq:Convolution_Direct_Integral}
    (V * W)_k = \int_G^\oplus d\mu\ V_{g}\otimes W_{g^{-1}k}\,,
\end{equation}
which is the direct-integral analogue of the convolution tensor product in $\Vect(G)$. We emphasize this convolution structure because it is the common ingredient that survives across the different realizations of the would-be symmetry category, and it is the only structural input from section~\ref{sec:Symmetry_Category} that will be used in the later kernel computations.

To study candidate gaugings associated to categorical continuous symmetry corresponding to a group $G$, we look for candidate commutative connected algebra data in the center $Z(\CC^k(G)) = \CC^{k}_G(G)$ with $k\in H^4(BG,\mathbb{Z})$. For the present kernel-theoretic analysis, there is no essential difference between keeping $\CC^k(G)$ heuristically as $\QC^k(G)$ or as $\Hilb^k(G)$, because later sections use only the common convolution structure~\footnote{The reason that we use the notation $\CC^{k}_G(G)$ is because in these cases the Drinfeld center is the category of $G$-equivariant quasi-coherent sheaves on $G$, or the category of $G$-equivariant fields of Hilbert spaces over $G$. We will leave a detailed investigation of $\CC^{k}_G(G)$ in future work.}. When $k=0$, this reduces to the direct-integral convolution~(\ref{eq:Convolution_Direct_Integral}). For non-zero $k$, the convolution tensor product is twisted by a multiplicative bundle gerbe $\mathscr{M}$ on $G\times G$:
\begin{equation}\label{eq:Convolution_tensor_product}
    \mathscr{F}_1 * \mathscr{F}_2 = m_*(p_1^*\mathscr{F}_1 \otimes p_2^*\mathscr{F}_2 \otimes \mathscr{M})
\end{equation}
for $\mathscr{F}_i\in \CC^k(G)$~\cite{waldorf2017transgressive, Jia:2025vrj}. Depending on the realization, $\mathscr{F}_i$ may be quasi-coherent sheaves or measurable/Hilbert modules, but the same convolution idea underlies both pictures. This is the structure that later descends to the candidate modular kernels of the center.

Though analyzing the center directly is difficult, there is a simpler physical route. By the Cobordism Hypothesis, an $n$-dimensional fully extended TQFT $\mathcal{Z}$ corresponds to an $n$-category~\cite{Baez:1995xq, Lurie:2009keu}. It is therefore natural to expect that the SymTFT associated to a QFT with global $G$-symmetry and 't Hooft anomaly $k$ satisfies $\mathcal{Z}(\text{pt})=\CC^k(G)$ and $\mathcal{Z}(S^1)=\CC^k_G(G)$~\footnote{This can also be seen from the perspective of factorization homology~\cite{amabel2021differential, Costello:2021jvx}. For $\CC = \QC$ this is discussed in~\cite{Ben-Zvi:2008vtm, ben2012morita}.}. In the quasi-coherent picture one has
\begin{equation*}
    \QC^k_G(G) \cong \QC^k([G/G]) \cong \bigsqcup_{[g]\in Cl(G)} \Rep^k(C_G(g))\,,
\end{equation*}
so a simple object is labeled by the pair $([g],\rho)$ where $\rho$ is an irreducible representation of $C_G(g)$. The same label set is expected in the measurable/Hilbert picture by the direct-integral generalization of the familiar finite-group computation; we defer that derivation to future work~\cite{fiveguys_HilbG}. Accordingly, throughout the rest of the paper we take pairs $([g],\rho)$ as the labels for simple objects in the Drinfeld center (for development on this issue see~\cite{fiveguys_HilbG}). If $\mathcal{Z}(S^1)$ is indeed identified with that center, then in physical terms the TQFT $\mathcal{Z}$ should contain loop operators $W_{[g], \rho}(\ell)$ labeled by $([g],\rho)$ for a loop $\ell\subset M$. From now on, we focus on the 3d SymTFT associated to a 2d QFT and use the following correlator on $M = S^3$ as the candidate $S$-kernel of the model:
\begin{equation}\label{eq:TQFT_Corr}
    \begin{split}
        S^{(k)}_{([g],\sigma),([h],\rho)} &= \int \mathcal{D}\Phi\ e^{i\int_{S^3} S_{\mathcal{Z}}[\Phi]}\ W_{[g],\sigma}(\ell_1)\ W_{[h],\rho}(\ell_2)\,,
    \end{split}
\end{equation}
for a Hopf link $\ell_1\cup \ell_2 \subset S^3$ where $S_{\mathcal{Z}}[\Phi]$ is the action functional of $\mathcal{Z}$. In the present framework, the correlator $S^{(k)}_{([g],\sigma),([h],\rho)}$ serves as the candidate $S$-kernel of $\CC^k_G(G)$. One can also consider the ``self-link'' correlator of a single $W_{[g],\sigma}(\ell)$ by calculating its correlation with a slight deformation of it; this serves as the candidate $T$-kernel of $\CC^k_G(G)$~\footnote{For an MTC, these correlators define the usual $S$- and $T$-matrices~\cite{turaev2016quantum, bakalov2001lectures}. For a continuous $G$, however, $\CC^k_G(G)$ is expected to be neither finite nor semisimple. We nevertheless use the same TQFT correlators as candidate continuous kernels and then ask whether they satisfy analogues of the modular relations. For developments on TQFTs associated to non-semisimple categories, see~\cite{Lyubashenko:1994tm, lyubashenko1995modular, DeRenzi:2019iwu, Hofer:2024cnu}.}.

The computation of~(\ref{eq:TQFT_Corr}) becomes accessible once a suitable Lagrangian description of the TQFT model is chosen. It was shown in~\cite{Bonetti:2024cjk, Jia:2025jmn} that when $k = 0$ the SymTFT is a $G$-gauge $BF$-theory. For non-zero $k$, the $G$-symmetry is anomalous, and the anomaly is captured by a level $k$ Chern-Simons theory via rearrangements of topological defects~\cite{Jia:2025uun}. This suggests that we consider the TQFT with action $\int\text{tr}B\wedge F + \frac{k}{4\pi}\text{CS}[A]$, where $\text{CS}[A]$ is the Chern-Simons action. Thus we will need to compute the correlator
\begin{equation}\label{eq:BFkCS_Corr}
    S^{(k)}_{([g],\sigma),([h],\rho)} = \int \mathcal{D}B\mathcal{D}A\ e^{i\int_{S^3} \text{tr} B\wedge F + \frac{k}{4\pi}\text{CS}[A]}\ W_{[g],\sigma}(\ell_1)\ W_{[h],\rho}(\ell_2)\,,
\end{equation}
in order to obtain candidate $S$- and $T$-kernels associated to $\CC^k_G(G)$.

In the above reasoning, we have effectively made the important assumption that $\mathcal{Z}(\text{pt}) = \CC^k(G)$ when $\mathcal{Z}$ is the $BF+k$CS TQFT. There is evidence in favor of this identification. First, recall that the action of a 3d Dijkgraaf-Witten theory for $G = \mathbb{Z}_n$ with twist $k$ is~\cite{Kapustin:2014gua}
\begin{equation}
    S = \frac{in}{2\pi} \int B\wedge dA + \frac{ik}{4\pi} \int A\wedge dA\,,
\end{equation}
which is a discrete $BF + k$CS TQFT assigning to a point the category $\Vect^k(G)$~\cite{Dijkgraaf:1989pz, RT1991}. This strongly suggests that a continuous analogue of the same theory should assign to a point a convolution-type category such as $\CC^k(G)$. Second, in $\mathcal{Z}_{BF+k\text{CS}}$ one can indeed define loop operators labeled by the pair $([g], \rho)$ as above~\cite{Cattaneo:2002tk, Cordova:2022rer, Jia:2025jmn}. We use these facts as evidence for the identification $\mathcal{Z}_{BF+k\text{CS}}(\text{pt})=\CC^k(G)$, rather than as a theorem. Assuming this identification, in section~\ref{sec:Modular_Trans} we compute candidate $S$- and $T$-kernels using~(\ref{eq:BFkCS_Corr}) for general $G$ and check that they reduce to the known cases in~\cite{Freed:2009qp, Jia:2025vrj}.

The candidate $S$- and $T$-kernels are crucial for the study of the gaugings of the categorical symmetry. For an MTC, a Lagrangian algebra object in it can be written as $\mathcal{L} = \bigoplus v_i s_i$ where $s_i$'s are simple objects of the MTC labeled by $i$. Define the column vector $V_{\mathcal{L}}$ with components $v_i$, we have~\cite{Ji:2019eqo, Kaidi:2021gbs, You:2023gay}:
\begin{equation}\label{eq:SV=TV=V}
    S\cdot V_{\mathcal{L}} = V_{\mathcal{L}}\,,\quad T\cdot V_{\mathcal{L}} = V_{\mathcal{L}}\,.
\end{equation}
Thus for an MTC, a Lagrangian algebra object corresponds to a common eigenvector of the $S$- and $T$-matrices with eigenvalue $1$. Motivated by this theorem, for $\CC^k_G(G)$ we study the analogous common $+1$ eigenspace of the continuous kernels. A candidate Lagrangian algebra datum can then be written as
\begin{equation}\label{eq:L_Direct_Integral}
    \mathcal{L} = \int^\oplus_{Cl(G)} d\mu' \bigoplus_{\rho\in \text{Irrep}(C_G(g))} v_{[g], \rho} W_{[g], \rho}
\end{equation}
with the measure $d\mu'$ on $Cl(G)$ and simple objects $W_{[g], \rho}$ labeled by the pair $([g], \rho)$. Hence in this case the condition~(\ref{eq:SV=TV=V}) can be written concretely as
\begin{equation}\label{eq:ST_Direct_Integral}
    \int_{Cl(G)}^\oplus d\mu' \bigoplus_{\rho\in \text{Irrep}(C_G(g))} S_{([h], \sigma), ([g], \rho)} v_{[g], \rho} = \int_{Cl(G)}^\oplus d\mu' \bigoplus_{\rho\in \text{Irrep}(C_G(g))} T_{([h], \sigma), ([g], \rho)} v_{[g], \rho} = v_{[h],\sigma}\,.
\end{equation}
A rigorous proof that~(\ref{eq:SV=TV=V}) continues to characterize candidate Lagrangian algebra data for non-semisimple $\CC^k_G(G)$ is not presently available. Nevertheless, the expressions~(\ref{eq:L_Direct_Integral}) and~(\ref{eq:ST_Direct_Integral}) are precisely the direct-integral analogues of the familiar MTC formulas, and in the examples studied later, they reproduce the expected candidate data. We therefore use~(\ref{eq:SV=TV=V}) as a working criterion for candidate Lagrangian algebra data in the continuous setting.

\section{SymTFT for anomalous $U(1)$ symmetry and $\widehat{u(1)}_k$ modular invariance}\label{sec:U(1)_k}

In this section, we review the SymTFT data for the 2d $U(1)$ symmetry with an anomaly labeled by $k\in H^4(BU(1),\mathbb{Z})$ \cite{Jia:2025vrj, Brennan:2024fgj, Antinucci:2024zjp}, and discuss the resulting candidate Lagrangian algebra data and the corresponding topological boundary states on the torus within the working criterion adopted in this paper. As an application, we study 2d compact bosons at radius $R$, whose partition functions are built from the $\widehat{u(1)}_k$ characters when $R^2$ is rational. We then compare the resulting candidate data with the modular invariants of the $\widehat{u(1)}_k$ algebra and find agreement in this case.

\subsection{3d SymTFT for anomalous $U(1)$ and candidate boundary data}

The simple line operators $W_{(x,n)}$ of the 3d SymTFT of $U(1)$ symmetry with anomaly $k\in H^4(BU(1),\mathbb{Z})$ are labeled by a pair $(x,n)$ with $x=e^{i\theta_x}\in U(1)$ $(\theta_x\in[0,2\pi))$ and $n\in \mathbb{Z}$, subject to the equivalency $(\theta_x,n)\sim (\theta_x+2\pi,n+2k)$. The fusion rule is
    \begin{equation}
        W_{(x,n)}\times W_{(y,n')} = W_{\left(xy,n_1+n_2- \frac{2k}{2\pi}(\theta_x+\theta_y-[\theta_x+\theta_y])\right)}\,,
    \end{equation}
and the candidate $S$- and $T$-kernels are
\begin{equation}\label{eq:U(1)-modular-matrices}
S_{(x,n),(y,n')}=e^{-i(n\theta_y+n'\theta_x-\frac{2k}{2\pi}\theta_x\theta_y)}\,, \quad
T_{(x,n),(y,n')}=\delta_{2\pi} (\theta_x-\theta_y)\delta_{n,n'}\exp\left(i\left(n\theta_x-\frac{k\theta_x^2}{2\pi}\right)\right)\,,
\end{equation}
where $\delta_{2\pi} (\theta_x-\theta_y)$ is defined as
    \begin{equation}
        \delta_{2\pi} (\theta_x-\theta_y)= \sum_{n\in \mathbb{Z}} \delta(\theta_x-\theta_y+2\pi n)\,.
    \end{equation}
In the familiar finite or semisimple setting, a symmetry boundary is determined by a Lagrangian algebra. In the present model we therefore organize the corresponding candidate symmetry-boundary data by the following universal object
    \begin{equation}\label{eq:LDir}
        \mathcal{L}_{\textrm{Dir}} = \bigoplus_n W_{(1,n)}\,,
    \end{equation}
for any $k$, and the corresponding symmetry boundary supports the (anomalous) $U(1)$ symmetry. There also exist
    \begin{equation}
        \mathcal{L}^k_q = \bigoplus_m \bigoplus_{n=0}^{q-1} W_{\left(e^{\frac{2\pi i}{q}n},\frac{k}{q}n+qm\right)}\,,
    \end{equation}
for any positive integer $q$ satisfying $\frac{k}{q}\in \mathbb{Z}$. One can check the coefficient vectors $V_{\mathcal{L}}$ are invariant under the action of $S/T$-matrices
    \begin{equation}
        S\cdot V_{\mathcal{L}}=V_{\mathcal{L}}\,,\quad T\cdot V_{\mathcal{L}}=V_{\mathcal{L}}\,,
    \end{equation}
for $\mathcal{L}=\mathcal{L}_{\rm Dir}$ and $\mathcal{L}=\mathcal{L}^k_q$, see App.~\ref{app:check-of-Lagrangian-algebra-U(1)} for the proof.

In the following, we will put the SymTFT on $T^2\times[0,1]$ with the symmetry boundary $B_{\textrm{sym}}:T^2\times \{0\}$ and physical boundary $B_{\textrm{phys}}:T^2 \times \{1\}$. The $S$-matrix implies the commutation relations of line operators
    \begin{equation}\label{eq:U(1)-commutator-of-lines}
        W_{(x,n)} [\Gamma_2] W_{(y,n')}[\Gamma_1] = e^{-i(n\theta_y+n'\theta_x-\frac{2k}{2\pi}\theta_x\theta_y)}W_{(y,n')}[\Gamma_1]W_{(x,n)} [\Gamma_2]\,,
    \end{equation}
where $\Gamma_1,\Gamma_2$ are temporal and spatial circles of $T^2$, respectively. The algebra spans an infinite-dimensional Hilbert space $\mathcal{H}$ on $T^2$, and the two boundaries $B_{\rm sym}$ and $B_{\rm phys}$ can be described by state vectors in $\mathcal{H}$.

\subsubsection*{Symmetry boundary for $\mathcal{L}_{\textrm{Dir}}$}

Let us begin with the vacuum $|0,0\rangle$ of the symmetry boundary corresponding to $\mathcal{L}_{\textrm{Dir}}$, such that all line operators in $\mathcal{L}_{\textrm{Dir}}$ act trivially
    \begin{equation}
        W_{(1,n)}[\Gamma_1] |0,0\rangle = W_{(1,n)}[\Gamma_2]|0,0\rangle=|0,0\rangle\,.
    \end{equation}
Then $W_{(x,0)}$ operators serve as $U(1)$ generators and we define the generic state $|\theta_r,\theta_s\rangle$ with spatial and temporal holonomy $(\theta_r,\theta_s)$ as
    \begin{equation}\label{eq:U(1)-holonomy-states}
        |\theta_r,\theta_s\rangle \equiv e^{-2\pi i k \frac{\theta_s}{2\pi} \frac{ \theta_r}{2\pi}}W_{(x_r,0)}[-\Gamma_1] W_{(x_s,0)}[\Gamma_2] |0,0\rangle\,,
    \end{equation}
with $x_r = e^{i \theta_r}, x_s = e^{i\theta_s}$. It is easy to check they are the eigenstates of $W_{(1,n)}$ operators
    \begin{equation}
        W_{(1,n)}[\Gamma_1]|\theta_r,\theta_s\rangle= e^{2\pi i n \frac{\theta_s}{2\pi}} |\theta_r,\theta_s\rangle\,,\quad W_{(1,n)}[\Gamma_2]|\theta_r,\theta_s\rangle = e^{2\pi i n \frac{\theta_r}{2\pi}}|\theta_r,\theta_s\rangle\,,
    \end{equation}
and they transform under the modular transformation
    \begin{equation}
        S: \left(\begin{array}{c}
            \Gamma_1\\ \Gamma_2
        \end{array} \right)\rightarrow \left(\begin{array}{c}
            -\Gamma_2\\ \Gamma_1
        \end{array} \right)\,,\quad T: \left(\begin{array}{c}
            \Gamma_1\\ \Gamma_2
        \end{array} \right)\rightarrow \left(\begin{array}{c}
            \Gamma_1+\Gamma_2\\ \Gamma_2
        \end{array} \right)\,,
    \end{equation}
according to
    \begin{equation}\label{eq:U(1)-boundary-states-modular-transformation}
        \hat{S} |\theta_r,\theta_s\rangle = |-\theta_s,\theta_r\rangle\,,\quad \hat{T}|\theta_r,\theta_s\rangle = |\theta_r,\theta_s-\theta_r\rangle\,,
    \end{equation}
where the proof can be found in App.~\ref{app:check-of-Lagrangian-algebra-U(1)}. The anomaly is reflected by the additional phase factors when shifting the holonomies by $2\pi$
    \begin{equation}\label{anomalous-phase-state}
        |\theta_r+2\pi,\theta_s\rangle = e^{2\pi i k  \frac{\theta_s}{2\pi}} |\theta_r,\theta_s\rangle\,,\quad |\theta_r,\theta_s+2\pi\rangle = e^{-2\pi i k \frac{\theta_r}{2\pi}} |\theta_r,\theta_s\rangle\,.
    \end{equation}
We further impose the orthogonal relation
    \begin{equation}
        \langle \theta_{r'},\theta_{s'}| \theta_r,\theta_s\rangle = \delta_{2\pi}( \theta_{r'}-\theta_r) \delta_{2\pi}(\theta_{s'}-\theta_s)\,.
    \end{equation}

\subsubsection*{Symmetry boundary for $\mathcal{L}^k_q$ and $\mathbb{Z}_q$-gauging}

Although the global symmetry $U(1)$ is anomalous, the subgroup $\mathbb{Z}_q\in U(1)$ with $q$ a divisor of $k$ is still anomaly-free. In particular, if we set 
    \begin{equation}
        \theta_r = \frac{2\pi}{q} r\,,\quad \theta_s = \frac{2\pi}{q}s\,,\quad (r,s\in \mathbb{Z}_q)\,,
    \end{equation}
the anomalous phases in \eqref{anomalous-phase-state} vanish, and we have
    \begin{equation}
        |\frac{2\pi}{q}(r+q),\frac{2\pi}{q}s\rangle = |\frac{2\pi}{q}r,\frac{2\pi}{q}(s+q)\rangle\,.
    \end{equation}
The vacuum state associated with the candidate boundary datum $\mathcal{L}_q^k$ is defined as
    \begin{equation}
        |\Omega\rangle_{q}^k = \frac{1}{q} \sum_{r,s\in \mathbb{Z}_q} |\frac{2\pi}{q}r , \frac{2\pi}{q}s\rangle\,,
    \end{equation}
where the summation over $\mathbb{Z}_q$ is unambiguous due to the vanishing of anomalous phases. One can check all line operators in $\mathcal{L}_q^k$ act trivially
    \begin{equation}
        W_{\left(e^{\frac{2\pi i}{q}n},\frac{k}{q}n+qm\right)}[\Gamma_1] |\Omega\rangle_q^k = W_{\left(e^{\frac{2\pi i}{q}n},\frac{k}{q}n+qm\right)}[\Gamma_2] |\Omega\rangle_q^k = |\Omega\rangle_q^k\,,
    \end{equation}
for any $n\in \mathbb{Z}_q,m\in\mathbb{Z}$. 

\subsubsection*{Physical boundary and partition vector $|\chi\rangle$}

For any 2d theory $\mathfrak{T}$ whose global symmetry is $U(1)$ with anomaly labeled by $k$, denote $Z[\theta_r,\theta_s]$ as its partition function on $T^2$ with the background $U(1)$ holonomies $(\theta_r,\theta_s)$. The physical boundary $B_{\rm phys}$ is characterized by the partition vector $|\chi\rangle$
    \begin{equation}\label{eq:physical-boundary-state}
        |\chi\rangle = \int_0^{2\pi} d\theta_r \int_0^{2\pi} d\theta_s Z[\theta_r,\theta_s] |\theta_r,\theta_s\rangle\,,
    \end{equation}
where we assume
\begin{equation}\label{eq:partition-function-anomaly}
        Z[\theta_r+2\pi,\theta_{s}] = e^{-2\pi i k \frac{\theta_s}{2\pi}} Z[\theta_r,\theta_s]\,,\quad Z[\theta_r,\theta_{s}+2\pi] = e^{2\pi i k \frac{\theta_r}{2\pi}} Z[\theta_r,\theta_s]\,,
    \end{equation}
under the shift of holonomies, so that $|\chi\rangle$ is well-defined since the anomalous phase of $Z[\theta_r,\theta_s]$ exactly compensates with that of the state vector $|\theta_r,\theta_s\rangle$\footnote{Physically, the anomaly of the boundary theory on $B_{\rm phys}$ is canceled by the inflow from the bulk SymTFT.}. The modular transformation acting on $|\chi\rangle$ reads
    \begin{equation}
        \begin{split}
        \hat{S} |\chi\rangle =&  \int_0^{2\pi} d\theta_r \int_0^{2\pi} d\theta_s Z[\theta_s,-\theta_r] |\theta_r,\theta_s\rangle\,,\\
        \hat{T} |\chi\rangle =&  \int_0^{2\pi} d\theta_r \int_0^{2\pi} d\theta_s Z[\theta_r,\theta_s+\theta_r] |\theta_r,\theta_s\rangle\,,
        \end{split}
    \end{equation}
which implies the modular transformations of $Z[\theta_r,\theta_s]$
\begin{equation}
    T: Z[\theta_r,\theta_s]\rightarrow Z[\theta_r,\theta_s+\theta_r]\,,\quad S:Z[\theta_r,\theta_s]\rightarrow Z[\theta_s,-\theta_r]\,.
\end{equation}

The partition function $Z[\theta_r,\theta_r]$ is trivially obtained by projecting $|\chi\rangle$ onto $|\theta_r,\theta_s\rangle$
    \begin{equation}
        Z[\theta_r,\theta_s]=\langle \theta_r,\theta_s|\chi\rangle\,.
    \end{equation}
Switching symmetry boundary to $|\Omega\rangle_q^k$, we obtain other partition functions
    \begin{equation}
        Z_q^k = \,_q^k\langle\Omega| \chi\rangle = \frac{1}{q} \sum_{m,n\in \mathbb{Z}_q} Z\left[\frac{2\pi}{q}m,\frac{2\pi}{q}n \right]\,,
    \end{equation}
which corresponds to gauging an anomaly-free $\mathbb{Z}_q$ subgroup of $U(1)$.

\subsection{$\widehat{u(1)}_k$ characters and modular invariants}


The 2d compact boson theory with radius $R$ has a momentum $U(1)_n$ symmetry and a winding $U(1)_m$ symmetry, with a mixed 't Hooft anomaly between them. As shown in \cite{Lin:2021udi} and reviewed below, by choosing an appropriate $U(1) \in U(1)_n \times U(1)_m$, the mixed 't Hooft anomaly can induce an arbitrary anomaly labeled by $k\in \mathbb{Z}$ on $U(1)$. This makes the compact boson a natural arena for the application of the $U(1)$ SymTFT model. We focus on rational $R^2$ and show that the resulting candidate Lagrangian algebra data are in one-to-one correspondence with the modular invariants of the $\widehat{u(1)}_k$ algebra within the criterion adopted in this paper.

\subsubsection*{Compact boson and $\widehat{u(1)}_k$ characters}
We begin with a review of $\widehat{u(1)}_k$ characters. Consider the boson compactified on a circle of radius $R$ with partition function\cite{DiFrancesco:1997nk}
    \begin{equation}
        Z_{\rm bos}(R) = \sum_{m,n\in \mathbb{Z}} \frac{1}{|\eta(q)|^2} q^{\frac{1}{2} \left(\frac{n}{R}+\frac{mR}{2} \right)^2} \bar{q}^{\left(\frac{n}{R}-\frac{mR}{2} \right)^2}\,.
    \end{equation}
with $q=e^{2\pi i \tau}$ and $n,m$ are momentum and winding numbers. Define the character $\chi_{n,m}$ as
    \begin{equation}
        \chi_{n,m}(q) = \frac{1}{\eta(q)} q^{\frac{1}{2} \left(\frac{n}{R}+\frac{mR}{2} \right)^2}\,.
    \end{equation}
Suppose $R^2$ is a rational number parametrized by
    \begin{equation}
        R= \sqrt{\frac{2p'}{p}}\,,
    \end{equation}
with $p,p'>0$ and $\gcd(p,p')=1$. Introduce the new variables
    \begin{equation}
        \begin{split}
            n= 2p' n' +r\,,\quad (0\leq r \leq 2p'-1\,, n'\in \mathbb{Z})\\
            m=2pm'+s\,,\quad (0\leq s \leq 2p-1\,, m'\in\mathbb{Z})
        \end{split}
    \end{equation}
and setting
    \begin{equation}
        u=n'+m'\,,\quad l=pr+p's\,,
    \end{equation}
the characters $\chi_{m,n}$ are relabeled by the pair $u,l$ and becomes
    \begin{equation}
        \chi_{u,l}(q)=\frac{1}{\eta(q)} q^{k \left(u+\frac{l}{2k} \right)^2}\,,
    \end{equation}
with the level $k =pp'$ and the range of $l$ can be restricted within $0\leq l<2k$. The $\widehat{u(1)}_k$ character is defined as
    \begin{equation}
        \chi^{(k)}_{l} (q) = \sum_{u\in \mathbb{Z}}\chi_{u,l}(q)=\sum_{u\in \mathbb{Z}}\frac{1}{\eta(q)} q^{k \left(u+\frac{l}{2k} \right)^2}\,.
    \end{equation}
We will begin with the partition function written as the diagonal modular invariant
    \begin{equation}\label{eq:U(1)k-modular-invariant}
        Z_{\textrm{bos}}(R=\sqrt{2k})=\sum_{l=0}^{2k-1} \chi^{(k)}_{l}(q) \overline{\chi^{(k)}_l (q)}\,,
    \end{equation}
which corresponds to the compact boson with radius $R=\sqrt{2k}=\sqrt{2p'p}$. The partition function for $R=\sqrt{2p'/p}$ is obtained via gauging as shown below.

\subsubsection*{Anomalous $U(1)$ symmetry}
The compact boson has a momentum $U(1)_n$ symmetry and a winding $U(1)_m$ symmetry, generated by the currents\cite{Lin:2021udi}
\begin{equation}
    \begin{gathered}
        \textrm{Momentum $U(1)_n$}: \quad J_n (z)=\frac{iR}{\sqrt{2}} \partial X(z)\,,\quad \bar{J}_n(z) =\frac{i R}{\sqrt{2}}\bar{\partial} X(\bar{z})\,, \\
        \textrm{Winding $U(1)_m$}: \quad J_m (z)=\frac{i\sqrt{2}}{R} \partial X(z)\,,\quad \bar{J}_m(z) =-\frac{i\sqrt{2}}{R}\bar{\partial} X(\bar{z})\,.
    \end{gathered}
\end{equation}
The $U(1)_n$ and $U(1)_m$ symmetries are each non-anomalous by themselves, but they have a mixed anomaly. For a pair of coprime integers $(a,b)$, we can defined the compact $U(1)_{a,b}$ subgroup generated by the Noether currents
    \begin{equation}
        \begin{gathered}
            J_{a,b}(z)=a J_n(z) + bJ_{m}(z)=\left( \frac{aR}{\sqrt{2}}+\frac{b\sqrt{2}}{R}\right)i \partial X(z)\,,\\
            \bar{J}_{a,b}(z)=a \bar{J}_n(z) - b\bar{J}_{m}(z)=\left( \frac{aR}{\sqrt{2}}-\frac{b\sqrt{2}}{R}\right)i \bar{\partial} X(\bar{z})\,,
        \end{gathered}
    \end{equation}
and its anomaly is given by
    \begin{equation}
        k_{U(1)} = \frac{1}{2} \left(\frac{\left( \frac{aR}{\sqrt{2}}+\frac{b\sqrt{2}}{R}\right)^2}{2} + \frac{\left( \frac{aR}{\sqrt{2}}-\frac{b\sqrt{2}}{R}\right)^2}{2}\right) = ab\,.
    \end{equation}
For $R=\sqrt{2p'/p}$, if we choose $a=p,b=p'$ then
    \begin{equation}
        J_{p,p'}(z) = 2\sqrt{pp'} i\partial X(z)\,,\quad J_{p,p'}(\bar{z})=0\,,
    \end{equation}
where the anti-holomorphic part vanishes and the anomaly is $k_{U(1)} = pp'=k$. For the operator $\mathcal{O}_{n,m}(0,0)$ carrying momentum and winding number $(n,m)$, the OPE reads
    \begin{equation}
        i\partial X(z) \mathcal{O}_{n,m}(0,0) \sim \frac{\left(\frac{\sqrt{2}n}{R}+\frac{mR}{\sqrt{2}} \right)}{2z} \mathcal{O}_{n,m}(0,0)\,,\quad i \bar{\partial} X(\bar{z}) \mathcal{O}_{n,m}(0,0)\sim \frac{\left(\frac{\sqrt{2}n}{R}-\frac{mR}{\sqrt{2}} \right)}{2z} \mathcal{O}_{n,m}(0,0)\,,
    \end{equation}
the $J_{p,p'}(z)$ current gives
    \begin{equation}
        J_{p,p'}(z) \mathcal{O}_{n,m}(0,0)\sim \frac{2\sqrt{pp'} \times\frac{1}{\sqrt{2}}\left(\frac{n}{R}+\frac{mR}{2} \right)}{z}\mathcal{O}_{n,m}(0,0)=\frac{2k(u+\frac{l}{2k})}{z}\mathcal{O}_{u,l}(0,0)\,,
    \end{equation}
where we change the labels to $u,l$ as we did before. We see the $U(1)_{p,p'}$ charge of the operator $\mathcal{O}_{u,l}$ is $2k(u+\frac{l}{2k})$. In the following, we will omit the subscripts of $U(1)_{p,p'}$ and denote it as $U(1)$.

\subsubsection*{Partition functions and modular invariants}

To apply the SymTFT method to the holomorphic $U(1)$, we will turn on the background holonomy of $U(1)$ and define the partition function
\begin{equation}\label{eq:U(1)-partition-function}
    Z[\theta_r,\theta_s] = e^{-2\pi i k \frac{\theta_r}{2\pi} \frac{\theta_s}{2\pi}} \sum_{l=0}^{2k-1}\frac{1}{|\eta(\tau)|^2}\Theta^{(k)}_l\left(\begin{array}{c}
            \theta_r\\\theta_s
        \end{array} \right) \overline{\Theta}^{(k)}_l\left(\begin{array}{c}
            0\\0
        \end{array} \right)\,,
\end{equation}
with $Z[0,0]=Z_{\rm bos}(R=\sqrt{2k})$ in \eqref{eq:U(1)k-modular-invariant}, and we introduce the theta functions
    \begin{equation}\label{eq:Theta-function}
        \Theta^{(k)}_l\left(\begin{array}{c}
            \theta_r\\ \theta_s
        \end{array} \right) (\tau) = \sum_{u\in \mathbb{Z}} e^{2\pi i k\left(u+\frac{l}{2k} + \frac{\theta_r}{2\pi} \right)^2\tau+ 4\pi i k \left(u+\frac{l}{2k} + \frac{\theta_r}{2\pi}\right) \frac{\theta_s}{2\pi}}\,,  
    \end{equation}
with $\theta_s,\theta_r\in [0,2\pi)$ are the $U(1)$ holonomies along temporal and spatial directions. One can read off the $U(1)$ charge $q_{u,l}$ from the exponent of $Z[\theta_r,\theta_s]$ that depends on $\theta_s$
    \begin{equation}
        q_{u,l}(\theta_r) = 2k \left(u+\frac{l}{2k} \right)+ k \times \frac{\theta_r}{2\pi}\,,
    \end{equation}
where the first term is the integer-valued charge of the operator $\mathcal{O}_{u,l}$, and the non-integer-valued shift $k \times \frac{\theta_r}{2\pi}$ is because $U(1)$ acts projectively in the twist sectors when $k\neq 0$. 

One can check the modular properties for the $\Theta$-functions
    \begin{equation}\label{eq:Theta-function-modular-transformation}
        \begin{gathered}
        \Theta^{(k)}_l\left(\begin{array}{c}
            \theta_r\\ \theta_s
        \end{array} \right) (\tau+1) = e^{\frac{\pi i l^2}{2k}} e^{-2\pi i k \left(\frac{\theta_r}{2\pi}\right)^2} \Theta^{(k)}_l\left(\begin{array}{c}
            \theta_r\\ \theta_s+\theta_r
        \end{array} \right) (\tau)\,,\\
        \Theta^{(k)}_l\left(\begin{array}{c}
            \theta_r\\ \theta_s
        \end{array} \right) (-\frac{1}{\tau}) = \sqrt{\frac{-i\tau}{2k}} e^{4\pi i k \frac{\theta_s}{2\pi} \frac{\theta_r}{2\pi}}\sum_{l'}  e^{-2\pi i \frac{l l'}{2k}} \Theta^{(k)}_{l'}\left(\begin{array}{c}
            \theta_s\\ -\theta_r
        \end{array} \right) (\tau)\,,
        \end{gathered}
    \end{equation}
and the $2\pi$-shift of holonomies
    \begin{equation}
        \Theta^{(k)}_l\left(\begin{array}{c}
            \theta_r+2\pi\\ \theta_s
        \end{array} \right) (\tau) = \Theta^{(k)}_l\left(\begin{array}{c}
            \theta_r\\ \theta_s
        \end{array} \right) (\tau)\,,\quad \Theta^{(k)}_l\left(\begin{array}{c}
            \theta_r\\ \theta_s+2\pi
        \end{array} \right) (\tau) = e^{4\pi i k \frac{\theta_r}{2\pi}}\Theta^{(k)}_l\left(\begin{array}{c}
            \theta_r\\ \theta_s
        \end{array} \right) (\tau)\,.
    \end{equation}
Furthermore, the partition function $Z[\theta_r,\theta_s]$ is modular covariant
    \begin{equation}\label{modular-transformation-ZU(1)_k}
        Z[\theta_r,\theta_s](\tau+1) = Z[\theta_r,\theta_s+\theta_r](\tau)\,, \quad Z[\theta_r,\theta_s](-1/\tau) = Z[\theta_s,-\theta_r](\tau)\,,
    \end{equation}
and we also have

\begin{equation}\label{anomalous-phase-ZU(1)_k}
        Z[\theta_r+2\pi,\theta_{s}] = e^{-2\pi i k \frac{\theta_s}{2\pi}} Z[\theta_r,\theta_s]\,,\quad Z[\theta_r,\theta_{s}+2\pi] = e^{2\pi i k \frac{\theta_r}{2\pi}} Z[\theta_r,\theta_s]\,,
    \end{equation}
where the addition phases reflect the anomaly of $U(1)$. We collect the proofs of modularities in App.~\ref{app:Theta-functions}.

Notice that the anomalous phases due to the shift of holonomies are the same as \eqref{eq:partition-function-anomaly}, therefore we can use the partition function \eqref{eq:U(1)-partition-function} to construct the partition vector $|\chi\rangle$ in \eqref{eq:physical-boundary-state} and obtain other partition functions by projecting $|\chi\rangle$ onto the topological boundary states associated with the candidate boundary data.

To compare the results in \cite{Gepner:1986hr}, let us introduce
    \begin{equation}
        Z_{r,s} = Z\left[\frac{2\pi}{k}r,\frac{2\pi}{k}s \right] = e^{-2\pi i \frac{r s}{k}} \sum_{l=0}^{2k-1} \frac{1}{|\eta(\tau)|^2}\Theta^{(k)}_l\left(\begin{array}{c}
            \frac{2\pi r}{k}\\ \frac{2\pi s}{k}
        \end{array} \right) \overline{\Theta}^{(k)}_l\left(\begin{array}{c}
            0\\0
        \end{array} \right)\,,
    \end{equation}
and define
    \begin{equation}
        \Theta_{l,k}(\tau) = \sum_{u\in \mathbb{Z}+l/2k} e^{2\pi i k u^2 \tau}\,,
    \end{equation}
then one has
    \begin{equation}
        Z_{(r,s)} = e^{2\pi i \frac{r s}{k}} \sum_{l=0}^{2k-1} \frac{1}{|\eta(\tau)|^2}  e^{2\pi i \frac{ls}{k}}\Theta_{l+2r,k} \overline{\Theta}_{l,k}=e^{-2\pi i \frac{rs}{k}} \sum_{l=0}^{2k-1} \frac{1}{|\eta(\tau)|^2} e^{2\pi i \frac{ls}{k}}\Theta_{l,k} \overline{\Theta}_{l-2r,k}\,,
    \end{equation}
where we shift $l\rightarrow l-2r$. The projection of $|\chi\rangle$ on $ \,_q^k\langle\Omega| \chi\rangle$ gives
    \begin{equation}
        Z_q^k = \,_q^k\langle\Omega| \chi\rangle = \frac{1}{q} \sum_{m,n\in \mathbb{Z}_q} Z_{\left(\frac{k}{q}m,\frac{k}{q}n\right)}\,.
    \end{equation}
In fact, it can be written as
\begin{equation}
    Z_q^k = \sum_{\substack{\gamma\\q|\gamma ,\gamma|k}} T_{\gamma}\,, \quad \textrm{with} \quad T_{\gamma} = \sum_{A\in \Gamma_k} Z_{A(0,\gamma)}\,,
\end{equation}
where $\Gamma_k$ is the finite modular group $\rm SL(2,\mathbb{Z})/\Gamma(k)$ with 
    \begin{equation}
        \Gamma(k)=\left.\left\{ \left( \begin{array}{cc}
            a&b\\c&d
        \end{array} \right)\in \Gamma \right| a=d=1\ \textrm{mod}\ k\,, c=d=0\ \textrm{mod}\ k \right\}\,,
    \end{equation}
which, as discussed in \cite{Gepner:1986hr,Gannon:1996hp}, gives a full classification of the modular invariants of $\widehat{u(1)}_k$ algebra. Therefore, within the working criterion adopted here, the candidate Lagrangian algebra data of the anomalous $U(1)$ SymTFT are in one-to-one correspondence with the modular invariants of $\widehat{u(1)}_k$.

To be concrete, consider the partition function $Z^k_{p}$ written as
    \begin{equation}
        Z_p^k = \frac{1}{p} \sum_{r,s\in \mathbb{Z}_p} e^{-\frac{2\pi i}{p}\left(p'r\right)s}\sum_{l=0}^{2k-1}\frac{1}{|\eta(\tau)|^2}\Theta^{(k)}_l\left(\begin{array}{c}
            \frac{2\pi r}{p}\\\frac{2\pi s}{p}
        \end{array} \right) \overline{\Theta}^{(k)}_l\left(\begin{array}{c}
            0\\0
        \end{array} \right)\,,
    \end{equation}
with $pp'=k$, where 
\begin{equation}
        \Theta^{(k)}_l\left(\begin{array}{c}
            \frac{2\pi r}{p}\\ \frac{2\pi s}{p}
        \end{array} \right) (\tau) = \sum_{u\in \mathbb{Z}} e^{2\pi i k\left(u+\frac{l}{2k} + \frac{p' r}{k} \right)^2\tau+ \frac{2\pi i}{p} \left(2ku+l + 2p' r\right)s}\,.  
    \end{equation}
Summing over $s$ will restrict $l$ to be
    \begin{equation}
        l = p s'-p' r\,, \quad (s'=0,\cdots,2p'-1)\,,
    \end{equation}
for given $r=0,\cdots,p-1$. Then the partition function is
    \begin{equation}
        Z_q^k = \frac{1}{2} \sum_{r=0}^{2p-1}\sum_{s'=0}^{2p'-1}\frac{1}{|\eta(\tau)|^2} \sum_{u,u'} e^{2\pi i k \left(u + \frac{ps'+p'r}{2k} \right)^2\tau} e^{-2\pi i k \left(u + \frac{ps'-p'r}{2k} \right)^2\bar{\tau}}\,,
    \end{equation}
which is the partition function of the compact boson of radius $R=\sqrt{2p/p'}$\cite{Thorngren:2021yso}.

\section{The $S$ and $T$ Matrices for General Categorical Continuous Symmetries}\label{sec:Modular_Trans}

In section~\ref{sec:U(1)_k}, we found that the candidate Lagrangian algebra data of the $U(1)$ SymTFT with level $k$ are in one-to-one correspondence with the modular invariants of $\widehat{u(1)}_k$ within the working criterion adopted in this paper. It is then natural to ask whether an analogous picture persists for a general compact Lie group $G$ with anomaly $k\in H^4(BG,\mathbb{Z})$. Addressing that question requires a detailed study of the $k$-twisted SymTFT for $G$, in particular of the modular $S$- and $T$-kernels (which become integral kernels for continuous $G$) and of the resulting invariant-vector problem.

Before turning to the explicit kernel computations, let us make the logical status of the results in this section precise. Once the two assumptions stated in the introduction are granted, the derivation of the Hopf-link and framing kernels from the $BF{+}kCS$ path integral is a calculation within the model described above. By contrast, the subsequent interpretation of common $+1$ eigenvectors of $S$ and $T$ as candidate Lagrangian algebra data, and hence as candidate gaugings, remains conditional on the extension of the familiar semisimple MTC criterion to the present continuous and non-semisimple setting. Finally, any claim that these kernel computations determine the true categorical continuous symmetry of an arbitrary QFT with compact Lie-group symmetry should be read only as evidence for that broader conjectural picture.

In previous work \cite{Jia:2025vrj}, candidate $S$- and $T$-kernels for $G=SU(2)$ and $k=0$ were proposed by generalizing the finite-group formulas. Here we take a step forward and derive semiclassical candidate modular kernels by explicitly evaluating the topological correlator of the Hopf link in $S^3$, for generic simple line operators and arbitrary Lie group $G$ with $k\in H^4(BG,\mathbb{Z})$. We present explicit expressions for these kernels and check several key consistency properties. We then specialize to $G=SU(2)$ and $k=0$ and study the resulting candidate invariant-vector data under the action of the kernels.

The derivation in this section should be read as a semiclassical reduction rather than as a measure-theoretically complete path-integral evaluation. We formally integrate out the $B$ field, reduce the remaining path integral to flat connections with prescribed holonomy data, and then evaluate the Chern-Simons phase on a singular-flat-connection ansatz adapted to the link complement. This procedure leads to explicit formulas for the candidate kernels and reproduces the expected expressions in previously understood cases, but several analytic steps are used at a formal level. We therefore present the resulting formulas as semiclassical candidate modular kernels for the model introduced above.

More specifically, the discussion below should be read as consisting of three separate steps. First, we formally reduce the path integral to flat-connection data on the Hopf-link complement after formally integrating out the $B$ field. Second, we replace the loop insertions by the corresponding holonomy and character data and evaluate the Chern-Simons phase on a singular-flat-connection ansatz. Namely, we fix representatives $e^{i\alpha}$ and $e^{i\beta}$ for the two holonomies, parametrize the residual gauge freedom by an orbit variable $x\in G$ (modulo the relevant stabilizers), and evaluate the Chern-Simons phase on a singular-flat-connection ansatz adapted to these boundary holonomies. Third, in the regular sector we pass from the resulting group integral to a Weyl-orbit expression by a formal localization argument. At each of these steps, normalization factors coming from measures, orbit volumes and Jacobians are treated only to the extent needed to write down the candidate kernels and compare them with known cases and a more complete analytic treatment of these factors is left for future work. The resulting formulas should be understood as explicit reduced-moduli-space expressions in a semiclassical sense.

Recall from section~\ref{sec:Symmetry_Category} that, within the model adopted here, the relevant SymTFT action is taken to be:
    \begin{equation}
        S_{BF} = \int_{M_3} d^3x  \,\textrm{tr}(B\wedge F) + \frac{k}{4\pi} CS[A]\,,
    \end{equation}
where $CS[A]$ is the 3d Chern-Simons term. The Wilson and 't Hooft (Gukov-Witten) loop operators are, respectively~\cite{Cattaneo:2002tk, Jia:2025jmn}:
    \begin{equation}\label{eq:Wilson-and-tHooft-operator}
        W_{\sigma} =  \textrm{tr}_{\sigma}\mathcal{P} \exp \left(i \int_{\ell} A \right)\,, \quad U_{[g]}= \int_G dx\, U_{xgx^{-1}}\,,
    \end{equation}
Here $U_g$ is defined as a twist operator, which impose the holonomy of $G$ surrounding $U_g$ to be $g$. The integral $\int_G dx$ over $x\in G$ reflects that we need to sum over all group elements in the conjugacy class $[g]$, so that $U_{[g]}$ is gauge invariant and depends on the entire conjugacy class $[g]$ rather than a specified group element. A generic loop operator $W_{[g],\sigma}$ is labeled by the pair $([g],\sigma)$, for a given conjugacy class $[g]\in Cl(G)$ and an irreducible representation $\sigma \in \mathbf{Rep}(C_G([g]))$ of the corresponding centralizer group $C_{G}([g])$. In particular, we write $W_{[e],\sigma}=W_{\sigma}$ and $W_{[g],\textrm{id}}=U_{[g]}$ where ${\rm id}$ denotes the trivial representation.

\subsection{Hopf link and modular $S$ and $T$-kernels}\label{sec:Modular_ST_Calculation}

\paragraph{$S$-kernel}

\begin{figure}[h]
\begin{equation}
    \begin{tikzpicture}
        \draw[thick,-<-=.4] (0,0) arc (45:360+30:1);
        \draw[thick,-<-=.4] (0.15,-1.415) arc (45+180:360+30+180:1);
        \node at (-2.8,-0.5) {$W_{[g],\sigma}(\ell_1)$};
        \node at (2.8,-0.5) {$W_{[h],\rho}(\ell_2)$};
\end{tikzpicture}\nonumber
\end{equation}
\caption{The candidate $S$-kernel is evaluated via a pair of loop operators linked with each other.}
\label{Fig_Hopf_Link}
\end{figure}
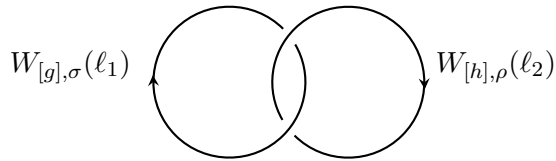

It will be useful to have an expression of the $S$-matrix for general Lie groups. For this, recall that by definition, we have
\begin{equation}\label{eq:Def_S}
    S^{(k)}_{([g],\sigma),([h],\rho)} = \int \mathcal{D}B \mathcal{D}A\ e^{i\int_{S^3} \textrm{tr}BF + \frac{k}{4\pi}\text{CS}[A]}\ W_{[g],\sigma}(\ell_1)\ W_{[h],\rho}(\ell_2)\,,
\end{equation}
where $\ell_1\cup \ell_2$ is a Hopf link in $S^3$ as shown in Figure~\ref{Fig_Hopf_Link},
and $\sigma\in \Rep(C_G([g]))$, $\rho\in \Rep(C_G([h]))$. The ``Wilson-'t Hooft'' operator in~(\ref{eq:Def_S}) is defined as:
\begin{equation}
    W_{[g],\sigma}(\ell_1) = \int_G dx\ \tr_\sigma \mathcal{P}\exp\left(i \int_{\ell_1} A \right)\times U_{xgx^{-1}}\,,
\end{equation}
where $A$ should be understood as the gauge field restricted to the Lie algebra of $C_G(xgx^{-1})$. Thus the $S$-kernel is:
\begin{equation}
    \begin{split}
        S^{(k)}_{([g],\sigma),([h],\rho)} &= \int \mathcal{D}B \mathcal{D}A\ e^{i\int_{S^3} \textrm{tr}BF + \frac{k}{4\pi}\text{CS}[A]}\\
        &\quad \times \int dxdy\ \tr_\sigma\mathcal{P} e^{i\int_{\ell_1} A} U_{\text{Ad}_x g}(\ell_1) \ \tr_\rho\mathcal{P} e^{i\int_{\ell_2} A} U_{\text{Ad}_y h}(\ell_2)\,,
    \end{split}
\end{equation}
for $(x,y)\in G\times G$. To evaluate the above integral, we will integrate out $B$ and replace $U_{xgx^{-1}},U_{ygy^{-1}}$ as boundary conditions of holonomies along $\ell_2$ and $\ell_1$:
\begin{equation}\label{eq:S_kernel_1}
    \begin{split}
        S^{(k)}_{([g],\sigma),([h],\rho)} &= \int dxdy \ \int_{\text{Hol}_A(\ell_2)= \text{Ad}_x g,\text{Hol}_A(\ell_1)=\text{Ad}_y h}\mathcal{D}A\ \delta(F) \\
        &\qquad \times e^{i\int_{S^3} \frac{k}{4\pi}\text{CS}[A]}\ \tr_\sigma\mathcal{P} e^{i\int_{\ell_1} A} \ \tr_\rho\mathcal{P} e^{i\int_{\ell_2} A}\,.
    \end{split}
\end{equation}
We note that in the above integral, we have exchanged the order of integration since the holonomy constraints set the boundary condition for $A$, and physically we are simply summing over all contributions from configurations with allowed boundary conditions. It is also worth noting that in the above formulation, the non-compactness of $B$ is crucial, since otherwise integrating out $B$ fails to lead to a simple $\delta$-function.

The flatness connection $F=0$ actually implies
\begin{equation}\label{eq:Commuting_Hols}
    \text{Hol}_{A}(\ell_1) \times \text{Hol}_{A}(\ell_2) = \text{Hol}_{A}(\ell_2) \times \text{Hol}_{A}(\ell_1)\,,
\end{equation}
so that a Hopf link imposes commuting monodromies. The reason is as follows. Consider the boundary of a small tubular region of $\ell_1$, which is a torus with two $1$-cycles $\gamma_1$ and $\gamma_2$. We choose $\gamma_1$ to be parallel with $\ell_1$, and $\gamma_2$ is the small circle linking $\ell_1$, which is homotopic to $\ell_2$. The holonomies along $\gamma_1$ and $\gamma_2$ are separately $\text{Hol}_A (\gamma_1) = \text{Hol}_A (\ell_1)$ and $\text{Hol}_A (\gamma_2) = \text{Hol}_A (\ell_2)$.
Since the combination $\gamma_1 \gamma_2 \gamma_1^{-1} \gamma_{2}^{-1}$ is homotopic to a trivial 1-cycle, the holonomy along $\gamma_1 \gamma_2 \gamma_1^{-1} \gamma_{2}^{-1}$ must be trivial due to the flatness condition, which further implies \eqref{eq:Commuting_Hols}.

At this stage, we can evaluate the Wilson loops and obtain
\begin{equation}
    \begin{split}
        S^{(k)}_{([g],\sigma),([h],\rho)} &= \int dxdy \ \int_{\text{Hol}_A(\ell_2)= \text{Ad}_x g,\text{Hol}_A(\ell_1)=\text{Ad}_y h}\mathcal{D}A\ \delta(F) \\
        &\qquad \times e^{i\int_{S^3} \frac{k}{4\pi}\text{CS}[A]}\ \chi^*_{\sigma(\text{Ad}_x g)}\left(\text{Ad}_y h\right)\ \chi^*_{\rho(\text{Ad}_y h)}\left(\text{Ad}_x g\right)\,,
    \end{split}
\end{equation}
where we write $\sigma(\text{Ad}_x g),\rho(\text{Ad}_y h)$ to make explicit the dependence of the representations on the corresponding centralizer groups $C_G(\text{Ad}_x g)$ and $C_G(\text{Ad}_y h)$. To simplify the integral, notice that a constant gauge transformation acting by conjugation
\begin{equation}
    \text{Hol}_{A}(\ell_1) \rightarrow z \text{Hol}_{A}(\ell_1) z^{-1}\,, \quad \text{Hol}_{A}(\ell_2) \rightarrow z \text{Hol}_{A}(\ell_2) z^{-1}\,,
\end{equation}
leave both the characters\footnote{
The character is invariant due to the identity $\chi_{\rho(h_1)} (h_2) = \chi_{\rho(k h_1 k^{-1})} (k h_2 k^{-1})$ for any $k\in G$ with $h_1h_2=h_2h_1$.
} and the Chern-Simons term $\textrm{CS}$ invariant. Therefore we can write
\begin{equation}\label{eq:Def_S_hol}
    \begin{split}
        S^{(k)}_{([g],\sigma),([h],\rho)} &= \int dxdy \ \int_{\text{Hol}_A(\ell_2)=  g,\text{Hol}_A(\ell_1)=\text{Ad}_{x^{-1}y} h}\mathcal{D}A\ \delta(F) \\
        &\qquad \times e^{i\int_{S^3} \frac{k}{4\pi}\text{CS}[A]}\ \chi^*_{\sigma(g)}\left(\text{Ad}_{x^{-1}y} h\right)\ \chi^*_{\rho(\text{Ad}_{x^{-1}y} h)}\left(g\right)\\
        &= \left(\int dx \right) \int du \ \int_{\text{Hol}_A(\ell_2)=  g,\text{Hol}_A(\ell_1)=\text{Ad}_{u} h}\mathcal{D}A\ \delta(F) \\
        &\qquad \times e^{i\int_{S^3} \frac{k}{4\pi}\text{CS}[A]}\ \chi^*_{\sigma(g)}\left(\text{Ad}_{u} h\right)\ \chi^*_{\rho(\text{Ad}_{u} h)}\left(g\right)\\
        &= \int du \ \int_{\text{Hol}_A(\ell_2)=  g,\text{Hol}_A(\ell_1)=\text{Ad}_{u} h}\mathcal{D}A\ \delta(F) \\
        &\qquad \times e^{i\int_{S^3} \frac{k}{4\pi}\text{CS}[A]}\ \chi^*_{\sigma(g)}\left(\text{Ad}_{u} h\right)\ \chi^*_{\rho(\text{Ad}_{u} h)}\left(g\right)\\
        &= \int du \ \int_{\text{Hol}_A(\ell_2)=  g,\text{Hol}_A(\ell_1)=\text{Ad}_{u} h}\mathcal{D}A\ \delta(F) \\
        &\qquad \times e^{i\int_{S^3} \frac{k}{4\pi}\text{CS}[A]}\ \chi^*_{\sigma(g)}\left(\text{Ad}_{u} h\right)\ \chi^*_{\rho( h)}\left(\text{Ad}_{u^{-1}} g\right)\,,
    \end{split}
\end{equation}
where we set $\text{Vol}(G) = 1$. Given the commuting pair $(g,\text{Ad}_u h)$, one can always use the constant gauge transformation to conjugate $g$ into the fundamental chamber of the Cartan torus, namely $g\in T/W$, where $W$ denotes the Weyl group of $G$. Similarly, we can also set $h\in T/W$ using the $u$-integral. Without loss of generality, we will always choose $g,h\in T/W$ as the representative of $[g],[h]$.

To further evaluate the \eqref{eq:Def_S_hol}, we expand $A$ perturbatively as $A = A_0 + \epsilon \widetilde{A}$, where $A_0$ is a saddle point solution of $F=0$ satisfying the boundary conditions. We consider the semi-classical approximation and discard the loop contributions from $\mathcal{D} \widetilde{A}$. The delta function $\delta(F)$ should be replaced by the $\delta(g (\text{Ad}_u h) g^{-1} (\text{Ad}_u h)^{-1})$, which is a Dirac delta function on the group $G$, imposing the commuting of holonomies. After that, we have
\begin{equation}\label{eq:S_Simplify}
    S^{(k)}_{([g],\sigma),([h],\rho)} = \int_G du \ \delta(g (\text{Ad}_u h) g^{-1} (\text{Ad}_u h)^{-1}) \,e^{i\int_{S^3} \frac{k}{4\pi}\text{CS}[A]}\ \chi^*_{\sigma(g)}\left(\text{Ad}_{u} h\right)\ \chi^*_{\rho( h)}\left(\text{Ad}_{u^{-1}} g\right)\,.
\end{equation}
To proceed, recall that for an integral of the form:
\begin{equation}
    I = \int_X d\mu\ \delta(f(x)) F(x)\,,
\end{equation}
we have:
\begin{equation}
    I = \int_{f^{-1}(0)} d\mu_{f^{-1}(0)}\ \frac{1}{|J_f(x)|} F(x)\,.
\end{equation}
Therefore,~(\ref{eq:S_Simplify}) can be further evaluated to be:
\begin{equation}\label{eq:S_Simplify_2}
    S^{(k)}_{([g],\sigma),([h],\rho)} = \int_{\mathcal{M}(g,h)}  \ \frac{d\mu_{\mathcal{M}}(u)}{|J(u)|}   
    \,e^{i\int_{S^3} \frac{k}{4\pi}\text{CS}[A]}\ \chi^*_{\sigma(g)}\left(\text{Ad}_{u} h\right)\ \chi^*_{\rho( h)}\left(\text{Ad}_{u^{-1}} g\right)\,,
\end{equation}
where $J(u)$ is the Jacobian associated to $f(u) := g (\text{Ad}_u h) g^{-1} (\text{Ad}_u h)^{-1}$ and
    \begin{equation}
    \mathcal{M}(g,h) := \{u\in G|  = g (\text{Ad}_u h) g^{-1} (\text{Ad}_u h)^{-1} = e\}\,,
\end{equation}
is a submanifold in $G$.

We will analyse the structure of $\mathcal{M}(g,h)$. Recall that we have chosen $g,h \in T/W$. Given any $x\in \mathcal{M}(g,h)$, since $\text{Ad}_x h \in C_G(g)$, we can always consider a $C_G(g)$ conjugation to set $\text{Ad}_x h \in T$, where $W_g$ is the Weyl group of $C_G(g)$, and $T$ is also the Cartan torus of $C_G(g)$. In other words, we can write $x = c_g x'$ for certain $c_g \in C_G(g)$ and $\text{Ad}_{x'} h \in T$. Since $\text{Ad}_{x} h \in T$ implise $\text{Ad}_{x} h$ commute with all elements in $T$, then $x^{-1} T x$ should commute with $h$ and must be a maximal torus in $C_G(h)$. Since both $x^{-1}Tx$ and $T$ are maximal torus of $C_G(\beta)$, they must be conjugated to each other by an element of $C_G(\beta)$ since $G$ is compact semi-simple. Thus there exists $c_h\in C_G(h)$ such that $c_h T c_h^{-1} = x^{-1} T x$, or equivalently:
\begin{equation}
	c_h T c_h^{-1} = x'^{-1} c_g^{-1} T c_g x' \Rightarrow \text{Ad}_{c_g x' c_h} T = T \,.
\end{equation}
This further implies that $c_g x' c_h \in N_G(T)$, the normalizer of $T$ in $G$. Thus for any $x' \in \mathcal{M}(g,h)$ we have:
\begin{equation}
	x' \in C_G(g) \cdot N_G(T) \cdot C_G(h)\,.
\end{equation}
Since $W = N_G(T)/T$, we can write any $n\in N_G(T)$ as $n = w t$ for $w \in W$ and $t \in T$, where $w$ should be understood as a representative of the coset $wT$. Since $T\in C_G(h)$ we have:
\begin{equation}
	\mathcal{M}(g,h) = C_G(g)\cdot W\cdot C_G(h).
\end{equation}
It is clear that for any $w \in W$ we have:
\begin{equation}
	C_G(g)\cdot w\cdot C_G(h) = C_G(g)\cdot w_g w w_h \cdot C_G(h)
\end{equation}
for $w_g \in N_{C_G(g)}(T)/T$ and $w_h \in N_{C_G(h)}(T)/T$ since $w_g$ and $w_h$ can be absorbed into $C_G(g)$ and $C_G(h)$, respectively. Therefore, we have:
\begin{equation}
	\mathcal{M}(g,h) = \bigsqcup_{w\in W_g \backslash W/W_h} C_G(g)\cdot w \cdot C_G(h) := \bigsqcup_{w\in W_g\backslash W/W_h} Z_w
\end{equation}
where we have defined $Z_w := C_G(g)\cdot w \cdot C_G(h)$ for each $w\in W_g\backslash W/W_h$. Given the structure of $\mathcal{M}(g,h)$, \eqref{eq:S_Simplify_2} becomes
\begin{equation}
\begin{split}
    S^{(k)}_{([g],\sigma),([h],\rho)} =& \sum_{w\in W_g\backslash W/W_h}\int_{Z_w}  \ \frac{d\mu_{\mathcal{M}}(u)}{|J(u)|}   
    \,e^{i\int_{S^3} \frac{k}{4\pi}\text{CS}[A]}\ \chi^*_{\sigma(g)}\left(\text{Ad}_{u} h\right)\ \chi^*_{\rho( h)}\left(\text{Ad}_{u^{-1}} g\right)\,,\\
    =&\sum_{w\in W_g\backslash W/W_h}\int_{Z_w}  \ \frac{d\mu_{\mathcal{M}}(u)}{|J(u)|}   
    \,e^{i\int_{S^3} \frac{k}{4\pi}\text{CS}[A]}\ \chi^*_{\sigma}\left(\text{Ad}_{w} h\right)\ \chi^*_{\rho}\left(\text{Ad}_{w^{-1}} g\right)\,,
\end{split}
\end{equation}
where we drop the label of centralizer in the representation $\rho$ and $\sigma$, and stick to the choice that $\rho$ is a representation of $C_G(h)$ and $\sigma$ is a representation of $C_G(g)$ throughout.

To evaluate the Chern-Simons phase, we need to write down the expression of gauge field $A$ satisfying the boundary condition. However, for a given commuting pair $(g,\text{Ad}_u h)$, the choice of gauge fields is not unique due to the compactness of $G$. For example, the central element $-1\in SU(2)$ can be expressed as $e^{\frac{2\pi i}{2}\sigma_a}$ for any Pauli matrix $\sigma_a$, and more generally it is clear that the exponential map from $\mathfrak{g}$ to $G$ is not injective. To proceed, we will adopt the following ansatz for $A$. In the sector $Z_w = C_G(g)\cdot w \cdot C_G(h)$, we can apply a constant gauge transformation to set $(g,\text{Ad}_u h)$ to the representative $(g,\text{Ad}_w h)$, we then choose $A$ to be
\begin{equation}
    A = \alpha \delta^{(1)}_{D_1}+ \text{Ad}_w\beta \delta^{(1)}_{D_2}\,,
\end{equation}
which produces holonomies $g,\text{Ad}_w h$ around $\ell_2$ and $\ell_1$. Here we choose $\alpha,\beta \in \mathfrak{t}$
and $\partial D_i = \ell_i$, namely $D_i$ is the disk bounded by $\ell_i$, and $\delta^{(1)}_{D_i}$ is the Poincar\'e dual of $D_i$. Using the above ansatz, we have: 
\begin{equation}
    F = dA +iA\wedge A = \alpha \delta_{\ell_1}^{(2)} + \text{Ad}_w\beta \delta_{\ell_2}^{(2)}\,,
\end{equation}
where $\delta^{(2)}_{\ell_i} = d \delta^{(1)}_{D_i}$ is the Poincar\'e dual of the boundary $\ell_i$ and $A\wedge A=0$ due to $[\alpha,\text{Ad}_w\beta]=0$, one has
\begin{equation}
    AdA = \alpha \text{Ad}_w \beta (\delta^{(1)}_{D_1}  \wedge \delta_{l_2}^{(2)} + \delta_{D_2}^{(1)} \wedge \delta^{(2)}_{l_1})\,.
\end{equation}
and the integral of CS evaluates to
\begin{equation}
\begin{split}
    \int_{S^3} \text{CS} =& \tr(\alpha  \text{Ad}_w\beta) \int_{S^3}(\delta^{(1)}_{D_1}  \wedge \delta_{l_2}^{(2)} + \delta_{D_2}^{(1)} \wedge \delta^{(2)}_{l_1}) \\ =& \tr(\alpha \text{Ad}_w\beta) (\ell_1\cdot_{S^3} D_2 + \ell_2\cdot_{S^3} D_1) = -2\tr(\alpha \text{Ad}_w\beta)\,,
\end{split}
\end{equation}
where $A\wedge A\wedge A = 0$ since $A\wedge A=0$, and the minus sign is related to the orientation we choose in Figure~\ref{Fig_Hopf_Link}.

With the above ansatz for the Chern-Simons term, we arrive at
\begin{equation}
\begin{split}
    S^{(k)}_{([g],\sigma),([h],\rho)}
    =&\sum_{w\in W_g\backslash W/W_h} \left(\int_{Z_w}  \ \frac{d\mu_{\mathcal{M}}(u)}{|J(u)|}\right)   
    \,e^{-\frac{ik}{2\pi}\tr(\alpha \text{Ad}_{w}\beta)}\ \chi^*_{\sigma}\left(\text{Ad}_{w} h\right)\ \chi^*_{\rho}\left(\text{Ad}_{w^{-1}} g\right)\,.
\end{split}
\end{equation}
To calculate $|J(u)|$, let $u = e^{i \sum_a \epsilon_a \eta_a} u_0$ where $u_0\in \mathcal{M}(g,h)$ and $\eta_a$ are the generators in the normal direction $N_{\mathcal{M}(\alpha,\beta)}$ at $u_0$ of the tangent space of $\mathcal{M}(\alpha,\beta)$ in $\mathfrak{g}$. Denote
    \begin{equation}
        f(\epsilon_a) = g e^{i \sum_a \epsilon_a \eta_a}(\text{Ad}_{u_0} h)e^{-i \sum_a \epsilon_a \eta_a}g^{-1} e^{i \sum_a \epsilon_a \eta_a}(\text{Ad}_{u_0} h)^{-1}e^{-i \sum_a \epsilon_a \eta_a}\,,
    \end{equation}
and we have $f(0)=e$. When $\epsilon_a$ are small, we can expand $f(\epsilon_a)$ using the Lie algebra $\mathfrak{g}$. Notice that $e^{sX}Ye^{-sX} = Y+s[X,Y]+\mathcal{O}(s^2)$, we have
\begin{equation}
\begin{split}
    f(\epsilon_a) =&e+ i g [\sum_a\epsilon_a \eta_a,\text{Ad}_{u_0}h]g^{-1}(\text{Ad}_{u_0} h)^{-1}+ig(\text{Ad}_{u_0} h)g^{-1}[\sum_a\epsilon_a \eta_a,\text{Ad}_{u_0}h^{-1}]\\
    =& e -i (1-\text{Ad}_g)(1-\text{Ad}_{\text{Ad}_{u_0}h}) \circ (\sum_a \epsilon_a \eta_a)\,.
\end{split}
\end{equation}
At each $u_0$, we can extend $g,\text{Ad}_{u_0}h$ to a Cartan torus $T_{u_0}$, and all $T_{u_0}$ are isomorphic to $T_{w}$ by $C_G(g)$ conjugation. We note that on each root plane $P_\gamma := (X_{\gamma}\oplus X_{-\gamma})\cap\mathfrak{g}|_{\mathbb{C}}$ for a positive root $\gamma\in \Lambda_+$, we have:
\begin{equation}
    \text{ad}_\theta X_\gamma = \gamma(\theta) X_\gamma,\ \text{ad}_\theta X_{-\gamma} = -\gamma(\theta)X_{-\gamma}\,.
\end{equation}
Since we choose $g=e^{i \alpha},h=e^{i\beta}$ with both $\alpha,\beta\in \mathfrak{t}$, we have:
\begin{equation}
    \text{Ad}_g|_{P_\gamma} = \begin{pmatrix}
        e^{i\gamma(\alpha)} & 0 \\
        0 & e^{-i \gamma(\alpha)}
    \end{pmatrix}\,,\quad \text{Ad}_{\text{Ad}_wh}|_{P_\gamma} = \begin{pmatrix}
        e^{i\gamma(\text{Ad}_w\beta)} & 0 \\
        0 & e^{-i \gamma(\text{Ad}_w\beta)}
    \end{pmatrix}\,.
\end{equation}
Hence, we have
\begin{equation}
    |J(u)| = \prod_{\substack{\gamma\in\Lambda_+\\ e^{i\gamma(\alpha)},e^{i\gamma(\text{Ad}_w\beta)}\neq 1}}|(1- e^{i\gamma(\alpha)})(1- e^{-i\gamma(\alpha)})(1-e^{i\gamma(\text{Ad}_w\beta)})(1-e^{-i\gamma(\text{Ad}_w\beta)})|\,.
\end{equation}
Thus the $S$-kernal can be further written as
\begin{equation}\label{eq:S_kernel_final}
\begin{split}
    &S^{(k)}_{([g],\sigma),([h],\rho)}
    \\=&\sum_{w\in W_g\backslash W/W_h} \frac{\text{Vol}(Z_w)  
    \,e^{-\frac{ik}{2\pi}\tr(\alpha \text{Ad}_{w}\beta)}\ \chi^*_{\sigma}\left(\text{Ad}_{w} h\right)\ \chi^*_{\rho}\left(\text{Ad}_{w^{-1}} g\right)}{\prod_{\substack{\gamma\in\Lambda_+\\ e^{i\gamma(\alpha)},e^{i\gamma(\text{Ad}_w\beta)}\neq 1}}|(1- e^{i\gamma(\alpha)})(1- e^{-i\gamma(\alpha)})(1-e^{i\gamma(\text{Ad}_w\beta)})(1-e^{-i\gamma(\text{Ad}_w\beta)})|} \\
    =& \sum_{w\in W_g\backslash W/W_h} \frac{\text{Vol}(Z_w)  
    \,e^{-\frac{ik}{2\pi}\tr(\alpha \text{Ad}_{w}\beta)}\ \chi^*_{\sigma}\left(\text{Ad}_{w} h\right)\ \chi^*_{\rho}\left(\text{Ad}_{w^{-1}} g\right)}{ 16 \prod_{\substack{\gamma\in\Lambda_+\\ e^{i\gamma(\alpha)},e^{i\gamma(\text{Ad}_w\beta)}\neq 1}} \sin^2\frac{\gamma(\alpha)}{2} \sin^2\frac{\gamma(\text{Ad}_w\beta)}{2} } 
\end{split}
\end{equation}

The $S$-kernel simplifies greatly when both $g$ and $h$ are regular, namely when $C_G(g) \cong C_G(h) \cong T$. In this case since $\gamma$ runs over all the positive roots and $w$ permutes $\gamma$, the denominator of~\eqref{eq:S_kernel_final} becomes:
\begin{equation}\label{eq:Zw_Regular}
\begin{split}
    16 \prod_{\gamma\in\Lambda_+} \sin^2\frac{\gamma(\alpha)}{2} \sin^2\frac{\gamma(\text{Ad}_w\beta)}{2} = \prod_{\gamma\in\Lambda_+} 4 \sin^2\frac{\gamma(\alpha)}{2} \prod_{\gamma\in\Lambda_+} 4 \sin^2\frac{\gamma(\beta)}{2} = |\Delta(g)|^2 |\Delta(h)|^2
\end{split}
\end{equation}
where $\Delta(e^{ix}) := \prod_{\gamma\in\Lambda_+} 2 \sin \frac{\gamma(x)}{2}$ is the Weyl denominator. Moreover, we have $Z_w = T\cdot w \cdot T = T$ for $w \in W$. Therefore both the volume factor and the Jacobian factor are independent from $w$. Hence we have:
\begin{equation}
    S^{(k)}_{([g],\sigma),([h],\rho)} \propto \sum_{w\in W} e^{-\frac{ik}{2\pi}\tr(\alpha w(\beta))} \chi^*_\sigma(e^{iw(\beta)}) \chi^*_\rho(e^{iw^{-1}(\alpha)})\,,
\end{equation}
for regular $g$ and $h$, where $w(\beta) \equiv \text{Ad}_{w} (\beta)$. We note that in this case $\mathcal{M}(g,h) = N(T)$.

One key property of $S$-kernel is that we should require $S S^\dagger = 1$ in a measurable sense, and later we will find that the $S$-kernel satisfying the relation in the regular sector is the normalized one as follows~\footnote{Actually, the reduced moduli-space calculation determines the bare amplitude only up to measure-dependent half-density factors. Hence we choose the normalization for which the resulting $S$-kernel is unitary on $L^2(Cl(G))$, and we will see in section~\ref{sec:More_Examples} that in the $SU(2)_k$ lattice reduction this reproduces the discrete Weyl measure and the Kac-Peterson matrix.}:
\begin{equation}
    \frac{|\Delta(g)|^2 |\Delta(h)|^2}{\text{Vol}(T)} S^{(k)}_{([g],\sigma),([h],\rho)}\,.
\end{equation}
With the hindsight that the unitary $S$-kernel shall be given by the above expression in the regular sector, from now on we will stick to the above expression, rather than the unnormalized~\eqref{eq:S_kernel_final}, as the semi-classical $S$-kernel when checking various properties of the $S$- and $T$-kernels in the regular sector.

\paragraph{$T$-kernel}

The $T$-kernel can be computed in a similar fashion. Concretely, one replaces the framed unknot by a pair of nearby loops $\ell$ and $\ell'$ so that $\ell\cup\ell'$ forms a Hopf link with linking number $+1$. In this local model the same reduction to holonomy data applies, now with both holonomies constrained to lie in the same conjugacy class. Schematically one is therefore led to:
\begin{equation}
    T_{([g],\sigma),([h],\rho)}^{(k)} \propto \delta_{[g],[h]}\ \delta_{\sigma,\rho} \ \int \mathcal{D}B\mathcal{D}A\ e^{i\int_{S^3} BF + \frac{k}{4\pi}\text{CS}}\ W_{[g],\sigma}(\ell)\,,
\end{equation}
where $W_{[g],\sigma}(\ell)$ should be understood as a Wilson loop with $(+1)$-framing. Integrating out $B$, we have:
\begin{equation}
    T_{([g],\sigma),([h],\rho)}^{(k)} \propto \delta_{[g],[h]}\ \delta_{\sigma,\rho} \ \int_{F = 0,\ \text{Hol}_A(\ell') \in [g]} \mathcal{D}A\ e^{i\int_{S^3}\frac{k}{4\pi}\text{CS}}\ \chi_\sigma\left(\text{Hol}_A(\ell')\right)\,.
\end{equation}
In this framed local model we then use the ansatz:
\begin{equation}
    A = x\alpha x^{-1} \delta^{(1)}_{D}\,,
\end{equation}
with $g=e^{i \alpha}$ and $x e^{i\alpha} x^{-1}\in [g]$, hence:
\begin{equation}
    F=dA = x \alpha x^{-1} \delta^{(2)}_{\ell}\,,
\end{equation}
for $\partial D = \ell$. Therefore, we have:
\begin{equation}
    \int_{S^3}\text{CS} = \tr\ \alpha^2 \int_{S^3} \delta^{(1)}_{D} \wedge \delta^{(2)}_{\ell} = \tr \alpha^2\,,
\end{equation}
where $\int_{S^3} \delta^{(1)}_{D} \wedge \delta^{(2)}_{\ell}$ is the self-linking number of $\ell$ and is computed by deforming $\ell$ to $\ell'$ and considering the linking number between $\ell$ and $\ell'$. Thus, we have:
\begin{equation}
    T_{([g],\sigma),([h],\rho)}^{(k)} \propto \delta_{[g],[h]}\ \delta_{\sigma,\rho} \ \int_G d\mu\ e^{\frac{ik}{4\pi}\tr \alpha^2}\ \chi_\sigma(g)\,,
\end{equation}
where the residual integral over $x\in G$ again parameterizes the remaining conjugation orbit. Again choosing a representative $g\in T$, the same semiclassical reduction gives:
\begin{equation}
    T_{([g],\sigma),([h],\rho)}^{(k)} \propto \delta_{[g],[h]}\ \delta_{\sigma,\rho} \ e^{\frac{ik}{4\pi}\tr\ \alpha^2}\ \chi_\sigma(g)\,.
\end{equation}
After normalizing by the dimension $d_{\sigma}$ of the representation $\sigma$, we have:
\begin{equation}\label{eq:T_matrix}
    T_{([g],\sigma),([h],\rho)}^{(k)} = \delta_{[g],[h]}\ \delta_{\sigma,\rho}\ e^{\frac{ik}{4\pi}\tr\ \alpha^2}\ \frac{\chi_\sigma(g)}{d_\sigma}\,
\end{equation}
for $\alpha \in \mathfrak{t}$ and $g=e^{i\alpha}$.

We note that the self-linking of a Wilson-'t Hooft operator $W_{[g],\sigma}(\ell)$ can equivalently be viewed as the linking of a Wilson operator $W_{\sigma}(\ell) = \mathcal{P}\tr_\sigma e^{i\oint_\ell A}$ and a 't Hooft operator (or more precisely a Gukov-Witten operator) $U_{[g]}(\ell') = e^{i\oint_{\ell'} (\alpha, B)}$ defined in~\cite{Cattaneo:2002tk, Jia:2025jmn}, where $\ell'$ is a slight deformation of $\ell$ and $\ell\cup \ell'$ forms a Hopf link. Indeed, when there is no anomaly, i.e. $k = 0$, one can compute that~\cite{Cordova:2022rer, Jia:2025jmn}:
\begin{equation}
    \langle W_{\sigma}(\ell) U_{[g]}(\ell') \rangle \propto \frac{\chi_\sigma(g)}{d_\sigma}\,,
\end{equation}
where the VEV is evaluated in pure $BF$ theory. This matches the result~(\ref{eq:T_matrix}).

As a final remark, we recall that one expects candidate modular kernels $S$ and $T$ to satisfy
\begin{equation}\label{eq:Test_ST_Relations}
    S S^{\dagger} = TT^{\dagger}=1\,,\quad S^2=(ST)^3=C\,,
\end{equation}
where $C$ is the charge conjugation matrix. We check these properties for regular $g,h$ and provide an expression for the corresponding charge-conjugation kernel $C$ for $SU(N)$ in Appendix~\ref{app:proof-of-SU(2)-modular-properties}. The reason that we test the above relations only for the regular sector is because the set of regular points in $G$ constitutes a dense open subset of $G$, whereas the set of singular points is measure zero with respect to the Haar measure of $G$ (and furthermore with respect to the induced measure on $Cl(G)$). Thus as integral transformations with respect to Haar measure,~(\ref{eq:Test_ST_Relations}) is completely determined by the regular sectors whereas the singular sector is simply undetectable. In this sense~(\ref{eq:Test_ST_Relations}) is better understood a relation hold on $L^2$ Hilbert space, namely e.g. $SS^\dagger = 1$ should be understood as $\langle S^\dagger f, S^\dagger g\rangle = \langle f,g \rangle$ for $L^2$-functions $f$ and $g$ on $Cl(G)$. However this certainly does not mean that the singular sector is meaningless. It should better be viewed as certain extension of the of the regular sector of the $S$ and $T$-kernels. A complete analysis of these extensions would require additional harmonic analysis for the varying centralizers and is beyond the scope of the present work. In this paper we instead provide nontrivial evidence for the extension by singular sector by showing, in the $SU(2)_k$ example, that the singular sector reproduces the Kac-Peterson $S$-matrix.

\subsection{Examples}\label{sec:More_Examples}

\paragraph{$U(1)_k$}

The simplest example is $G = U(1)$. In that case, the reduced integral collapses to a single point $(g, h)$, and the formula becomes:
\begin{equation}
    S^{(k)}_{(g,n),(h,m)} = e^{i\int_{S^3} \frac{2k}{4\pi} AdA}\ e^{-im \theta_x}\ e^{-in \theta_y}
\end{equation}
where $g = e^{i\theta_x}$, $h = e^{i\theta_y}$. One can further evaluate the above integral to be:
\begin{equation}
    S^{(k)}_{(g,n),(h,m)} = e^{-i\frac{2k}{2\pi} \theta_x \theta_y}\ e^{-im \theta_x}\ e^{-in \theta_y} = e^{-i(m \theta_x + n \theta_y + \frac{2k}{2\pi}\theta_x \theta_y)}\,.
\end{equation}
Similarly, the $T$-matrix is evaluated as
    \begin{equation}
        T^{(k)}_{(g,n),(h,m)}= \delta_{2\pi}(\theta_x-\theta_y)\delta_{n,m} \exp\left(i\left(n \theta_x+\frac{k \theta_x^2}{2\pi} \right) \right)\,,
    \end{equation}
and they match the $U(1)$ case with $k\in H^4(BU(1),\mathbb{Z})$ given in \eqref{eq:U(1)-modular-matrices} up to the sign of $k$.

\paragraph{$SU(2)_k$}

For $G=SU(n)$ with regular $[g]$ and $[h]$, we can restrict the generators of $g$ and $h$ to the maximal torus of $G$. At the level of the present formula we suppress the Jacobian and orbit-volume factors that would appear in a more complete reduction from conjugation orbits to the maximal torus, since our main purpose here is to display the phase and character structure of the kernel. With that understood, the value of~(\ref{eq:S_kernel_final}) at regular $g$ and $h$ takes the form (cf. the argument around~(\ref{eq:Zw_Regular})):
\begin{equation}\label{S-matrix-su(N)-formal}
        S^{(k)}_{([g],\sigma),([h],\rho)} = \sum_{w\in W} e^{-i\frac{k}{2\pi} \textrm{tr}(\alpha w(\beta))} \ \chi_{\rho}(e^{iw^{-1}(\alpha)})\ \chi_\sigma(e^{iw(\beta)}),\ g = e^{i\alpha},\ h = e^{i\beta},\ \alpha,\beta\in \mathfrak{t}
\end{equation}
where $W=S_n$ is the Weyl group of $SU(n)$.

Now we specialize to $SU(2)_k$ with generic $[g]$ and $[h]$. One can parametrize the maximal torus $T$ as
    \begin{equation}
        \exp(i \alpha H) = \left(\begin{array}{cc}
            e^{\frac{i \alpha}{2}} &  \\
             & e^{-\frac{i \alpha}{2}}
        \end{array} \right)\,, \quad \textrm{with}\quad H=\frac{1}{2}\left(\begin{array}{cc}
            1 & 0 \\
            0 & -1
        \end{array} \right)\,,
    \end{equation}
and $\alpha \sim \alpha + 4\pi$. The Weyl reflection $S_2=\mathbb{Z}_2$ flip the sign of $\alpha$, and we can restrict $\alpha \in (0,2\pi)$ as a representative of the conjugacy class $ g=e^{i \alpha}\in[g]$. The representations of $C_G(g) = T$ are labeled by an integer, and the $S$-matrix in \eqref{S-matrix-su(N)-formal} when both $g$ and $h$ are regular reads:
\begin{equation}
    S^{(k)}_{([g],n),([h],m)} = e^{-i\frac{k}{4\pi} \alpha \beta} e^{-\frac{i}{2} m \alpha - \frac{i}{2} n \beta} + e^{i\frac{k}{4\pi} \alpha \beta} e^{\frac{i}{2} m \alpha +\frac{i}{2} n \beta} = 2\cos \frac{1}{2}\left(m\alpha + n\beta + \frac{k}{2\pi} \alpha \beta \right)\,.
\end{equation}
The $T$-matrix \eqref{eq:T_matrix} reads
    \begin{equation}
        T^{(k)}_{([g],n),([h],m)} = e^{i\frac{k}{8\pi}\alpha^2}e^{\frac{i}{2}n\alpha} \delta_{4\pi}(\alpha-\beta)\delta_{n,m}\,.
    \end{equation}

It is interesting to take a closer look at the singular sector using the general formula~(\ref{eq:S_kernel_final}). let $[g] = [e]$ and $h = e^{i\beta H}$, we consider the following sector:
\begin{equation}\label{eq:S_Full_Ring}
    S^{(k)}_{([e],n),([h], 0)} = \chi_n^*(h) = \frac{\sin \frac{n+1}{2}\beta}{\sin\frac{\beta}{2}}\,.
\end{equation}
A further restriction to the lattice $\beta = \frac{2\pi(r+1)}{k+2}$, $r=0,\cdots,k$ leads to:
\begin{equation}\label{eq:S_Full_Ring_2}
    S^{(k)}_{nr} := S^{(k)}_{([e],n),([h], 0)} = \frac{\sin \frac{(n+1)(r+1)\pi}{(k+2)}}{\sin\frac{(r+1)\pi}{k+2}}\,.
\end{equation}
Normalize by $\sqrt{\frac{2\pi}{k+2}}\sqrt{\Delta^2/4\pi}$ where $\Delta = 4\sin^2\beta/2$ is the Weyl denominator of $SU(2)$, we have:
\begin{equation}\label{eq:KP_matrix}
    \widetilde{S}^{(k)}_{nr} = \sqrt{\frac{2}{k+2}} \sin \frac{(n+1)(r+1)\pi}{k+2}
\end{equation}
which is exactly the Kac-Peterson $S$-matrix~\cite{Lynker:2004sw}.

Actually, from~(\ref{eq:S_Full_Ring}) we obtain the representation ring $R(SU(2))$ and the level-$k$ Verlinde ring is given by:
\begin{equation}\label{eq:R_k=R/I_k}
    R_k(SU(2)) = R(SU(2))/I_k(SU(2))
\end{equation}
where the level-$k$ fusion ideal $I_k(SU(2)) = \langle\chi_{k+1}\rangle$. The points on the Verlinde lattice is then obtained by solving $\chi_{k+1}(e^{i\beta}) = \sin \frac{k+2}{2}\beta/\sin\frac{\beta}{2} = 0$. This implies $\beta = 2\pi (r+1)/(k+2)$ for $r = 0,\cdots,k$, which is exactly the lattice to which we restrict to obtain~(\ref{eq:S_Full_Ring_2}). Thus we see the unnormalized $\widetilde{S}$ can be obtained via starting from the full representation ring using $BF+kCS$. Then~(\ref{eq:KP_matrix}) is obtained by orthonormalizing~(\ref{eq:S_Full_Ring_2}) on the $k+1$ Verlinde points $\beta_r=2\pi(r+1)/(k+2)$. Requiring $\widetilde S\widetilde S^\dagger=1$ fixes the diagonal discrete Weyl measure
\begin{equation}\label{eq:Wrr}
    W_{rr}=\frac{2}{k+2}\sin^2\frac{(r+1)\pi}{k+2}\,.
\end{equation}
This can alternatively be obtained by starting with the Weyl measure of $SU(2)$ (see Appendix~\ref{app:check-of-Lagrangian-algebra-SU(2)}) $\mu(\beta) d\beta = \frac{\sin^2 \frac{\beta}{2}}{\pi}d\beta$, then restricting to the Verlinde point $\beta_r = \frac{2\pi(r+1)}{k+2}$ with spacing $\Delta\beta_r = \frac{2\pi}{k+2}$, hence the discrete measure is:
\begin{equation}
    \mu(\beta_r) \Delta\beta_r = \frac{2}{k+2} \sin^2 \frac{(r+1)\pi}{k+2}
\end{equation}
which is exactly the above discrete Weyl measure. Having fixed $W_{rr}$, we have:
\begin{equation}
    \widetilde S^{(k)}_{nr} = \sqrt{W_{rr}}\, S^{(k)}_{nr} = \sqrt{\frac{2}{k+2}}\sin\frac{(n+1)(r+1)\pi}{k+2}
\end{equation}
reproducing~(\ref{eq:KP_matrix}). We also note that this discrete Weyl measure is exactly the factor that makes $\widetilde S^{(k)} \widetilde (S^{(k)})^\dagger = 1$. Therefore, we obtain a good sanity check as this characteristic data of level-$k$ CS can be extracted from our general formula~(\ref{eq:S_kernel_final}) obtained from $BF+kCS$ in a specific limit. We remark this also provides an example of the extension from the regular sector to the singular sector as we commented at the end of section~\ref{sec:Modular_ST_Calculation}.

\paragraph{$SU(3)_k$}

We then consider the more involved $SU(3)_k$ case with $ W=S_3$. The maximal torus $T$ of $SU(3)$ can be written as, for example:
    \begin{equation}
        \exp\left( i \alpha_1 H_1 +i \alpha_2 H_2 \right)=\left(\begin{array}{ccc}
            e^{i \frac{\alpha_1}{2}}&0&0\\
            0&e^{i\frac{\alpha_2}{2}}&0\\
            0&0&e^{-i\frac{\alpha_1+\alpha_2}{2}}
        \end{array} \right)
    \end{equation}
and Weyl group $S_3$ permute the diagonal elements. The generators are chosen as
    \begin{equation}
        H_1 = \frac{1}{2}\left(\begin{array}{ccc}
            1&0&0\\0&0&0\\0&0&-1
        \end{array} \right)\,,\quad H_2=\frac{1}{2}\left(\begin{array}{ccc}
            0&0&0\\0&1&0\\0&0&-1
        \end{array} \right)\,,
    \end{equation}
and we introduce the Killing metric $g_{ij} = 2\textrm{Tr} (H_i H_j)$ with $i,j=1,2$. The maximal torus $T$ is a torus with identification $\alpha_i \sim \alpha_i+4\pi$. The $U(1)$ charges are represented by a pair of integers $(n_1,n_2)$. Then the $S$-matrix \eqref{S-matrix-su(N)-formal} can be written as
\begin{equation}
    S_{([g],n),([h],m)} = \sum_{w \in S_3} e^{-i\frac{k}{4\pi} \alpha \cdot w(\beta)} e^{-\frac{i}{2} n \cdot w(\beta)-\frac{i}{2} m \cdot w^{-1}(\alpha)}\,,
\end{equation}
and the $T$-matrix \eqref{eq:T_matrix} is
    \begin{equation}
        T^{(k)}_{([g],n),([h],m)} = e^{i\frac{k}{8\pi}\alpha\cdot \alpha}e^{\frac{i}{2}n\cdot\alpha} \delta_{4\pi}(\alpha-\beta)\delta_{n,m}\,.
    \end{equation}
Here the inner products are defined by $n\cdot \alpha = n_1 \alpha_1 + n_2 \alpha_2$ and $\alpha \cdot \beta = \sum_{i,j} g_{ij} \alpha_i \beta_j$. The action of Weyl group on $\alpha$ is induced by
    \begin{equation}
        \alpha \cdot w(H) = w(\alpha) \cdot H\,,\quad \textrm{with} \quad \alpha \cdot H\equiv \alpha_1 H_1 + \alpha_2 H_2\,.
    \end{equation}

\subsection{Revisiting candidate $SU(2)$ data for $k=0$}

In this section, we revisit the candidate $SU(2)$ data $\mathcal{L}$ proposed in \cite{Jia:2025vrj} for the $k=0$ case and analyze them within the working criterion adopted in this paper. We fix the coefficients of the simple line operators in $\mathcal{L}$ and then examine whether the corresponding coefficient vectors $V_{\mathcal{L}}$ are invariant under the candidate $S$- and $T$-kernels.
\begin{equation}
    S\cdot V_{\mathcal{L}} = V_{\mathcal{L}}\,,\quad T\cdot V_{\mathcal{L}} = V_{\mathcal{L}}\,.
\end{equation}
We will summarize the results below and refer to App.~\ref{app:check-of-Lagrangian-algebra-SU(2)} for more details.

There are several types of candidate data. The Dirichlet one $\mathcal{L}_{\rm Dir}$ and the non-vanishing components of the coefficient vector $V_{\mathcal{L}_{\rm Dir}}$ are
    \begin{equation}
        \mathcal{L}_{\textrm{Dir}} = \bigoplus_{j \geq 0} (j+1) W_{(0,j)}\,,\quad (V_{\mathcal{L}_{\rm Dir}})_{(0,j)} = 4\pi \delta\left(\alpha\right){\large |}_{\alpha=0}  (j+1)\,,
    \end{equation}
where $(j+1)$ is the dimension of the irreducible representation of  $SU(2)$ labeled by $j$. The presence of the delta function indicates that all non-vanishing components of $V_{\mathcal{L}_{\rm Dir}}$ are supported at the point $\alpha=0$, namely $g=e$. The symmetry boundary corresponding to $\mathcal{L}_{\rm Dir}$ supports the $SU(2)$ global symmetry. 

We also have the Neumann-type candidate datum $\mathcal{L}_{\rm Neu}$ and the coefficient vector $V_{\rm Neu}$
    \begin{equation}
        \mathcal{L}_{\textrm{Neu}} = \bigoplus_{\alpha\in [0,2\pi]} W_{(\alpha,0)}\,,\quad (V_{\mathcal{L}_{\textrm{Neu}}})_{[g],n} = \left\{\begin{array}{l}
             \delta_{j,0}\,, \quad \alpha=0\,,2\pi\,,\\
            \delta_{n,0}\,, \quad 0<\alpha<2\pi\,,
        \end{array} \right. 
    \end{equation}
where the summation over the conjugacy label $\alpha\in [0,2\pi)$ is understood as an integral, and non-vanishing coefficients of $V_{\mathcal{L}_{\rm Neu}}$ are uniformly distributed over $\alpha\in [0,2\pi]$ with trivial representations. The corresponding symmetry boundary can be obtained by (flat) gauging the whole $SU(2)$ global symmetry, which supports the $\mathbf{Rep}(SU(2))$ categorical symmetry.

Then we have the $SO(3)$-type candidate datum $\mathcal{L}_{SO(3)}$
    \begin{equation}
        \mathcal{L}_{\textrm{SO(3)}} = \bigoplus_{j \geq 0} \left( (2j+1) W_{(0,2j)} \oplus (2j+1)W_{(2\pi,2j)}\right)\,,
    \end{equation}
where $(2j+1)$ is the dimension of the admissible irreducible representations for $SO(3)$. The symmetry associated with $\mathcal{L}_{SO(3)}$ is obtained from $SU(2)$ by gauging the center $\mathbb{Z}_2\in SU(2)$, which consists of the remaining $SO(3)$ symmetry and a dual $\mathbb{Z}_2$ symmetry. In fact, $\mathcal{L}_{SO(3)}$ is the special case of $\mathcal{L}_{A_q}$ for $q=2$ introduced below and the coefficient vector $V_{\mathcal{L}_{SO(3)}}$ is defined therein. 

Finally, we have the $A_q$-type candidate data $\mathcal{L}_{A_q}$
    \begin{equation}
        \mathcal{L}_{A_{q,\rm odd}} = \left(\bigoplus_{j\geq 0} d_q(j)W_{(0,j)}\right) \oplus \left(\bigoplus_m \bigoplus_{n=1}^{\left[ \frac{q}{2}\right]} 2W_{\left(\frac{4\pi}{q}n,q m \right)}\right)\,,
    \end{equation}
for odd $q$ and
    \begin{equation}
        \mathcal{L}_{A_{q,\rm even}} = \left(\bigoplus_{j\geq 0} d_q(j)W_{(0,j)}\right) \oplus \left(\bigoplus_m \bigoplus_{n=1}^{\left[ \frac{q}{2}\right]-1} 2W_{\left(\frac{4\pi}{q}n,q m \right)}\right)\oplus \left(\bigoplus_{j\geq 0} d_q(j)W_{(2\pi,j)}\right)\,,
    \end{equation}
for even $q$, where $d_q(j)$ is defined as
    \begin{equation}
        d_q(j) = \frac{1}{q}\sum_{n=0}^{q-1} \chi_j\left(\frac{4\pi}{q} n\right)\,,
    \end{equation}
which counts the components in the irreducible representation labeled by $j$ that are invariant under the action of $\mathbb{Z}_q$ subgroup
    \begin{equation}\label{eq:Zq-subgroup-of-SU2}
        \mathbb{Z}_q = \left\{ \left.\left(\begin{array}{cc}
            e^{\frac{2\pi i n}{q}} & 0  \\
            0 & e^{-\frac{2\pi i n}{q}} 
        \end{array}\right)\right|  n=0,1,\cdots,q-1\right\}\,.
    \end{equation}
The coefficient vectors $V_{\mathcal{L}_{q}}$ are
    \begin{equation}
        (V_{\mathcal{L}_{A_{q,\rm odd}}})_{([g],n)} = \left\{\begin{array}{l}
            d_q(j) 4\pi \delta(\alpha){\large |}_{\alpha=0}\,,\quad (\alpha=0)\,,\\
            \frac{2}{\mu(\alpha)}\sum_{r=1}^{\left[\frac{q}{2}\right]}\delta(\alpha-\frac{4\pi}{q}r) \sum_k\delta_{n,qk}\,, \quad (\alpha\neq 0)\,,
        \end{array} \right.
    \end{equation}
for odd $q$, where we have a normalization factor $2/\mu(\alpha)$ in the second line with $\mu(\alpha) = \frac{\sin^2 \frac{\alpha}{2}}{\pi}$ the measure of conjugacy class.
Similarly, one has
    \begin{equation}
        (V_{\mathcal{L}_{A_{q,\rm even}}})_{([g],n)} = \left\{\begin{array}{l}
            d_q(j) 4\pi\delta(\alpha){\large |}_{\alpha=0}\,,\quad (\alpha=0)\,,\\
            \frac{2}{\mu(\alpha)}\sum_{r=1}^{\left[\frac{q}{2}\right]-1}\delta(\alpha-\frac{4\pi}{q}r)\sum_k \delta_{n,qk}\,, \quad (0<\alpha<2\pi)\,,\\
            d_q(j) 4\pi\delta(\alpha-2\pi){\large |}_{\alpha=2\pi}\,,\quad (\alpha=2\pi)
        \end{array} \right.
    \end{equation}
for even $q$, where we include the contribution from $\alpha=2\pi$. The symmetry associated with $\mathcal{L}_{A_{q}}$ is obtained from $SU(2)$ by gauging the $\mathbb{Z}_q$ subgroup \eqref{eq:Zq-subgroup-of-SU2}.

\section{Conclusion and Outlook}\label{sec:Conclusion}

Within the kernel-theoretic model derived above, we analyzed candidate gaugings of categorical continuous symmetries by studying candidate Lagrangian algebra data in the would-be Drinfeld center of the symmetry category. The main structural input used in the analysis is the convolution tensor product and its transgression in the corresponding Drinfeld center, so the discussion does not require choosing uniquely between possible analytic realizations such as $\QC^k(G)$, $\Hilb^k(G)$, or related models. Under the two explicit assumptions stated in the introduction, we derived candidate $S$- and $T$-kernels from the proposed $BF{+}k$CS SymTFT, used them to obtain candidate modular invariants and candidate gaugings, and checked that the resulting formulas reproduce the established examples studied in this paper.

The results should be read with a precise scope. We have not established from first principles either a unique categorical realization of $\CC^k(G)$ for general compact Lie groups or a theorem identifying candidate Lagrangian algebra data in this continuous setting with the common $+1$ eigenspace of the modular kernels. Accordingly, statements about the true symmetry category of a general QFT, or about a full non-semisimple continuous analogue of the familiar MTC theorems, remain conjectural. What the present paper provides is evidence that the proposed model is internally consistent and physically informative.

Recent papers have developed broader continuous SymTFT and categorical frameworks~\cite{Bonetti:2024cjk, Jia:2025jmn, Jia:2025vrj, Stockall:2025ngz}. Rather than developing a general framework for classifying gaugings of continuous categorical symmetries, the scope of the present work is more limited: we extract explicit semiclassical modular kernels from the $BF{+}k$CS model and use them to analyze candidate gaugings in concrete benchmark cases. The next steps are to justify the point-assigned category of the corresponding TQFT more directly (see~\cite{fiveguys_HilbG}), clarify the relation between the quasi-coherent and measurable/Hilbert realizations, and establish the continuous counterpart of the modular-invariant criterion as a theorem rather than a working assumption.

\acknowledgments

The authors would like to thank Hank Chen, Wei Gu, Ran Luo, Yi-Nan Wang, Hao Xu, Wenbin Yan and Yi Zhang for illuminating discussions. QJ is supported by National Research Foundation of Korea (NRF) Grant No. RS-2024-00405629 and Jang Young-Sil Fellow Program at the Korea Advanced Institute of Science and Technology. JT is supported by National Natural Science Foundation of China under Grant No. 12405085 and by the Natural Science Foundation of Shanghai (Grant No. 24ZR1419300). 


\appendix

\section{Modular properties of $\Theta$-functions }\label{app:Theta-functions}

In this appendix, we will derive the $S$ and $T$-transformations \eqref{eq:Theta-function-modular-transformation} of the $\Theta$-functions defined in \eqref{eq:Theta-function}. Then we will derive the modular transformation \eqref{modular-transformation-ZU(1)_k} of the partition function $Z[\theta_r,\theta_s]$ defined in \eqref{eq:U(1)-partition-function}.

We begin with the definition of the $\Theta$-function 
    \begin{equation}
        \Theta^{(k)}_l\left(\begin{array}{c}
            \theta_r\\ \theta_s
        \end{array} \right) (\tau) = \sum_{u\in \mathbb{Z}} e^{2\pi i k\left(u+\frac{l}{2k} + \frac{\theta_r}{2\pi} \right)^2\tau+ 4\pi i k \left(u+\frac{l}{2k} + \frac{\theta_r}{2\pi}\right) \frac{\theta_s}{2\pi}}\,.
    \end{equation}
It is straightforward to check that the $T$-transformation gives
    \begin{equation}
    \begin{split}
        &\Theta^{(k)}_l\left(\begin{array}{c}
            \theta_r\\ \theta_s
        \end{array} \right) (\tau+1)\\ =& \sum_{u\in \mathbb{Z}} e^{2\pi i k\left(u+\frac{l}{2k} + \frac{\theta_r}{2\pi} \right)^2\tau+ 4\pi i k \left(u+\frac{l}{2k} + \frac{\theta_r}{2\pi}\right) \frac{\theta_s}{2\pi}} \times e^{2\pi i k\left(u+\frac{l}{2k} + \frac{\theta_r}{2\pi} \right)^2}\,,  \\
        =& \sum_{u\in \mathbb{Z}} e^{2\pi i k\left(u+\frac{l}{2k} + \frac{\theta_r}{2\pi} \right)^2\tau+ 4\pi i k \left(u+\frac{l}{2k} + \frac{\theta_r}{2\pi}\right) \frac{\theta_s}{2\pi}} e^{2\pi i k \left(u+\frac{l}{2k} \right)^2} e^{4\pi i k \left(u+\frac{l}{2k}\right) \frac{\theta_r}{2\pi}} e^{2\pi i k \left(\frac{\theta_r}{2\pi}\right)^2}\\
        =&\sum_{u\in \mathbb{Z}} e^{2\pi i k\left(u+\frac{l}{2k} + \frac{\theta_r}{2\pi} \right)^2\tau+ 4\pi i k \left(u+\frac{l}{2k} + \frac{\theta_r}{2\pi}\right) \frac{\theta_s}{2\pi}} e^{\frac{\pi i l^2}{2k}} e^{4\pi i k \left(u+\frac{l}{2k}\right) \frac{\theta_r}{2\pi}} e^{2\pi i k \left(\frac{\theta_r}{2\pi}\right)^2}\\
        =&e^{\frac{\pi i l^2}{2k}} e^{-2\pi i k \left(\frac{\theta_r}{2\pi}\right)^2}\sum_{u\in \mathbb{Z}} e^{2\pi i k\left(u+\frac{l}{2k} + \frac{\theta_r}{2\pi} \right)^2\tau+ 4\pi i k \left(u+\frac{l}{2k} + \frac{\theta_r}{2\pi}\right) \left(\frac{\theta_s+\theta_r}{2\pi}\right)} \\
        =&e^{\frac{\pi i l^2}{2k}} e^{-2\pi i k \left(\frac{\theta_r}{2\pi}\right)^2}\Theta^{(k)}_l\left(\begin{array}{c}
            \theta_r\\ \theta_s+\theta_r
        \end{array} \right) (\tau)\,.
    \end{split}
    \end{equation}
For $S$-transformation, we have
    \begin{equation}
        \begin{split}
            &\Theta^{(k)}_l\left(\begin{array}{c}
            \theta_r\\ \theta_s
        \end{array} \right) (-1/\tau)\\
        =&\sum_{u\in \mathbb{Z}} e^{2\pi i k\left(u+\frac{l}{2k} + \frac{\theta_r}{2\pi} \right)^2(-1/\tau)+ 4\pi i k \left(u+\frac{l}{2k} + \frac{\theta_r}{2\pi}\right) \frac{\theta_s}{2\pi}}\\
        =& \int_{-\infty}^{+\infty} dx \sum_{u\in \mathbb{Z}} \delta(x-u)  e^{2\pi i k\left(x+\frac{l}{2k} + \frac{\theta_r}{2\pi} \right)^2(-1/\tau)+ 4\pi i k \left(x+\frac{l}{2k} + \frac{\theta_r}{2\pi}\right) \frac{\theta_s}{2\pi}}\\
        =&\int_{-\infty}^{+\infty} dx \sum_{u'\in \mathbb{Z}} e^{2\pi i u' x}e^{2\pi i k\left(x+\frac{l}{2k} + \frac{\theta_r}{2\pi} \right)^2(-1/\tau)+ 4\pi i k \left(x+\frac{l}{2k} + \frac{\theta_r}{2\pi}\right) \frac{\theta_s}{2\pi}}\\
        =&\int_{-\infty}^{+\infty} dx \sum_{u'\in \mathbb{Z}} e^{2\pi i k \left(x+\frac{l}{2k} + \frac{\theta_r}{2\pi} - \frac{\tau}{2k} \left(u'+k\frac{\theta_s}{2\pi} \right)\right)^2(-1/\tau)} e^{2\pi i k \frac{\tau}{4k^2} \left(u'+2k \frac{\theta_s}{2\pi}\right)^2} e^{-2\pi i u' \left(\frac{l}{2k}+\frac{\theta_r}{2\pi} \right)}\\
        =& \sqrt{\frac{-i\tau}{2k}}\sum_{u' \in \mathbb{Z}} e^{2\pi i k \frac{\tau}{4k^2} \left(u'+2k \frac{\theta_s}{2\pi}\right)^2} e^{-2\pi i u' \left(\frac{l}{2k}+\frac{\theta_r}{2\pi} \right)}\,.
        \end{split}
    \end{equation}
Relabel $u'=2k u+l'$, we then have
    \begin{equation}
        \begin{split}
            &\sqrt{\frac{-i\tau}{2k}}\sum_{u' \in \mathbb{Z}} e^{2\pi i k \frac{\tau}{4k^2} \left(u'+2k \frac{\theta_s}{2\pi}\right)^2} e^{-2\pi i u' \left(\frac{l}{2k}+\frac{\theta_r}{2\pi} \right)}\\
            =&\sqrt{\frac{-i\tau}{2k}} \sum_{u\in \mathbb{Z}} \sum_{l'=0}^{2k-1}e^{2\pi i k \frac{\tau}{4k^2} \left(2k u +l' +2k \frac{\theta_s}{2\pi}\right)^2} e^{-2\pi i \left(2ku +l' \right) \left(\frac{l}{2k}+\frac{\theta_r}{2\pi} \right)}\\
            =&\sqrt{\frac{-i\tau}{2k}}  \sum_{l'=0}^{2k-1}\sum_{u\in \mathbb{Z}} e^{2\pi i k \tau \left(u+\frac{l'}{2k} + \frac{\theta_s}{2\pi} \right)^2} e^{-4\pi i k \left(u+\frac{l'}{2k} \right)\left(\frac{l}{2k}+\frac{\theta_r}{2\pi} \right)}\\
            =&\sqrt{\frac{-i\tau}{2k}}  \sum_{l'=0}^{2k-1}\sum_{u\in \mathbb{Z}} e^{2\pi i k \tau \left(u+\frac{l'}{2k} + \frac{\theta_s}{2\pi} \right)^2} e^{4\pi i k \left(u+\frac{l'}{2k} +\frac{\theta_s}{2\pi}\right)\left(-\frac{\theta_r}{2\pi}\right)} e^{4\pi i k \frac{\theta_s}{2\pi} \frac{\theta_r}{2\pi}} e^{-2\pi i \frac{l l'}{2k}}\\
            =& \sqrt{\frac{-i\tau}{2k}} e^{4\pi i k \frac{\theta_s}{2\pi} \frac{\theta_r}{2\pi}}\sum_{l'}  e^{-2\pi i \frac{l l'}{2k}} \Theta^{(k)}_{l'}\left(\begin{array}{c}
            \theta_s\\ -\theta_r
        \end{array} \right) (\tau) \,.
        \end{split}
    \end{equation}

Then we will check the modular transformation of the partition function \eqref{eq:U(1)-partition-function} using the modular properties of $\Theta$-functions. To do so, we need the following modular transformation for $\eta$-functions
    \begin{equation}
        \eta(\tau+1) = e^{\frac{2\pi i}{24}}\eta(\tau)\,,\quad \eta(-1/\tau) = \sqrt{-i\tau} \eta(\tau)\,.
    \end{equation}
Under the $T$-transformation, one has
    \begin{equation}
    \begin{split}
        &Z[\theta_r,\theta_s](\tau+1)\\
        =& e^{-2\pi i k \frac{\theta_r}{2\pi} \frac{\theta_s}{2\pi}} \sum_{l=0}^{2k-1}\frac{1}{|\eta(\tau+1)|^2}\Theta^{(k)}_l\left(\begin{array}{c}
            \theta_r\\\theta_s
        \end{array} \right)(\tau+1) \overline{\Theta}^{(k)}_l\left(\begin{array}{c}
            0\\0
        \end{array} \right)(\tau+1)\\
        =&e^{-2\pi i k \frac{\theta_r}{2\pi} \frac{\theta_s}{2\pi}} \sum_{l=0}^{2k-1}\frac{1}{|\eta(\tau)|^2}e^{-2\pi i k \left(\frac{\theta_r}{2\pi} \right)^2} \Theta^{(k)}_l  \left(\begin{array}{c}
            \theta_r\\\theta_s+\theta_r
        \end{array} \right)(\tau) \overline{\Theta}^{(k)}_l\left(\begin{array}{c}
            0\\0
        \end{array} \right)(\tau)\\
        =&e^{-2\pi i k \frac{\theta_r}{2\pi} \left(\frac{\theta_s+\theta_r}{2\pi}\right)} \sum_{l=0}^{2k-1}\frac{1}{|\eta(\tau)|^2}\Theta^{(k)}_l \left(\begin{array}{c}
            \theta_r\\\theta_s+\theta_r
        \end{array} \right)(\tau) \overline{\Theta}^{(k)}_l\left(\begin{array}{c}
            0\\0
        \end{array} \right)(\tau)\\
        =&Z[\theta_r,\theta_s+\theta_r](\tau)\,.
    \end{split}
    \end{equation}
And under the $S$-transformation, one has
\begin{equation}
    \begin{split}
        &Z[\theta_r,\theta_s](-1/\tau)\\
        =&e^{-2\pi i k \frac{\theta_r}{2\pi} \frac{\theta_s}{2\pi}} \sum_{l=0}^{2k-1}\frac{1}{|\eta(-1/\tau)|^2}\Theta^{(k)}_l\left(\begin{array}{c}
            \theta_r\\\theta_s
        \end{array} \right)(-1/\tau) \overline{\Theta}^{(k)}_l\left(\begin{array}{c}
            0\\0
        \end{array} \right)(-1/\tau)\\
        =&e^{-2\pi i k \frac{\theta_r}{2\pi} \frac{\theta_s}{2\pi}} \sum_{l=0}^{2k-1}\frac{1}{|\eta(\tau)|^2|\sqrt{-i\tau}|^2}\times \frac{|\sqrt{-i\tau}|^2}{2k}e^{4\pi i k \frac{\theta_s}{2\pi}\frac{\theta_r}{2\pi}}\\
        &\times \sum_{l'=0}^{2k-1} \sum_{l''=0}^{2k-1}e^{-2\pi i \frac{l l'}{2k}} e^{2\pi i \frac{l l''}{2k}}\Theta^{(k)}_{l'}\left(\begin{array}{c}
            \theta_s\\-\theta_r
        \end{array} \right)(\tau) \overline{\Theta}^{(k)}_{l''}\left(\begin{array}{c}
            0\\0
        \end{array} \right)(\tau)\\
        =&e^{2\pi i k \frac{\theta_r}{2\pi} \frac{\theta_s}{2\pi}} \sum_{l'=0}^{2k-1}\sum_{l''=0}^{2k-1}\frac{1}{|\eta(\tau)|^2}\frac{1}{2k}\sum_{l=0}^{2k-1} e^{\frac{2\pi i}{2k} l(l''-l')}\Theta^{(k)}_{l'}\left(\begin{array}{c}
            \theta_s\\-\theta_r
        \end{array} \right)(\tau) \overline{\Theta}^{(k)}_{l''}\left(\begin{array}{c}
            0\\0
        \end{array} \right)(\tau)\\
        =&e^{2\pi i k \frac{\theta_r}{2\pi} \frac{\theta_s}{2\pi}} \sum_{l'=0}^{2k-1}\sum_{l''=0}^{2k-1}\frac{1}{|\eta(\tau)|^2} \delta_{l',l''} \Theta^{(k)}_{l'}\left(\begin{array}{c}
            \theta_s\\-\theta_r
        \end{array} \right)(\tau) \overline{\Theta}^{(k)}_{l''}\left(\begin{array}{c}
            0\\0
        \end{array} \right)(\tau)\\
        =&e^{2\pi i k \frac{\theta_r}{2\pi} \frac{\theta_s}{2\pi}} \sum_{l'=0}^{2k-1}\frac{1}{|\eta(\tau)|^2} \Theta^{(k)}_{l'}\left(\begin{array}{c}
            \theta_s\\-\theta_r
        \end{array} \right)(\tau) \overline{\Theta}^{(k)}_{l'}\left(\begin{array}{c}
            0\\0
        \end{array} \right)(\tau)\\
        =&Z[\theta_s,-\theta_r](\tau)\,.
    \end{split}
\end{equation}

\section{Candidate boundary data and topological boundary states for $U(1)$ SymTFT}\label{app:check-of-Lagrangian-algebra-U(1)}

In this appendix, we check the candidate boundary data $\mathcal{L}$ for $U(1)$ SymTFT, whose coefficients are written as a column vector $V_{\mathcal{L}}$ satisfying
    \begin{equation}
        S\cdot V_{\mathcal{L}}=V_{\mathcal{L}}\,,\quad T\cdot V_{\mathcal{L}}=V_{\mathcal{L}}\,,
    \end{equation}
where $S,T$ are modular matrices. We then prove the modular properties of the topological boundary states on torus introduced in \eqref{eq:U(1)-holonomy-states}.

\subsection*{Candidate data for $U(1)$ SymTFT with anomaly $k\in H^4(BU(1),\mathbb{Z})$}

First of all, the candidate $S$- and $T$-kernels for $U(1)$ SymTFT are \eqref{eq:U(1)-modular-matrices}
\begin{equation}
\begin{split}
&S_{(x,n),(y,n')}=e^{-i(n\theta_y+n'\theta_x-\frac{2k}{2\pi}\theta_x\theta_y)}\,,\\
&T_{(x,n),(y,n')}=\delta_{x,y}\delta_{n,n'}\exp\left(i\left(n\theta_x-\frac{k\theta_x^2}{2\pi}\right)\right)\,,
\end{split}
\end{equation}
and there are two kinds of candidate data. The Dirichlet-type one
    \begin{equation}
        \mathcal{L}_{\textrm{Dir}} = \bigoplus_n W_{(1,n)}\,,
    \end{equation}
for any $k$, and also
    \begin{equation}
        \mathcal{L}^k_q = \bigoplus_m \bigoplus_{n=0}^{q-1} W_{\left(e^{\frac{2\pi i}{q}n},\frac{k}{q}n+qm\right)}\,,
    \end{equation}
for any positive integer $q$ satisfying $\frac{k}{q}\in \mathbb{Z}$. One can check the topological spin for each component in $\mathcal{L}_{\textrm{Dir}}$ and $\mathcal{L}^k_q$ is one, therefore one trivially has
    \begin{equation}
        T \cdot V_{\mathcal{L}_{\textrm{Dir}}} = V_{\mathcal{L}_{\textrm{Dir}}}\,, \quad T \cdot V_{\mathcal{L}^k_q} = V_{\mathcal{L}^k_q}\,,
    \end{equation}
and we only need to check $S$-matrix acting on them
    \begin{equation}
        (S\cdot V)_{(x,n)} = \frac{1}{2\pi} \int_0^{2\pi} d\theta_y\sum_{n'} S_{(x,n),(y,n')} V_{(y,n')}\,.
    \end{equation}
For $\mathcal{L}_{\textrm{Dir}}$, we write the vector
    \begin{equation}
        (V_{\mathcal{L}_{\rm Dir}})_{(x,n)} = \delta_{2\pi}(\theta_x)\,,
    \end{equation}
and we have
    \begin{equation}
        (S\cdot V_{\mathcal{L}_{\textrm{Dir}}})_{x,n} = \frac{1}{2\pi}\sum_{n'} S_{(x,n),(1,n')} =\sum_{n'} \frac{1}{2\pi} e^{-in' \theta_x} = \delta_{2\pi}(\theta_x)\,.
    \end{equation}
For $\mathcal{L}^k_q$, one writes
    \begin{equation}
        (V_{\mathcal{L}^k_{q}})_{(x,n)}= \sum_{m=0}^{q-1}\sum_{m'} \delta_{2\pi}(\theta_x - \frac{2\pi}{q}m) \delta_{n,\frac{k}{q}m+qm'}\,,
    \end{equation}
and one has
    \begin{equation}
        \begin{split}
            &(S\cdot V_{\mathcal{L}^k_q})_{x,n}= \frac{1}{2\pi}\sum_{n'=0}^{q-1}\sum_{m}  S_{(x,n),(e^{\frac{2\pi i}{q}n'},\frac{k}{q}n'+qm)}\\
            =&\sum_{n'=0}^{q-1}\sum_{m} \frac{1}{2\pi} e^{-i \left( \frac{2\pi}{q}n n' + \left(\frac{k}{q}n'+qm \right)\theta_x-\frac{2k}{2\pi}\left(\frac{2\pi}{q}n' \right)\theta_x\right)}\\
            =& \frac{1}{2\pi} \sum_{m} e^{-im q\theta_x} \sum_{n'=0}^{q-1} e^{-i \frac{2\pi}{q}n' \left(n - \frac{k}{2\pi}\theta_x \right)}\\
            =& q \delta_{2\pi}(q \theta_x) \sum_{m'}\delta_{n,\frac{k}{2\pi}\theta_x + qm'}=\sum_{m=0}^{q-1}\sum_{m'} \delta_{2\pi}(\theta_x - \frac{2\pi}{q}m) \delta_{n,\frac{k}{q}m+qm'}\,.
        \end{split} 
    \end{equation}

\subsection*{Modular properties of topological boundary state $|\theta_r,\theta_s\rangle$}
We then derive the modular transformation \eqref{eq:U(1)-boundary-states-modular-transformation} of the topological boundary state $|\theta_r,\theta_s\rangle$ defined in \eqref{eq:U(1)-holonomy-states}
\begin{equation}
        |\theta_r,\theta_s\rangle \equiv e^{-2\pi i k \frac{\theta_s}{2\pi} \frac{ \theta_r}{2\pi}}W_{(x_r,0)}[-\Gamma_1] W_{(x_s,0)}[\Gamma_2] |0,0\rangle\,,
    \end{equation}
using the commutation relations of line operators \eqref{eq:U(1)-commutator-of-lines}
    \begin{equation}
        W_{(x,n)} [\Gamma_2] W_{(y,n')}[\Gamma_1] = e^{-i(n\theta_y+n'\theta_x-\frac{2k}{2\pi}\theta_x\theta_y)}W_{(y,n')}[\Gamma_1]W_{(x,n)} [\Gamma_2]\,.
    \end{equation}
First, under the $S$-transformation, one has
    \begin{equation}
    \begin{split}
        &\hat{S}|\theta_r,\theta_s\rangle\\
        =& e^{-2\pi i k \frac{\theta_s}{2\pi} \frac{ \theta_r}{2\pi}}W_{(x_r,0)}[\Gamma_2] W_{(x_s,0)}[\Gamma_1] |0,0\rangle\\
        =& e^{-2\pi i k \frac{\theta_s}{2\pi} \frac{ \theta_r}{2\pi}} \times e^{i\frac{2k}{2\pi}\theta_x \theta_y}W_{(x_s,0)}[\Gamma_1]W_{(x_r,0)}[\Gamma_2]|0,0\rangle\\
        =&e^{2\pi i k \frac{\theta_s}{2\pi} \frac{ \theta_r}{2\pi}}W_{(x_s,0)}[\Gamma_1]W_{(x_r,0)}[\Gamma_2]|0,0\rangle\\
        =&e^{-2\pi i k \left(\frac{-\theta_s}{2\pi}\right) \frac{ \theta_r}{2\pi}}W_{(x_s^{-1},0)}[-\Gamma_1]W_{(x_r,0)}[\Gamma_2]|0,0\rangle = |-\theta_s,\theta_r\rangle\,.
    \end{split}
    \end{equation}
Next, under the $T$-transformation, one has
    \begin{equation}
        \hat{T}|\theta_r,\theta_s\rangle = e^{-2\pi i k \frac{\theta_s}{2\pi} \frac{ \theta_r}{2\pi}}W_{(x_r,0)}[-\Gamma_1-\Gamma_2] W_{(x_s,0)}[\Gamma_2] |0,0\rangle\,,
    \end{equation}
and we need to resolve $W_{(x_r,0)}[-\Gamma_1-\Gamma_2]$. Notice that in the Lagrangian description of the $U(1)$ SymTFT, which is the BF theory with twist term $kCS(A)$, the line operator $W_{(x,0)}$ is the Wilson loop of $B$-field
    \begin{equation}
        W_{(x,0)}[\Gamma] = \exp\left(i\frac{\theta}{2\pi}\oint_{\Gamma}  B\right)\,,
    \end{equation}
and we can introduce the holonomy operator $\hat{b}_1,\hat{b}_2$ along $\Gamma_1$ and $\Gamma_2$ according to
    \begin{equation}
        W_{(x,0)}[\Gamma_i]=\exp\left(i\frac{\theta}{2\pi} \hat{b}_i \right)\,,\quad \textrm{with}\quad i=1,2\,.
    \end{equation}
Since the BF theory is quadratic, the commutator $\left[\hat{b}_1,\hat{b}_2\right]$ is a constant and can be deduced from the BCH-formula
    \begin{equation}
    \begin{split}
        &W_{(x_1,0)}[\Gamma_1]W_{(x_2,0)}[\Gamma_2]\\
        =& \exp\left(i\frac{\theta_1}{2\pi} \hat{b}_1 +i\frac{\theta_2}{2\pi} \hat{b}_2 +\frac{1}{2} \left[i\frac{\theta_1}{2\pi} \hat{b}_1 ,i\frac{\theta_2}{2\pi} \hat{b}_2  \right] \right)\\
        =&\exp\left(i\frac{\theta_1}{2\pi} \hat{b}_1 +i\frac{\theta_2}{2\pi} \hat{b}_2\right) \exp\left(-\frac{1}{2}\frac{\theta_1}{2\pi}\frac{\theta_2}{2\pi} \left[\hat{b}_1 , \hat{b}_2  \right] \right)\,,
    \end{split}
    \end{equation}
and \eqref{eq:U(1)-commutator-of-lines} then implies $\left[\hat{b}_1,\hat{b}_2\right]=4\pi i k $. Setting $x_1=x_2=x_r$, one has
\begin{equation}
    W_{(x_r,0)}[\Gamma_1]W_{(x_r,0)}[\Gamma_2] =W_{(x_r,0)}[\Gamma_1+\Gamma_2] e^{-2\pi i k \left(\frac{\theta_r}{2\pi}\right)^2}\,.
\end{equation}
Therefore one has
    \begin{equation}
    \begin{split}
        &\hat{T}|\theta_r,\theta_s\rangle\\
        =& e^{-2\pi i k \frac{\theta_s}{2\pi} \frac{ \theta_r}{2\pi}}W_{(x_r,0)}[-\Gamma_1-\Gamma_2] W_{(x_s,0)}[\Gamma_2] |0,0\rangle\\
        =& e^{-2\pi i k \frac{\theta_s}{2\pi} \frac{ \theta_r}{2\pi}} e^{2\pi i k \left(\frac{\theta_r}{2\pi}\right)^2}W_{(x_r,0)}[-\Gamma_1]W_{(x_r,0)}[-\Gamma_2]W_{(x_s,0)}[\Gamma_2] |0,0\rangle\\
        =&e^{-2\pi i k \left(\frac{\theta_s-\theta_r}{2\pi}\right) \frac{ \theta_r}{2\pi}}W_{(x_r,0)}[-\Gamma_1]W_{(x_sx^{-1}_r,0)}[\Gamma_2] = |\theta_r,\theta_s-\theta_r\rangle\,.
    \end{split}
    \end{equation}

\section{Properties of candidate modular kernels for continuous $G$}\label{app:proof-of-SU(2)-modular-properties}

In this appendix, we examine the candidate modular kernels introduced in \eqref{eq:S_kernel_final} and \eqref{eq:T_matrix} and check that they satisfy
\begin{equation}
    SS^{\dagger}=TT^{\dagger}=1\,,\quad S^2 = (ST)^3 =C\,,
\end{equation}
in the regular sector where $C$ is the charge conjugation matrix, which will be introduced later. Here, the summation over a continuous index, which labels the conjugacy class, should be understood as an integral. For the particular integrated identities checked in this appendix, it is sufficient to restrict to regular conjugacy classes $[g]$ such that $C_G([g])=T$ is the Cartan torus, because the irregular classes have measure zero with respect to the integration measure used here. This simplification should not be confused with the treatment of the full kernels in the main text, where singular sectors are kept explicitly when needed. As an illustration, we will mainly focus on $G=SU(N)$, although the derivation should also apply to other gauge groups.

We parametrize the maximal torus $T$ of $SU(N)$ according to
    \begin{equation}
        \exp \left( i \sum_{i=1}^{N-1} \alpha_i H_i \right)= \left(\begin{array}{ccccc}
            e^{i\alpha_1/2} &&&&  \\
             & e^{i\alpha_2/2}&&&\\
             && \ddots &&\\
             &&&e^{i\alpha_{N-1}/2}&\\
             &&&&e^{-i(\alpha_1+\cdots+\alpha_{N-1})/2}
        \end{array}\right)\,,
    \end{equation}
with the Cartan generators $H_i$ given by
    \begin{equation}
        (H_i)_{ab} = \left\{ \begin{array}{l}
             \frac{1}{2}\,, \quad a=b=i\,,\\
             -\frac{1}{2}\,, \quad a=b=N\,,\\
             0\,, \quad \textrm{others}\,,
        \end{array}\right. \quad (i,j=1,\cdots,N-1)
    \end{equation}
and the Killing metric $g_{ij}$ is defined as
    \begin{equation}\label{eq:Killing-metric-SU(N)}
        g_{ij} = 2\textrm{tr} (H_i H_j) =  \delta_{i,j} + \frac{1}{2}\delta_{i,j+1}+\frac{1}{2}\delta_{i+1,j}\,.
    \end{equation}
The maximal torus $T$ is parametrized by $\alpha=(\alpha_1,\cdots,\alpha_{N-1})$ with $\alpha_i \sim \alpha_i+4\pi$. The representations of $T$ are also labeled by $N-1$ integers $n=(n_1,\cdots,n_{N-1})$. The Weyl group $W=S_N$ acts as an automorphism on $T$ and there exists a canonical homeomorphism 
    \begin{equation}
        T/W \cong Cl(G)\,,
    \end{equation}
where the action of Weyl group on the vector $\alpha$ is induced by
    \begin{equation}\label{eq:Weyl-action-on-vectors}
        \alpha \cdot w(H) = w(\alpha) \cdot H\,,\quad \textrm{with} \quad \alpha \cdot H\equiv \sum_{i=1}^{N-1} \alpha_i H_i\,.
    \end{equation}
For example, we can choose the generators of $S_N$ as transpositions $(1i)$ of the diagonal elements with $i=2,\cdots,N$. If $i\neq N$, then $(1i)$ will simply swap $H_1$ and $H_i$, therefore exchange $\alpha_1$ with $\alpha_i$; if $i=N$, then one finds the action of $(1N)$ as
    \begin{equation}
        (1N)(H_i) = \left\{\begin{array}{l}
            -H_1\,,\quad i=1\,,\\
            H_i-H_1\,,\quad 1<i<N\,,
        \end{array} \right.
    \end{equation}
so that it maps $\alpha_1\rightarrow -\alpha_1-\cdots -\alpha_{N-1}$. As a byproduct, one can check the periodic Delta function $\delta_{4\pi}(\alpha)$ defined as
    \begin{equation}
        \delta_{4\pi}(\alpha) = \sum_{k_1,\cdots,k_{N-1}\in \mathbb{Z}}\prod_{i=1}^{N-1} \delta(\alpha_i-4\pi k_i)
    \end{equation}
is invariant under the Weyl transformation
    \begin{equation}
        \delta_{4\pi}(w(\alpha)) = \delta_{4\pi} (\alpha)\,.
    \end{equation}

Suppose we pick a region $\mathcal{C}$ on the $\alpha$-plane that parametrize the conjugacy class $Cl(SU(N))$, then we introduce the region $\mathcal{F}$ on the $\alpha$-plane as the disjoint union of the Weyl orbit of $\mathcal{C}$
    \begin{equation}
        \mathcal{F} = \bigsqcup_{w\in S_N} w[\mathcal{C}]\,.
    \end{equation}
Upon identifying $\alpha_i \sim \alpha_i+4\pi$, $\mathcal{F}$ parametrize the maximal torus $T$.

With the above notations, the candidate kernels \eqref{eq:S_kernel_final} and \eqref{eq:T_matrix} are written as
\begin{equation}
    \begin{split}
        S^{(k)}_{([g],n),([h],m)} =& \sum_{w\in S_N} e^{-i\frac{k}{4\pi} \alpha \cdot w(\beta)} e^{-\frac{i}{2} n \cdot w(\beta)-\frac{i}{2} m\cdot w^{-1}(\alpha)}\,,\\
        T^{(k)}_{([g],n),([h],m)} =& \delta_{4\pi} (\alpha -\beta) \delta_{n,m} e^{\frac{ik}{8\pi} \alpha \cdot \alpha} e^{\frac{i}{2} n\cdot \alpha}\,,
    \end{split}
\end{equation}
with $g = e^{i\alpha\cdot H},\ h = e^{i\beta\cdot H}$ and $\alpha,\beta \in \mathcal{C}$.

\subsection{Proof of $S S^{\dagger}=TT^{\dagger}=1$}

To check the unitarity of $S$-matrix, let us compute
\begin{equation}
    \begin{split}
&\sum_{\beta\in \mathcal{C},m}S_{(\alpha,n),(\beta,m)}  S^*_{(\beta,m),(\gamma,p)} \\
=&\sum_{\beta\in \mathcal{C},m} \sum_{w,w'} e^{-i\frac{k}{8\pi} (\alpha \cdot w(\beta) + w^{-1}(\alpha)\cdot \beta)} e^{-\frac{i}{2} n \cdot w(\beta)-\frac{i}{2} m \cdot w^{-1}(\alpha)} e^{i \frac{k}{8\pi} (\beta \cdot w'(\gamma)+w'^{-1}(\beta) \cdot \gamma)} e^{\frac{i}{2}m \cdot w'(\gamma)+\frac{i}{2}p\cdot w'^{-1}(\beta)}\\
=&\sum_{\beta\in \mathcal{C},w,w'}e^{-i\frac{k}{8\pi} (\alpha \cdot w(\beta) + w^{-1}(\alpha)\cdot \beta)}  e^{-\frac{i}{2} n \cdot w(\beta)} e^{i \frac{k}{8\pi} (\beta \cdot w'(\gamma)+w'^{-1}(\beta) \cdot \gamma)} e^{\frac{i}{2} p\cdot w'^{-1}(\beta)} \delta_{4\pi}(w^{-1}(\alpha)-w'(\gamma))\,,
    \end{split}
\end{equation}
where $\sum_{\beta \in \mathcal{C}}$ should be understood as integral $\int_{\mathcal{C}} d^{N-1}\beta$ in the region $\mathcal{C}$.
Since we restrict both $\alpha$ and $\beta$ to the conjugacy class chamber $\mathcal{C}$, then one must have $w^{-1} = w'$ for the delta function to be non-vanishing. We then have
\begin{equation}
    \begin{split}
        &\sum_{\beta\in \mathcal{C},m}S_{(\alpha,n),(\beta,m)}  S^*_{(\beta,m),(\gamma,p)}\\
        =&\sum_{\beta\in \mathcal{C},w}e^{-i\frac{k}{8\pi} (\alpha \cdot w(\beta) + w^{-1}(\alpha)\cdot \beta)} e^{-\frac{i}{2} n \cdot w(\beta)} e^{i \frac{k}{8\pi} (\beta \cdot w^{-1}(\gamma)+w(\beta) \cdot \gamma)} e^{\frac{i}{2} p\cdot w(\beta)} \delta_{4\pi}(w^{-1}(\alpha)-w^{-1}(\gamma))\\
        =&\sum_{\beta\in \mathcal{C},w}e^{-i\frac{k}{8\pi} (\alpha \cdot w(\beta) + w^{-1}(\alpha)\cdot \beta)} e^{-\frac{i}{2} n \cdot w(\beta)} e^{i \frac{k}{8\pi} (\beta \cdot w^{-1}(\gamma)+w(\beta) \cdot \gamma)} e^{\frac{i}{2} p\cdot w(\beta)} \delta_{4\pi}(\alpha-\gamma)\\
        =&\sum_{\beta\in \mathcal{C},w} e^{-\frac{i}{2} n \cdot w(\beta)}  e^{ip\cdot w(\beta)} \delta_{4\pi}(\alpha-\gamma)\\
        =& \sum_{\beta\in \mathcal{F}}e^{-\frac{i}{2} n \cdot \beta}  e^{\frac{i}{2}p\cdot \beta} \delta_{4\pi}(\alpha-\gamma)\\
        =& \delta_{n,p} \delta_{4\pi}(\alpha-\gamma)\,,     
    \end{split}
\end{equation}
so that we have proved $SS^{\dagger}=1$. The proof for $TT^{\dagger}=1$ is trivial since $T$ is a diagonal and each element is a pure phase.

\subsection{Calculation of $S^2=C$}
Moreover, one can check $S^2$ is
\begin{equation}
    \begin{split}
&\sum_{\beta\in \mathcal{C},m}S_{(\alpha,n),(\beta,m)}  S_{(\beta,m),(\gamma,p)} \\
=&\sum_{\beta\in \mathcal{C},m} \sum_{w,w'} e^{-i\frac{k}{4\pi} \alpha \cdot w(\beta)} e^{-\frac{i}{2} n \cdot w(\beta)-\frac{i}{2} m \cdot w^{-1}(\alpha)} e^{-i \frac{k}{4\pi} \beta \cdot w'(\gamma)} e^{-\frac{i}{2}m \cdot w'(\gamma)-\frac{i}{2}p\cdot w'^{-1}(\beta)}\\
=&\sum_{\beta\in \mathcal{C},w,w'}e^{-i\frac{k}{4\pi} \alpha \cdot w(\beta)} e^{-\frac{i}{2} n \cdot w(\beta)} e^{-i \frac{k}{4\pi} \beta \cdot w'(\gamma)} e^{-\frac{i}{2}p\cdot w'^{-1}(\beta)} \delta_{4\pi}(w^{-1}(\alpha)+w'(\gamma))\,,
    \end{split}
\end{equation}
where we define the charge conjugation matrix $C$ as
    \begin{equation}\label{eq:charge-conjugation-matrix}
        C_{(\alpha,n),(\gamma,p)} = \sum_{\beta,w,w'}e^{-i\frac{k}{4\pi} \alpha \cdot w(\beta)} e^{-\frac{i}{2} n \cdot w(\beta)} e^{-i \frac{k}{4\pi} \beta \cdot w'(\gamma)} e^{-\frac{i}{2}p\cdot w'^{-1}(\beta)} \delta_{4\pi}(w^{-1}(\alpha)+w'(\gamma))\,,
    \end{equation}
which can be further simplified as
\begin{equation}
    C_{(\alpha,n),(\gamma,p)} = \sum_{w'} \delta_{n+w'(p)+\frac{k}{2\pi}g(\alpha+w'(\gamma)),0}\delta_{4\pi}(\alpha+w'(\gamma))\,,
\end{equation}
and we will return to that later.

\subsection{Proof of $(ST)^3=C$}
As a non-trivial check, we will also show that $(ST)^3=C$ with $C$ defined in \eqref{eq:charge-conjugation-matrix}. First of all, $ST$ is given by
    \begin{equation}
    \begin{split}
        (ST)_{(\alpha,n),(\gamma\,p)} =& \sum_w \sum_{\beta\in \mathcal{C},m} e^{-i\frac{k}{4\pi} \alpha \cdot w(\beta) } e^{-\frac{i}{2} n \cdot w(\beta)-\frac{i}{2} m \cdot w^{-1}(\alpha)} e^{\frac{i}{2} m\cdot \beta} e^{i \frac{k}{8\pi} \beta \cdot \beta} \delta_{4\pi}(\beta-\gamma)\delta_{m,p}\\
        =&\sum_w e^{-i\frac{k}{4\pi} \alpha \cdot w(\gamma) } e^{-\frac{i}{2} n \cdot w(\gamma)-\frac{i}{2} m \cdot w^{-1}(\alpha)} e^{\frac{i}{2} p\cdot \gamma} e^{i \frac{k}{8\pi} \gamma \cdot \gamma}\,,
    \end{split}
    \end{equation}
and $(ST)^3$ reads
    \begin{equation}
        \begin{split}
            &(ST)^3_{(\alpha,n),{(\delta,q)}}\\
            =& \sum_{w_1,w_2,w_3} \sum_{m,p}\sum_{\beta,\gamma\in \mathcal{C}}e^{-\frac{ik}{4\pi}(\alpha \cdot w_1(\beta)+\beta\cdot w_2(\gamma)+\gamma\cdot w_3(\delta))} e^{\frac{ik}{8\pi}(\beta\cdot \beta + \gamma\cdot \gamma + \delta \cdot \delta) }\\
            &\times e^{-\frac{i}{2} n \cdot w_1(\beta)-\frac{i}{2} m \cdot w_2(\gamma)-\frac{i}{2} p \cdot w_3(\delta)} e^{-\frac{i}{2}m \cdot w_{1}^{-1}(\alpha) -\frac{i}{2} p \cdot w_{2}^{-1}(\beta) -\frac{i}{2} q\cdot w_{3}^{-1}(\gamma)} e^{\frac{i}{2}m\cdot \beta + \frac{i}{2} p\cdot \gamma + \frac{i}{2}q\cdot \delta}\\
            =&\sum_{w_1,w_2,w_3} \sum_{\beta,\gamma\in \mathcal{C}}e^{-\frac{ik}{4\pi}(\alpha \cdot w_1(\beta)+\beta\cdot w_2(\gamma)+\gamma\cdot w_3(\delta))} e^{\frac{ik}{8\pi}(\beta\cdot \beta + \gamma\cdot \gamma + \delta \cdot \delta)}\\
            &\times e^{-\frac{i}{2} n \cdot w_1(\beta) - \frac{i}{2} q \cdot w_{3}^{-1}(\gamma)+\frac{i}{2}q \cdot \delta} \delta_{4\pi}(\beta-w_{1}^{-1}(\alpha)-w_2(\gamma)) \delta_{4\pi}(\gamma -w_2^{-1}(\beta)-w_3(\delta))\,.
        \end{split}
    \end{equation}
    
To proceed, we use the identity $\alpha \cdot \beta = w(\alpha)\cdot w(\beta)$ for any $w\in S_3$. To show this, one has
\begin{equation}
    w(\alpha) \cdot w(\beta) = 2\textrm{Tr} \left(\sum_{i,j} w(\alpha)_i w(\beta)_j H_i H_j \right) = 2\textrm{Tr}\left(\sum_{i,j}\alpha_i \beta_j w(H_i) w(H_j)\right)\,.
\end{equation}
Since $w$ acts as permutation of the diagonal components, we can write $w(H) = U_w HU_w^{\dagger}$ for certain $U_w \in SU(N)$. We then have
\begin{equation}
    2\textrm{Tr}\left(\sum_{i,j}\alpha_i \beta_j w(H_i) w(H_j)\right)=2\textrm{Tr}\left(\sum_{i,j}\alpha_i \beta_j U_w H_i H_j U^{\dagger}_w\right)= 2\textrm{Tr}\left(\sum_{i,j}\alpha_i \beta_j H_i H_j\right) = \alpha \cdot \beta\,.
\end{equation}
Similarly, one can check $\alpha\cdot w(\beta)=w^{-1}(\alpha)\cdot \beta$.

Apply the identities, we can write
    \begin{equation}
    \begin{split}
        &e^{\frac{ik}{8\pi}(\beta\cdot \beta + \gamma\cdot \gamma + \delta \cdot \delta)}e^{-\frac{ik}{4\pi}(\beta\cdot w_2(\gamma)+\gamma\cdot w_3(\delta))}\\
        =&e^{\frac{ik}{8\pi}(w_2^{-1}(\beta)\cdot w_2^{-1}(\beta) + \gamma\cdot \gamma + w_3(\delta) \cdot w_3(\delta))}e^{-\frac{ik}{4\pi}(w_2^{-1}(\beta)\cdot \gamma+\gamma\cdot w_3(\delta))}\\
        =&e^{\frac{ik}{8\pi}(w_2^{-1}(\beta)-\gamma+w_3(\delta))^2} e^{-\frac{ik}{4\pi} w_2^{-1}(\beta)\cdot w_3(\delta)}=e^{-\frac{ik}{4\pi} w_2^{-1}(\beta)\cdot w_3(\delta)}\,,
    \end{split}
    \end{equation}
where the last step is because the delta function $\delta_{4\pi}(\gamma-w_2^{-1}(\beta)-w_3(\delta))$ impose 
    \begin{equation}
        \gamma-w_2^{-1}(\beta)-w_3(\delta) = 4\pi a \equiv 4\pi(a_1,a_2,\cdots,a_{N-1})\,,
    \end{equation}
for certain $a_i \in \mathbb{Z}$. We then have
    \begin{equation}
        (\gamma-w_2^{-1}(\beta)-w_3(\delta))^2 = 16\pi^2 \sum_{i,j}g_{ij} a_i a_j = 16\pi^2\left(\sum_{i\leq j} a_i a_j \right)\,,
    \end{equation}
where $g_{ij}$ is the Killing metric introduced in \eqref{eq:Killing-metric-SU(N)}, and one has
    \begin{equation}
        e^{\frac{ik}{8\pi}(w_2^{-1}(\beta)-\gamma+w_3(\delta))^2} = e^{2\pi i (\sum_{i\leq j} a_i a_j)} = 1\,.
    \end{equation}

Then $(ST)^3_{(\alpha,n),{(\delta,q)}}$ becomes
\begin{equation}
\begin{split}
            &\sum_{w_1,w_2,w_3} \sum_{\beta,\gamma\in \mathcal{C}}e^{-\frac{ik}{4\pi}(\alpha \cdot w_1(\beta)+w_2^{-1}(\beta)\cdot w_3(\delta))} e^{-\frac{i}{2} n \cdot w_1(\beta) - \frac{i}{2} q \cdot w_{3}^{-1}(\gamma)+\frac{i}{2}q \cdot \delta}\\ &\times \delta_{4\pi}(\beta-w_{1}^{-1}(\alpha)-w_2(\gamma)) \delta_{4\pi}(\gamma -w_2^{-1}(\beta)-w_3(\delta))\\
            =&\sum_{w_1,w_2,w_3} \sum_{\beta,\gamma \in \mathcal{C}}e^{-\frac{ik}{4\pi}(\alpha \cdot w_1(\beta)+w_2^{-1}(\beta)\cdot w_3(\delta))} e^{-\frac{i}{2} n \cdot w_1(\beta) - \frac{i}{2} q \cdot w_3^{-1} w_2^{-1} (\beta)}\\ &\times \delta_{4\pi}(\beta-w_{1}^{-1}(\alpha)-w_2(\gamma)) \delta_{4\pi}(\gamma -w_2^{-1}(\beta)-w_3(\delta))\\
            =&\sum_{w_1,w_2,w_3} \sum_{\beta,\gamma \in \mathcal{C}}e^{-\frac{ik}{4\pi}(\alpha \cdot w_1(\beta)+w_3^{-1}w_2^{-1}(\beta)\cdot \delta)} e^{-\frac{i}{2} n \cdot w_1(\beta) - \frac{i}{2} q \cdot w_3^{-1} w_2^{-1} (\beta)} \\ &\times \delta_{4\pi}(\beta-w_{1}^{-1}(\alpha)-w_2(\gamma)) \delta_{4\pi}(w_2(\gamma) -\beta-w_2 w_3(\delta))\,.
    \end{split}
\end{equation}
We can shift $w_3 \rightarrow w_2^{-1} w_3$ and get
\begin{equation}
\begin{split}
    &\sum_{w_1,w_2,w_3} \sum_{\beta,\gamma\in \mathcal{C}}e^{-\frac{ik}{4\pi}(\alpha \cdot w_1(\beta)+w_3^{-1}(\beta)\cdot \delta)} e^{-\frac{i}{2} n \cdot w_1(\beta) - \frac{i}{2} q \cdot w_3^{-1}(\beta)} \delta_{4\pi}(\beta-w_{1}^{-1}(\alpha)-w_2(\gamma)) \delta_{4\pi}(w_2(\gamma) -\beta-w_3(\delta))\\
    =& \sum_{w_1,w_3}\sum_{\beta\in \mathcal{C}}e^{-\frac{ik}{4\pi}(\alpha \cdot w_1(\beta)+w_3^{-1}(\beta)\cdot \delta)} e^{-\frac{i}{2} n \cdot w_1(\beta) - \frac{i}{2} q \cdot w_3^{-1}(\beta)} \delta_{4\pi}(w_{1}^{-1}(\alpha)+w_3(\delta))\,.
\end{split}
\end{equation}
After relabling the indices
    \begin{equation}
        w_1 \rightarrow w\,, \quad w_3 \rightarrow w'\,,\quad \delta\rightarrow \gamma\,,\quad q\rightarrow p\,,
    \end{equation}
we have exactly
\begin{equation}
    (ST)^3_{(\alpha,n),{(\gamma,p)}} = C_{(\alpha,n),(\gamma,p)}\,,
\end{equation}
with $C$ given in \eqref{eq:charge-conjugation-matrix}, which is the same to $S^2$.

\subsection{Charge conjugation matrix $C$}
We can further simplify the charge conjugation matrix 
\begin{equation}
        C_{(\alpha,n),(\gamma,p)} = \sum_{\beta,w,w'}e^{-i\frac{k}{4\pi} \alpha \cdot w(\beta)} e^{-\frac{i}{2} n \cdot w(\beta)} e^{-i \frac{k}{4\pi} \beta \cdot w'(\gamma)} e^{-\frac{i}{2}p\cdot w'^{-1}(\beta)} \delta_{4\pi}(w^{-1}(\alpha)+w'(\gamma))\,,
    \end{equation}
by replacing $\delta_{4\pi}(w^{-1}(\alpha)+w'(\gamma))=\delta_{4\pi}(\alpha+ww'(\gamma))$ and shifting $w'\rightarrow w^{-1}w'$
    \begin{equation}
    \begin{split}
        C_{(\alpha,n),(\gamma,p)} &= \sum_{\beta,w,w'}e^{-i\frac{k}{4\pi} \alpha \cdot w(\beta)} e^{-\frac{i}{2} n \cdot w(\beta)} e^{-i \frac{k}{4\pi} \beta \cdot w^{-1}w'(\gamma)} e^{-\frac{i}{2}p\cdot w'^{-1}w(\beta)} \delta_{4\pi}(\alpha+w'(\gamma))\\
        &=\sum_{\beta,w,w'}e^{-i\frac{k}{4\pi} \alpha \cdot w(\beta)} e^{-\frac{i}{2} n \cdot w(\beta)} e^{-i \frac{k}{4\pi} w(\beta) \cdot w'(\gamma)} e^{-\frac{i}{2}w'(p)\cdot w(\beta)} \delta_{4\pi}(\alpha+w'(\gamma)).
    \end{split}
    \end{equation}
Here the action of Weyl group on the charge $p$ is similarly induced by
    \begin{equation}
        \sum_{i=1}^{N-1}w^{-1}(p)_i \alpha_i = \sum_{i=1}^{N-1} p_i w(\alpha)_i\,,
    \end{equation}
and it is easy to check $w^{-1}(p)$ are still a collection of integers following the discussion below \eqref{eq:Weyl-action-on-vectors}. Then we can sum over $w$ and integrate $\beta$ to get
\begin{equation}
    C_{(\alpha,n),(\gamma,p)} = \sum_{w'} \delta_{n+w'(p)+\frac{k}{2\pi}g(\alpha+w'(\gamma)),0}\delta_{4\pi}(\alpha+w'(\gamma))\,,
\end{equation}
where $g(\alpha+w'(\gamma))$ is the $N-1$ vector contracted with the Killing metric \eqref{eq:Killing-metric-SU(N)}, whose components are
    \begin{equation}
        g(\alpha+w'(\gamma))_i = \sum_{j=1}^{N-1}g_{ij} (\alpha+w'(\gamma))_j\,,\quad \textrm{with}\quad i=1,\cdots,N-1\,,
    \end{equation}
which is $2\pi \mathbb{Z}$-valued due to the delta function $\delta_{4\pi}(\alpha+w'(\gamma))$.

\section{Candidate $SU(2)$ data for the SymTFT with $k=0$}\label{app:check-of-Lagrangian-algebra-SU(2)}

In this section, we check the candidate $SU(2)$ data against the working criterion used in the main text, namely that the coefficient vector $V_{\mathcal{L}}$ should satisfy
    \begin{equation}
        S\cdot V_{\mathcal{L}}=V_{\mathcal{L}}\,,\quad T\cdot V_{\mathcal{L}}=V_{\mathcal{L}}\,,
    \end{equation}
where $S,T$ are the candidate modular kernels for the SymTFT. This appendix should therefore be read as a consistency check within the model, not as an independent categorical proof that these vectors define genuine Lagrangian algebra objects.

\subsection{Basic setup}

We first review the candidate modular kernels for the SymTFT of $SU(2)$ with $k=0$, together with several auxiliary relations used in this appendix. 

\subsubsection*{Modular matrices}
The full candidate $S$- and $T$-kernels for the SymTFT of $SU(2)$ with $k=0$ proposed in \cite{Jia:2025vrj} are
\begin{itemize}
    \item When $0<\alpha,\beta<2\pi$ the $S$-matrix is
    \begin{equation}
        S_{([g],n),([h],m)} = e^{-\frac{i}{2}m\alpha-\frac{i}{2}n\beta} + e^{\frac{i}{2}m\alpha+\frac{i}{2}n\beta} = 2\cos\frac{1}{2}(m\alpha+n\beta)\,,
    \end{equation}
    and $T$-matrix is
    \begin{equation}
        T_{([g],n),([h],m)} = e^{\frac{i}{2}n \alpha} \delta_{4\pi}(\alpha-\beta) \delta_{n,m}\,.
    \end{equation}
    \item When $0<\alpha<2\pi$ and $\beta=0,2\pi$, the $S$-matrix is
        \begin{equation}
            S_{([g],n),(0,j)} = \frac{\sin \frac{j+1}{2} \alpha}{\sin \frac{\alpha}{2}}\,,\quad S_{([g],n),(2\pi,j)} = (-1)^{n} \frac{\sin \frac{j+1}{2} \alpha}{\sin \frac{\alpha}{2}}\,,
        \end{equation}
    with $j=0,1,2,\cdots$. The $T$-matrices vanish.
    \item When $0<\beta<2\pi$ and $\alpha=0,2\pi$, the $S$-matrices are
        \begin{equation}
            S_{(0,j),([h],n)} = \frac{\sin \frac{j+1}{2} \beta}{\sin \frac{\beta}{2}}\,,\quad S_{(2\pi,j),([h],n)} = (-1)^n\frac{\sin \frac{j+1}{2} \beta}{\sin \frac{\beta}{2}}\,,
        \end{equation}
    and $T$-matrices vanish.
    \item When both $\alpha,\beta=0,2\pi$, we have
        \begin{equation}
            \begin{gathered}
            S_{(0,j),(0,j')} = (j+1)(j'+1)\,,\quad S_{(2\pi,j),(0,j')} = (-1)^{j'}(j+1)(j'+1)\,,\\
            S_{(0,j),(2\pi,j')} = (-1)^{j}(j+1)(j'+1)\,,\quad S_{(2\pi,j),(2\pi,j')} = (-1)^{j+j'}(j+1)(j'+1)\,,
            \end{gathered}
        \end{equation}
    and the non-trivial $T$-matrices are
        \begin{equation}
            T_{(0,j),(0,j')} = \delta_{j,j'}\,, \quad T_{(2\pi,j),(2\pi,j')} = \delta_{j,j'}\,.
        \end{equation}
\end{itemize}

\subsubsection*{SU(2) characters and orthogonal relation}
The $SU(2)$ characters are given by
    \begin{equation}
        \chi_j(\alpha) = \frac{\sin\frac{j+1}{2}\alpha}{\sin\frac{\alpha}{2}}\,, \quad \textrm{with}\quad j=0,1,2,\cdots\,,
    \end{equation}
and one can check the orthogonal relation
    \begin{equation}\label{eq:su(2)-character-orthogonal-relation}
        \sum_{j \geq 0}\chi_j(\alpha) \chi_j(\beta) = \frac{1}{2\mu(\alpha)} \left(\delta_{2\pi} \left(\frac{\alpha}{2}-\frac{\beta}{2} \right) +\delta_{2\pi} \left(\frac{\alpha}{2}+\frac{\beta}{2} \right) \right)\,,
    \end{equation}
with $\mu(\alpha) = \frac{ \sin^2\frac{\alpha}{2}}{\pi}$ the measure of conjugacy class such that the group volume is normalized as one
    \begin{equation}
        \int_{0}^{2\pi} \mu(\alpha) d\alpha =1\,.
    \end{equation}
The expression is subtle when $\alpha,\beta$ approach $0$ or $2\pi$. If $[g],[h]$ are generic so that $\alpha,\beta \in (0,2\pi)$, the second delta function drops out and we have
    \begin{equation}\label{eq:SU(2)-character-orthogonal-relation}
        \sum_{j \geq 0}\chi_j(\alpha) \chi_j(\beta) = \frac{1}{2\mu(\alpha)} \delta_{2\pi} \left(\frac{\alpha}{2}-\frac{\beta}{2} \right) \,,\quad \alpha,\beta \in (0,2\pi)\,.
    \end{equation}
However, if one set $\alpha=\beta=0$ at the beginning, one has formally
    \begin{equation}
        \sum_{j \geq 0}\chi_j(0) \chi_j(0) = \frac{1}{2\mu(0)} \delta_{2\pi}(0) \times 2\,,
    \end{equation}
which is twice \eqref{eq:SU(2)-character-orthogonal-relation} after taking the limit $\alpha,\beta\rightarrow 0$. The reason is that the delta function $\delta_{2\pi}(\alpha)$ only contributes half at the boundary $\alpha=0,2\pi$. To show that, consider
    \begin{equation}
        \sum_{j\geq 0} \int_0^{2\pi}\mu(\alpha) d\alpha \chi_j(\alpha) \chi_j(\beta) = \sum_{j\geq 0} \delta_{j,0} \chi_j(\beta)=\chi_0(\beta)\xrightarrow{\beta\rightarrow 0} 1\,.
    \end{equation}
On the other hand, if we first set $\beta=0$, one also has
    \begin{equation}
         \int_0^{2\pi}\mu(\alpha) d\alpha \sum_{j\geq 0}\chi_j(\alpha) \chi_j(0) = \int_0^{2\pi} \mu(\alpha) d \alpha \frac{1}{2\mu(\alpha)} \delta\left(\frac{\alpha}{2}\right)\times 2=2\times\int_0^{4\pi} d\alpha \delta(\alpha)\,,
    \end{equation}
and one must have $\int_0^{4\pi} d\alpha \delta(\alpha)=1/2$. In the following, we will regularize the boundary at $\alpha=0,2\pi$ to $\alpha=\epsilon_+,2\pi-\epsilon_+$ for infinitesimal $\epsilon_+>0$, so that $\int_0^{2\pi} \delta(\alpha) d\alpha \equiv 1$ and we can extend \eqref{eq:SU(2)-character-orthogonal-relation} to $\alpha,\beta=0,2\pi$.

Before ending, we sketch a proof for the relation \eqref{eq:su(2)-character-orthogonal-relation}. Begin with
    \begin{equation}
        \sum_{j\geq 0}\chi_j(\alpha) \chi_j (\beta) = \frac{\sin \frac{j+1}{2}\alpha \sin\frac{j+1}{2}\beta}{\sin \frac{\alpha}{2}\sin \frac{\beta}{2}} =\frac{1}{2\sin \frac{\alpha}{2}\sin \frac{\beta}{2}} \sum_{j \geq 0}\left( \cos \frac{j+1}{2}(\alpha-\beta)  - \cos \frac{j+1}{2}(\alpha+\beta) \right)\,,
    \end{equation}
and for each cosine factor $\cos \frac{j+1}{2}\theta$, one has
    \begin{equation}
        \sum_{j\geq 0}\cos \frac{j+1}{2} \theta = \frac{1}{2}\sum_{j \geq 1} \left(e^{i\frac{j \theta}{2}} + e^{i\frac{j \theta}{2}} \right) = \frac{1}{2} \sum_{j \in \mathbb{Z}} e^{i \frac{j \theta}{2}} -\frac{1}{2} = \pi \delta_{2\pi}\left(\frac{1}{2}\theta\right) - \frac{1}{2}\,,
    \end{equation}
so that
    \begin{equation}
    \begin{split}
        \sum_{j\geq 0}\chi_j(\alpha) \chi_j (\beta)=& \frac{\pi}{2\sin \frac{\alpha}{2}\sin \frac{\beta}{2}} \left(\delta_{2\pi} \left(\frac{\alpha}{2}-\frac{\beta}{2} \right) -\delta_{2\pi} \left(\frac{\alpha}{2}+\frac{\beta}{2} \right) \right)\\
        =& \frac{\pi}{2\left(\sin \frac{\alpha}{2}\right)^2} \left(\delta_{2\pi} \left(\frac{\alpha}{2}-\frac{\beta}{2} \right) +\delta_{2\pi} \left(\frac{\alpha}{2}+\frac{\beta}{2} \right) \right)\,,      
    \end{split}
    \end{equation}
where we used $\sin \frac{-\alpha}{2}=-\sin \frac{\alpha}{2}$ in the last step.

\subsubsection*{Measure of integral}

In the following, we will consider the integral formally written as
    \begin{equation}
        (S\cdot V)_{([g],n)}=\int \rho([h]) d[h] \sum_{m} S_{([g],n),([h],m)} V_{([h],m)}\,,
    \end{equation}
where $\rho([h])$ is the measure and $m$ depends on the choice of $[h]$. We will fix the measure $\rho([h]) d[h]$ to be
    \begin{equation}
        \rho(h) dh = \left\{\begin{array}{l}
            \frac{1}{4\pi} d \beta\,,\quad g=1,-1\,,\\
            \frac{\sin^2 \frac{\beta}{2}}{\pi} d \beta\,, \quad g \neq 1,-1\,.
        \end{array} \right.
    \end{equation}
The reason is that, for $S_{(1,j),([h],m)}$ and $S_{(-1,j),([h],m)}$, the centralizers of $g=1,-1$ are the whole $SU(2)$, therefore the summation over conjugacy class $[h]$ should be considered as reduced from the measure of $SU(2)$. On the other hand, for generic $[g]$ the centralizer is $U(1)$ so that $[h]$ is reduced from the measure of $U(1)$.

\subsection{Candidate $SU(2)$ data for the SymTFT with $k=0$}
There are several types of candidate data, following the proposals of \cite{Jia:2025vrj}. Up to coefficients that should be properly chosen within the working criterion, they are presented as follows. The Dirichlet-type one is
    \begin{equation}
        \mathcal{L}_{\textrm{Dir}} = \bigoplus_{j \geq 0} W_{(0,j)}\,,
    \end{equation}
and the Neumann one is
    \begin{equation}
        \mathcal{L}_{\textrm{Neu}} = \bigoplus_{\alpha\in [0,2\pi]} W_{(\alpha,0)}\,,
    \end{equation}
and the SO(3) type is
    \begin{equation}
        \mathcal{L}_{\textrm{SO(3)}} = \bigoplus_{j \geq 0} \left( W_{(0,2j)} \oplus W_{(2\pi,2j)}\right)\,,
    \end{equation}
and the $A_q$-types are
    \begin{equation}
        \mathcal{L}_{A_q} = \bigoplus_m \bigoplus_{n=0}^{\left[ \frac{q}{2}\right]} W_{\left(\frac{4\pi}{q}n,q m \right)}\,,
    \end{equation}
and the SO(3) type is a special case for $q=2$.

It is easy to see that all components in these candidate data have topological spin zero. In the following, we will fix the coefficients between each component, encoded in the coefficient vector $V_{\mathcal{L}}$, and then examine the $S$-matrix condition $S\cdot V_{\mathcal{L}} = V_{\mathcal{L}}$ 

\subsubsection*{Dirichlet-type data: $\mathcal{L}_{\rm Dir}$}
First, let us consider $\mathcal{L}_{\rm Dir}$ given by
    \begin{equation}
        \mathcal{L}_{\textrm{Dir}} = \bigoplus_{j \geq 0} (j+1) W_{(0,j)}\,,
    \end{equation}
where the coefficients are chosen to be the dimension of the irreducible representations labeled by $j\in \mathbb{Z}_{\geq 0}$, and we propose the coefficient vector $V_{\rm Dir}$ as
    \begin{equation}
        (V_{\rm Dir})_{(0,j)} = 4\pi \delta\left(\alpha\right){\large |}_{\alpha=0}  (j+1)\,,
    \end{equation}
where the delta function is necessary to restrict the non-zero components of $V_{\rm Dir}$ to a single point $\alpha=0$.

When $\alpha \neq 0,2\pi$, one has
    \begin{equation}
        (S\cdot V_{\mathcal{L}_{\rm Dir}})_{([g],n)} = \sum_{j\geq 0}(j+1)\frac{\sin \frac{j+1}{2} \alpha}{\sin \frac{\alpha}{2}} \sim \sum_{j}\chi_j(0) \chi_j(\alpha)=0\,,
    \end{equation}
and when $\alpha=2\pi$, one has
    \begin{equation}
        (S\cdot V_{\mathcal{L}_{\rm Dir}})_{(2\pi,j)} = 4\pi\mu(0)\sum_{j'\geq 0} (-1)^{j'}(j+1)(j'+1)(j'+1) = 4\pi\mu(0)(j+1) \sum_{j'}\chi_{j'}(0) \chi_{j'}(2\pi) = 0\,,
    \end{equation}
and when $\alpha = 0$, one has
    \begin{equation}
        (S\cdot V_{\mathcal{L}_{\rm Dir}})_{(0,j)} = 4\pi\mu(0)\sum_{j'\geq 0} (j+1)(j'+1)(j'+1) =4\pi\mu(0)(j+1) \sum_{j'}\chi_{j'}(0) \chi_{j'}(0)\,,
    \end{equation}
where the summation gives
    \begin{equation}
        \sum_{j'}\chi_{j'}(0) \chi_{j'}(0) = \frac{1}{2\mu(0)} \delta\left(\frac{0}{2}\right) = \frac{1}{\mu(0)} \delta(0)\,,
    \end{equation}
so that we have
    \begin{equation}
        (S\cdot V_{\mathcal{L}_{\rm Dir}})_{(0,j)} = 4\pi\delta\left(\alpha\right){\large |}_{\alpha=0}  (j+1)\,,
    \end{equation}
where we assume two zero factors $\mu(0)$, both coming from the singular point of the group measure, will cancel with each other.

\subsubsection*{Neumann-type data: $\mathcal{L}_{\rm Neu}$}

Then we consider $\mathcal{L}_{\textrm{Neu}}$ given by
    \begin{equation}
        \mathcal{L}_{\textrm{Neu}} = \bigoplus_{\alpha\in [0,2\pi]} W_{(\alpha,0)}\,,
    \end{equation}
where we assume the coefficients are uniformly distributed over the conjugacy classes. The corresponding coefficient vector is
    \begin{equation}
        (V_{\mathcal{L}_{\textrm{Neu}}})_{[g],n} = \left\{\begin{array}{l}
             \delta_{j,0}\,, \quad \alpha=0\,,2\pi\,,\\
            \delta_{n,0}\,, \quad 0<\alpha<2\pi\,.
        \end{array} \right. 
    \end{equation}

 When $\alpha\neq 0,2\pi$, one has
    \begin{equation}
        (S\cdot V_{\mathcal{L}_{\rm Neu}})_{([g],n)} = \frac{1}{4\pi}\int_0^{2\pi} d\beta (e^{-\frac{i}{2}n \beta}+e^{\frac{i}{2}n\beta}) = \delta_{n,0}\,,
    \end{equation}
where we ignore the contributions from $\beta=0,2\pi$ since their measures are zero. For $\alpha =0$, one has
    \begin{equation}
        (S\cdot V_{\mathcal{L}_{\rm Neu}})_{(0,j)} = \int_0^{2\pi} \left(\frac{1}{\pi}\sin^2\frac{\beta}{2} \right) d\beta \frac{\sin \frac{j+1}{2} \beta}{\sin \frac{\beta}{2}} = \delta_{j,0}\,,
    \end{equation}
and for $\alpha=2\pi$, one similarly has
    \begin{equation}
        (S\cdot V_{\mathcal{L}_{\rm Neu}})_{(2\pi,j)} = \int_0^{2\pi} \left(\frac{1}{\pi}\sin^2\frac{\beta}{2} \right) d\beta \frac{\sin \frac{j+1}{2} \beta}{\sin \frac{\beta}{2}} = \delta_{j,0}\,.
    \end{equation}

\subsubsection*{Data for gauging the $\mathbb{Z}_q$ subgroup: $\mathcal{L}_{A_q}$}
Next we move to $\mathcal{L}_{\rm SO(3)}$ and $\mathcal{L}_{q}$. Since the $\mathcal{L}_{\rm SO(3)}$ is the special case of $\mathcal{L}_{A_q}$ for $q=2$, we will mainly focus on $\mathcal{L}_q$. 

Let us focus on $n=0$ and $n=\frac{q}{2}$ (if $q$ is even) to fix the coefficients before $W_{(0,j)}$ and $W_{(2\pi,j)}$. When the symmetry category $\mathcal{C}$ is finite, each simple object in a genuine Lagrangian algebra corresponds to an irreducible representation of the symmetry category $\mathscr{C}$, and the corresponding coefficient is the representation dimension. In the present model we use the same rule as a motivated ansatz for continuous $G$. For $W_{(0,j)}$, the corresponding representation has dimension $j+1$ and we list contribution to character for each component in Table~\ref{Table:Z3-gauge-table} for $q=3$.
\begin{table}[!h]
    \centering
    \begin{tabular}{c|c|c}
        $j$ & Characters $\chi_j(\alpha)$ & $d_3(j)$ \\
        0 & {\color{red}{1}} & 1\\
        1 & $e^{\frac{i}{2}\alpha} + e^{-\frac{i}{2}\alpha}$ & 0\\
        2 & $e^{i \alpha} +{\color{red}{1}} + e^{-i\alpha}$ & 1\\
        3 & ${\color{red}{e^{\frac{3i}{2}\alpha}}} +e^{\frac{i}{2}\alpha} + e^{-\frac{i}{2}\alpha}+{\color{red}{e^{-\frac{3}{2}i\alpha}}}$ & 2\\
        4&$e^{2i\alpha}+e^{i \alpha} +{\color{red}{1}} + e^{-i\alpha}+e^{-2i\alpha}$&1\\
        5&$e^{\frac{5i}{2}\alpha}+{\color{red}{e^{\frac{3i}{2}\alpha}}} +e^{\frac{i}{2}\alpha} + e^{-\frac{i}{2}\alpha}+{\color{red}{e^{-\frac{3}{2}i\alpha}}}+e^{-\frac{5i}{2}\alpha}$&2\\
        6&${\color{red}{e^{3i\alpha}}}+e^{2i\alpha}+e^{i \alpha} +{\color{red}{1}} + e^{-i\alpha}+e^{-2i\alpha}+{\color{red}{e^{-3i\alpha}}}$& 3\\
        7&$e^{\frac{7i}{2}\alpha}+e^{\frac{5i}{2}\alpha}+{\color{red}{e^{\frac{3i}{2}\alpha}}} +e^{\frac{i}{2}\alpha} + e^{-\frac{i}{2}\alpha}+{\color{red}{e^{-\frac{3}{2}i\alpha}}}+e^{-\frac{5i}{2}\alpha}+e^{-\frac{7i}{2}\alpha}$&2\\
        8&$e^{4i\alpha}+{\color{red}{e^{3i\alpha}}}+e^{2i\alpha}+e^{i \alpha} +{\color{red}{1}} + e^{-i\alpha}+e^{-2i\alpha}+{\color{red}{e^{-3i\alpha}}}+e^{-4i\alpha}$&3\\
        9&${\color{red}{e^{\frac{9i}{2}\alpha}}}+e^{\frac{7i}{2}\alpha}+e^{\frac{5i}{2}\alpha}+{\color{red}{e^{\frac{3i}{2}\alpha}}} +e^{\frac{i}{2}\alpha} + e^{-\frac{i}{2}\alpha}+{\color{red}{e^{-\frac{3}{2}i\alpha}}}+e^{-\frac{5i}{2}\alpha}+e^{-\frac{7i}{2}\alpha}+{\color{red}{e^{-\frac{9i}{2}\alpha}}}$&4\\
        $\cdots$ & $\cdots$ & $\cdots$
    \end{tabular}
    \caption{The invariant components for $q=3$}
    \label{Table:Z3-gauge-table}
\end{table}
Since $\mathcal{L}_q$ is related to gauging a $\mathbb{Z}_q$ subgroup of $SU(2)$, we expect the surviving components to be those invariant under the $\mathbb{Z}_q$. For example, when $q=3$, only the components colored in red in Table~\ref{Table:Z3-gauge-table} are invariant under the $\mathbb{Z}_3$ transformation, and the total number of invariant components is collected in an integer-valued function $d_q(j)$. And we will propose
    \begin{equation}
        \mathcal{L}_{q} = \left(\bigoplus_{j\geq 0} d_q(j)W_{(0,j)}\right) \oplus \left(\bigoplus_m \bigoplus_{n=1}^{\left[ \frac{q}{2}\right]} 2W_{\left(\frac{4\pi}{q}n,q m \right)}\right)\,,
    \end{equation}
for odd $q$ and
    \begin{equation}
        \mathcal{L}_{q} = \left(\bigoplus_{j\geq 0} d_q(j)W_{(0,j)}\right) \oplus \left(\bigoplus_m \bigoplus_{n=1}^{\left[ \frac{q}{2}\right]-1} 2W_{\left(\frac{4\pi}{q}n,q m \right)}\right)\oplus \left(\bigoplus_{j\geq 0} d_q(j)W_{(2\pi,j)}\right)\,,
    \end{equation}
for even $q$. In fact, $d_q(j)$ can be evaluated as
    \begin{equation}
        d_q(j) = \frac{1}{q}\sum_{n=0}^{q-1} \chi_j\left(\frac{4\pi}{q} n\right)\,,
    \end{equation}
and the reason is the following. For a single term $e^{\frac{ik}{2}\alpha}$ in the character, one has
    \begin{equation}
        \frac{1}{q}\sum_{n=0}^{q-1} e^{\frac{ik}{2}\left(\frac{4\pi}{q} \right)n} = \frac{1-e^{2\pi i k}}{1-e^{2\pi i \frac{k}{q}}}\,,
    \end{equation}
which vanishes unless $k=q\mathbb{Z}$, and we have
    \begin{equation}
        \frac{1}{q}\sum_{n=0}^{q-1} e^{\frac{ik}{2}\left(\frac{4\pi}{q} \right)n} = \left\{\begin{array}{l}
            1\,,\quad k=q\mathbb{Z}\,,\\
            0\,,\quad \textrm{others\,.}
        \end{array} \right.
    \end{equation}
Therefore by summing over $\alpha = \frac{4\pi}{q}n$, it will select all components with $k=q\mathbb{Z}$ in the character, and thus it produce $d_q(j)$.

Let us write down $V_{\mathcal{L}_q}$. When $q$ is odd, one has
    \begin{equation}
        (V_{\mathcal{L}_q})_{([g],n)} = \left\{\begin{array}{l}
            d_q(j) 4\pi \delta(\alpha){\large |}_{\alpha=0}\,,\quad (\alpha=0)\,,\\
            \frac{2}{\mu(\alpha)}\sum_{r=1}^{\left[\frac{q}{2}\right]}\delta(\alpha-\frac{4\pi}{q}r) \sum_k\delta_{n,qk}\,, \quad (\alpha\neq 0)\,,
        \end{array} \right. \qquad (q\textrm{ odd})
    \end{equation}
where we insert a normalization factor $2/\mu(\alpha)$ in the second line in order to produce the Lagrangian algebra based on \eqref{eq:L_Direct_Integral}. Similarly, when $q$ is even, one has
    \begin{equation}
        (V_{\mathcal{L}_q})_{([g],n)} = \left\{\begin{array}{l}
            d_q(j) 4\pi\delta(\alpha){\large |}_{\alpha=0}\,,\quad (\alpha=0)\,,\\
            \frac{2}{\mu(\alpha)}\sum_{r=1}^{\left[\frac{q}{2}\right]-1}\delta(\alpha-\frac{4\pi}{q}r)\sum_k \delta_{n,qk}\,, \quad (0<\alpha<2\pi)\,,\\
            d_q(j) 4\pi\delta(\alpha-2\pi){\large |}_{\alpha=2\pi}\,,\quad (\alpha=2\pi)
        \end{array} \right. \qquad (q\textrm{ even})
    \end{equation}
where we include the contribution from $\alpha=2\pi$.

Then let us check those coefficients vectors are invariant under the action of $S$-matrices. We first assume $\alpha \neq 0,2\pi$. When $q$ is odd, then $S\cdot V_{\mathcal{L}_q}$ has two part
    \begin{equation}
        (S\cdot V_{\mathcal{L}_q})_{([g],n)}= \left(\sum_{j}d_q(j)\frac{\sin \frac{j+1}{2} \alpha}{\sin \frac{\alpha}{2}}\right) + \left(\frac{1}{2\pi\mu(\alpha)}\sum_{m'=1}^{\left[\frac{q}{2}\right]}\sum_m e^{-\frac{i}{2}q m\alpha -\frac{i}{2}n \frac{4\pi}{q}m'}+e^{\frac{i}{2}q m\alpha +\frac{i}{2}n \frac{4\pi}{q}m'}\right)\,,
    \end{equation}
where the two contributions are separately from $\beta=0$ and $\beta\neq 0$. The first term reads
    \begin{equation}
    \begin{split}
        &\left(\sum_{j}d_q(j)\frac{\sin \frac{j+1}{2} \alpha}{\sin \frac{\alpha}{2}}\right)=\frac{1}{q}\sum_{n=0}^{q-1}\sum_j \chi_j(\frac{4\pi}{q}n) \chi_j (\alpha)\\
        =&\frac{1}{q}\sum_{n=0}^{q-1}\frac{1}{2\mu(\alpha)}\left(\delta_{2\pi}\left(\frac{\alpha}{2}-\frac{2\pi}{q}n \right) +\delta_{2\pi}\left(\frac{\alpha}{2}+\frac{2\pi}{q}n \right) \right)\,,\\
        =&\frac{1}{q}\sum_{n=0}^{q-1}\frac{1}{2\mu(\alpha)}\left(\delta_{2\pi}\left(\frac{\alpha}{2}-\frac{2\pi}{q}n \right) +\delta_{2\pi}\left(\frac{\alpha}{2}-\frac{2\pi}{q}(q-n) \right) \right)\,.
    \end{split}
    \end{equation}
Since we restrict $\alpha \in (0,2\pi)$, it then becomes
\begin{equation}
    \frac{1}{q}\sum_{n=1}^{\left[\frac{q}{2}\right]}\frac{1}{\mu(\alpha)}\delta\left(\frac{\alpha}{2}-\frac{2\pi}{q}n \right)\,.
\end{equation}
The second term read
\begin{equation}
    \begin{split}
        &\frac{1}{2\pi\mu(\alpha)}\sum_{m'=1}^{\left[\frac{q}{2}\right]}\sum_m e^{-\frac{i}{2}q m\alpha -\frac{i}{2}n \frac{4\pi}{q}m'}+e^{\frac{i}{2}q m\alpha +\frac{i}{2}n \frac{4\pi}{q}m'} \\
        =&\frac{1}{\mu(\alpha)}\sum_{m'=1}^{\left[\frac{q}{2}\right]} \left( e^{-\frac{i}{2}n \frac{4\pi}{q}m'}+e^{\frac{i}{2}n \frac{4\pi}{q}m'}\right) \delta_{2\pi}(\frac{q\alpha}{2})\\
        =&\frac{1}{\mu(\alpha)}\sum_{m'=1}^{\left[\frac{q}{2}\right]} \left( e^{-\frac{i}{2}n \frac{4\pi}{q}m'}+e^{\frac{i}{2}n \frac{4\pi}{q}m'}\right) \frac{1}{q}\sum_{k=1}^{\left[\frac{q}{2}\right]} \delta\left(\frac{\alpha}{2} - \frac{2\pi}{q}k\right)\,,
    \end{split}
\end{equation}
so that the two terms combine to
\begin{equation}
\begin{split}
    &\frac{1}{q}\left(\sum_{m'=-1}^{-\left[\frac{q}{2}\right]}  e^{i\frac{2\pi n}{q}m'}+1+\sum_{m'=1}^{\left[\frac{q}{2}\right]}e^{i \frac{2\pi n}{q}m'}\right) \frac{1}{\mu(\alpha)}\sum_{r=1}^{\left[\frac{q}{2}\right]} \delta\left(\frac{\alpha}{2} - \frac{2\pi}{q}r\right)\\
    =&\frac{1}{\mu(\alpha)}\sum_{r=1}^{\left[\frac{q}{2}\right]} \delta\left(\frac{\alpha}{2} - \frac{2\pi}{q}r\right)\sum_{k}\delta_{n,qk} = \frac{2}{\mu(\alpha)}\sum_{r=1}^{\left[\frac{q}{2}\right]} \delta\left(\alpha - \frac{4\pi}{q}r\right)\sum_{k}\delta_{n,qk}\,.
\end{split}
\end{equation}
When $q$ is even, then we have three parts
    \begin{equation}
    \begin{split}
        (S\cdot V_{\mathcal{L}_q})_{([g],n)}= &\left(\sum_{j}d_q(j)\frac{\sin \frac{j+1}{2} \alpha}{\sin \frac{\alpha}{2}}\right) + \left(\frac{1}{2\pi \mu(\alpha)}\sum_{m'=1}^{\left[\frac{q}{2}\right]-1}\sum_m e^{-\frac{i}{2}q m\alpha -\frac{i}{2}n \frac{4\pi}{q}m'}+e^{\frac{i}{2}q m\alpha +\frac{i}{2}n \frac{4\pi}{q}m'}\right)\\
        &+\left(\sum_{j}(-1)^n d_q(j)\frac{\sin \frac{j+1}{2} \alpha}{\sin \frac{\alpha}{2}}\right)\,,
    \end{split}
    \end{equation}
where the last term comes from $\beta=2\pi$ and it gives
\begin{equation}
    (-1)^n \times\frac{1}{q}\sum_{n=1}^{\left[\frac{q}{2}\right]-1}\frac{1}{\mu(\alpha)}\delta\left(\frac{\alpha}{2}-\frac{2\pi}{q}n \right)\,.
\end{equation}and it combines with the rest to give 
\begin{equation}
\begin{split}
    &\frac{1}{q}\left(\sum_{m'=-1}^{-\left[\frac{q}{2}\right]+1}  e^{i\frac{2\pi n}{q}m'}+1+\sum_{m'=1}^{\left[\frac{q}{2}\right]-1}e^{i \frac{2\pi n}{q}m'} + (-1)^n\right) \frac{1}{\mu(\alpha)}\sum_{r=1}^{\left[\frac{q}{2}\right]-1} \delta\left(\frac{\alpha}{2} - \frac{2\pi}{q}r\right)\\
    =&\frac{1}{\mu(\alpha)}\sum_{r=1}^{\left[\frac{q}{2}\right]-1} \delta\left(\frac{\alpha}{2} - \frac{2\pi}{q}r\right)\sum_{k}\delta_{n,qk}=\frac{2}{\mu(\alpha)}\sum_{r=1}^{\left[\frac{q}{2}\right]-1} \delta\left(\alpha - \frac{4\pi}{q}r\right)\sum_{k}\delta_{n,qk}\,.
\end{split}
\end{equation}

Now let us move to $\alpha=0$. When $q$ is odd one has
    \begin{equation}
        (S\cdot V_{\mathcal{L}_q})_{(1,j)} = \left(4\pi \mu(0)\sum_{j'\geq 0}(j+1)(j'+1)d_q(j')\right)+ \left(2\sum_{m=q\mathbb{Z}}\sum_{k=1}^{\left[\frac{q}{2}\right]} \frac{\sin \frac{j+1}{2}\left(\frac{4\pi}{q}k\right)}{\sin \frac{\left(\frac{4\pi}{q}k\right)}{2}}\right)\,,
    \end{equation}
and the first term is
    \begin{equation}
        4\pi \mu(0)\chi_j(0) \sum_{j'} \chi_{j'}(0) d_q(j') = 4\pi \mu(0) \chi_j(0) \frac{1}{q}\frac{1}{2\mu(0)}\delta\left(\frac{\alpha}{2}\right){\large |}_{\alpha=0}= \frac{4\pi}{q}\chi_j(0) \delta\left(\alpha\right){\large |}_{\alpha=0}\,,
    \end{equation}
and the second term is
    \begin{equation}
        \left(\sum_{k=-1}^{-\left[\frac{q}{2}\right]}\chi_j\left(\frac{4\pi}{q}k\right)+\sum_{k=1}^{\left[\frac{q}{2}\right]}\chi_j\left(\frac{4\pi}{q}k\right)\right) \times \left(\sum_{m\in q\mathbb{Z}} 1 \right)\,,
    \end{equation}
where we use $\chi_j(-\alpha) = \chi_j(\alpha)$. The infinite sum $\sum_{m\in q\mathbb{Z}} 1$ can be regularized using the $U(1)$ character~\footnote{For compact Lie group $G$ we have $\delta_G(g-e) = \sum_{\rho} d_\rho \chi_\rho(g)$~\cite{Witten:1992xu}. Thus when $g = e$ we have $\delta_G(0) = \sum_\rho d_\rho^2$. For $G = U(1)$ this becomes $\sum_{m\in\mathbb{Z}} 1 = \delta_{U(1)}(0)$. Similarly, for $\sum_{m\in q\mathbb{Z}} 1$ it evaluates to the result presented in the main text.}
    \begin{equation}
        \sum_{m\in q\mathbb{Z}} 1 = \sum_{k} e^{\frac{i}{2}qk \times 0} = 2\pi \delta\left(\frac{q \times 0}{2} \right) = \frac{4\pi}{q} \delta\left(0\right)\,,
    \end{equation}
so that the second term reads
    \begin{equation}
        \frac{4\pi}{q}\left(\sum_{k=-1}^{-\left[\frac{q}{2}\right]}\chi_j\left(\frac{4\pi}{q}k\right)+\sum_{k=1}^{\left[\frac{q}{2}\right]}\chi_j\left(\frac{4\pi}{q}k\right)\right) \delta \left(\alpha\right){\large |}_{\alpha=0}\,.
    \end{equation}
Combine the two terms, we get
    \begin{equation}
    \frac{4\pi}{q}\left(\sum_{k=-1}^{-\left[\frac{q}{2}\right]}\chi_j\left(\frac{4\pi}{q}k\right)+1+\sum_{k=1}^{\left[\frac{q}{2}\right]}\chi_j\left(\frac{4\pi}{q}k\right)\right) \delta \left(\alpha\right){\large |}_{\alpha=0}=4\pi d_q(j) \delta(\alpha){\large |}_{\alpha=0} \,.
    \end{equation}
When $q$ is even, one has
    \begin{equation}
    \begin{split}
        (S\cdot V_{\mathcal{L}_q})_{(1,j)} =& \left(4\pi\mu(0)\sum_{j'\geq 0}(j+1)(j'+1)d_q(j')\right)+ \left(2\sum_{m}\sum_{k=1}^{\left[\frac{q}{2}\right]-1} \frac{\sin \frac{j+1}{2}\left(\frac{4\pi}{q}k\right)}{\sin \frac{\left(\frac{4\pi}{q}k\right)}{2}}\right)\\
        &+\left(4\pi\mu(2\pi)\sum_{j'\geq 0}(-1)^j(j+1)(j'+1)d_q(j')\right)\,,
    \end{split}
    \end{equation}
where the third term is from $\beta= 2\pi$, which gives
    \begin{equation}
        4\pi \mu(2\pi)\chi_j(2\pi) \sum_{j'} \chi_{j'}(0) d_q(j') = 4\pi \mu(2\pi) \chi_j(0) \frac{1}{q}\frac{1}{2\mu(0)}\delta\left(\frac{\alpha}{2}\right){\large |}_{\alpha=0}= \frac{4\pi}{q}\chi_j(2\pi) \delta\left(\alpha\right){\large |}_{\alpha=0}\,,
    \end{equation}
where we use $\mu(\alpha) = \mu(2\pi-\alpha)$. The total contribution is again
    \begin{equation}
    \frac{4\pi}{q}\left(\sum_{k=-1}^{-\left[\frac{q}{2}\right]+1}\chi_j\left(\frac{4\pi}{q}k\right)+1+\sum_{k=1}^{\left[\frac{q}{2}\right]-1}\chi_j\left(\frac{4\pi}{q}k\right)+\chi_j(2\pi)\right) \delta \left(\alpha\right){\large |}_{\alpha=0}=4\pi d_q(j) \delta(\alpha){\large |}_{\alpha=0} \,.
    \end{equation}

Finally, we set $\alpha=2\pi$. When $q$ is odd one has
    \begin{equation}
        (S\cdot V_{\mathcal{L}_q})_{(2\pi,j)} = \left(4\pi \mu(0)\sum_{j'\geq 0}(j+1)(-1)^{j'}(j'+1)d_q(j')\right)+ \left(2\sum_{m}(-1)^{qm}\sum_{k=1}^{\left[\frac{q}{2}\right]} \frac{\sin \frac{j+1}{2}\left(\frac{4\pi}{q}k\right)}{\sin \frac{\left(\frac{4\pi}{q}k\right)}{2}}\right)\,,
    \end{equation}
the first term is
    \begin{equation}
        4\pi \mu(0)\chi_j(0) \sum_{j'} \chi_{j'}(2\pi) d_q(j') = 0\,,
    \end{equation}
since $\chi_{j'}(2\pi)$ is orthogonal to all components in $d_q(j')$ when $q$ is odd. The second term is also zero since
    \begin{equation}
        \sum_m (-1)^{qm} = \sum_m e^{\frac{i}{2} qm \times 2\pi} =\delta_{2\pi} (\pi)=0\,,
    \end{equation}
therefore $(S\cdot V_{\mathcal{L}_q})_{(2\pi,j)}=0$. On the other hand, when $q$ is even, one has
    \begin{equation}
    \begin{split}
        (S\cdot V_{\mathcal{L}_q})_{(2\pi,j)} =& \left(4\pi \mu(0)\sum_{j'\geq 0}(j+1)(-1)^{j'}(j'+1)d_q(j')\right)+ \left(2\sum_{m}(-1)^{qm}\sum_{k=1}^{\left[\frac{q}{2}\right]-1} \frac{\sin \frac{j+1}{2}\left(\frac{4\pi}{q}k\right)}{\sin \frac{\left(\frac{4\pi}{q}k\right)}{2}}\right)\\
        &+\left(4\pi \mu(2\pi)\sum_{j'\geq 0}(-1)^{j+j'}(j+1)(j'+1)d_q(j')\right)\,.
    \end{split}
    \end{equation}
Now since $\chi_{j'}(2\pi)$ is inside $d_q(j')$, then the first term gives
    \begin{equation}
    \begin{split}
        &4\pi \mu(0)\chi_j(0) \sum_{j'} \chi_{j'}(2\pi) d_q(j')\\
        =& 4\pi \mu(0) \chi_j(0) \frac{1}{q} \frac{1}{2\mu(2\pi)} \delta \left(\frac{\alpha-2\pi}{2}\right){\large |}_{\alpha=2\pi} = 4\pi \chi_j(0) \delta(\alpha-2\pi){\large |}_{\alpha=2\pi}\,,
    \end{split}
    \end{equation}
and the third term is similarly
    \begin{equation}
    \begin{split}
        &4\pi \mu(2\pi)\chi_j(2\pi) \sum_{j'} \chi_{j'}(2\pi) d_q(j')\\
        =& 4\pi \mu(2\pi) \chi_j(2\pi) \frac{1}{q} \frac{1}{2\mu(2\pi)} \delta \left(\frac{\alpha-2\pi}{2}\right){\large |}_{\alpha=2\pi} = 4\pi \chi_j(2\pi) \delta(\alpha-2\pi){\large |}_{\alpha=2\pi}\,.
    \end{split}
    \end{equation}
The middle term is
    \begin{equation}
        \left(\sum_{k=-1}^{-\left[\frac{q}{2}\right]+1}\chi_j\left(\frac{4\pi}{q}k\right)+\sum_{k=1}^{\left[\frac{q}{2}\right]-1}\chi_j\left(\frac{4\pi}{q}k\right)\right) \times \left(\sum_{m\in \mathbb{Z}} (-1)^{qm} \right)\,,
    \end{equation}
where one has
    \begin{equation}
        \sum_{m\in \mathbb{Z}} (-1)^{qm} = \sum_m 1 = \frac{4\pi}{q}\delta(0) = \frac{4\pi}{q} \delta(\alpha-2\pi){\large |}_{\alpha=2\pi}\,,
    \end{equation}
and we use the fact that $q$ is even. The total contribution is then
    \begin{equation}
    \begin{split}
    &\frac{4\pi}{q}\left(\sum_{k=-1}^{-\left[\frac{q}{2}\right]+1}\chi_j\left(\frac{4\pi}{q}k\right)+1+\sum_{k=1}^{\left[\frac{q}{2}\right]-1}\chi_j\left(\frac{4\pi}{q}k\right)+\chi_j(2\pi)\right) \delta \left(\alpha-2\pi\right){\large |}_{\alpha=2\pi}\\
    =&4\pi d_q(j) \delta(\alpha-2\pi){\large |}_{\alpha=2\pi} \,.
    \end{split}
    \end{equation}


\bibliographystyle{JHEP}
\bibliography{biblio.bib}






\end{document}